\documentclass{aa}  

\bibpunct{(}{)}{;}{a}{}{,} 

\usepackage{graphicx}
\usepackage{xcolor}
\usepackage{longtable}
\usepackage{txfonts}
\usepackage{soul}
\usepackage{hyperref}

\begin{document} 

   \title{Hubble Asteroid Hunter}

   \subtitle{II. Identifying strong gravitational lenses in HST images with crowdsourcing}

   \author{Emily O. Garvin\inst{1}\thanks{e-mail: egarvin@phys.ethz.ch}, Sandor Kruk \inst{2,3}\thanks{e-mail: kruksandor@gmail.com}, Claude Cornen \inst{4}, Rachana Bhatawdekar  \inst{3},  Raoul Ca\~nameras \inst{5}, Bruno Mer\'{\i}n \inst{6}
          }

   \institute{Institute for Particle Physics and Astrophysics, ETH Zürich, Wolfang-Pauli-Strasse 27, 8093 Zürich, Switzerland
   \and
    Max-Planck-Institut für Extraterrestrische Physik (MPE), Giessenbachstrasse 1, D-85748 Garching bei München, Germany
   \and
     European Space Agency (ESA), European Space Research and Technology Centre (ESTEC), Keplerlaan 1, 2201 AZ Noordwijk, The Netherlands
     \and
  Citizen Scientist, Zooniverse, Astrophysics Sub-department, University of Oxford, Keble Road, Oxford, OX1 3NP, UK
  \and
    Max-Planck-Institut für Astrophysik (MPA), Karl-Schwarzschild-Straße 1, D-85748 Garching bei München, Germany  
    \and   
  European Space Agency (ESA), European Space Astronomy Centre (ESAC), Camino Bajo del Castillo s/n, 28692 Villanueva de la Cañada, Madrid, Spain}


\titlerunning{Hubble Asteroid Hunter (HAH) : Strong gravitational lenses in {\it Hubble Space Telescope} images}
\authorrunning{Garvin et al.}

   \date{Received April 9, 2022; revised June 10, 2022; accepted June 28, 2022}

 
 \abstract
  {The {\it Hubble Space Telescope} (HST) archives constitute a rich dataset of high resolution images to mine for strong gravitational lenses. While many HST programs specifically target strong lenses, they can also be present by coincidence in other HST observations.}
   {We aim to identify non-targeted strong gravitational lenses in almost two decades of images from the ESA {\it Hubble Space Telescope} archive (eHST), without any prior selection on the lens properties.}
   {We used crowdsourcing on the Hubble Asteroid Hunter (HAH) citizen science project to identify strong lenses, alongside asteroid trails, in publicly available large field-of-view HST images. We visually inspected 2\,354 objects tagged by citizen scientists as strong lenses to clean the sample and identify the genuine lenses.}
   {We report the detection of 252 strong gravitational lens candidates, which were not the primary targets of the HST observations. 198 of them are new, not previously reported by other studies, consisting of 45 A grades, 74 B grades and 79 C grades. The majority are galaxy-galaxy configurations. The newly detected lenses are, on average, 1.3 magnitudes fainter than previous HST searches. This sample of strong lenses with high resolution HST imaging is ideal to follow-up with spectroscopy, for lens modelling and scientific analyses.}
   {This paper presents an unbiased search of lenses, which enabled us to find a high variety of lens configurations, including exotic lenses. We demonstrate the power of crowdsourcing in visually identifying strong lenses and the benefits of exploring large archival datasets. This study shows the potential of using crowdsourcing in combination with artificial intelligence for the detection and validation of strong lenses in future large-scale surveys such as ESA's future mission {\it Euclid} or in JWST archival images.}

   \keywords{Gravitational lensing: strong -- Galaxies: general -- Galaxies: individual}

   \maketitle
%

\section{Introduction}

Strong gravitational lensing is a powerful tool that has multiple applications in astrophysics and cosmology. Strong lenses are used to study the dark matter content and distribution in galaxies \citep[][]{Koopmans2006,Barnabe2009,Sonnenfeld2015,Oldham2018} and clusters \citep[][]{Richard2010,Jauzac2015,Caminha2019}, to study distant galaxies \citep[][]{Marshall2007,Dessauges2015,Swinbank2015,Canameras2017}, and to constrain cosmological parameters such as the Hubble constant and the dark energy equation of state from time delay observations \citep[e.g.][]{Suyu2010,Suyu2014,Sereno2014,Courbin2018,Wong2020,Millon2020}. With the help of strong gravitational lensing of massive clusters, we have also been able to probe galaxy evolution through Ultra-Violet luminosity functions (UV LF) \citep{Atek2014, Livermore2017}, stellar mass functions \citep{Bhatawdekar2019, Kikuchihara2020}, and UV slopes \citep{Bhatawdekar2021}, well into the epoch of reionisation. Strong gravitational lensing is a rare phenomenon which relies on the chance alignment of a foreground object with a large surface mass density (lens) with a bright background object (source) such as a galaxy, a quasar or a supernova \citep{kelly2015multiple}. Recently, other lensed sources have also been reported, such as magnified stars and stellar complexes \citep{welch2022star, vanzella2021muse}. Overall, since the discovery of the first gravitational lens \citep{Walsh1979} only approximately one thousand strong lenses have been confirmed.

Finding strong gravitational lenses is a difficult outlier detection problem. So far, numerous automated algorithms have been developed for the detection of strong lenses in large-scale surveys, for example through the identification of arcs and rings in multiband imaging \citep[e.g.][]{Alard2006,Marshall2009,Gavazzi2014,Sonnenfeld2018}, or through the blended signatures of lens and source galaxies in fibre spectroscopy \citep[e.g.][]{Bolton2008,Brownstein2012,Holwerda2015,Shu2017,Talbot2021}. With the latest developments of artificial intelligence, machine learning algorithms, and in particular convolutional neural networks (CNNs), have had an increasing number of applications in astronomy, from galaxy classification \citep{Dieleman2015, Huertas-Company2015, Walmsley2020, Walmsley2022} to estimating quantities such as photometric redshifts \citep{Samui2017,Disanto2018,Schuldt2021}. Strong lens searches have largely benefited from the use of CNNs \citep[e.g.][]{Bom2017,Schaefer2018,Petrillo2019a,Jacobs2019,Canameras2020,Huang2020}. The Bologna strong gravitational lens finding challenge has extensively compared the performance of these methods on simulated data resembling future ground- and space-based imaging surveys such as LSST and {\it Euclid} \citep{Metcalf2019}. While automated search algorithms are generally highly efficient on real data, they remain affected by false-positives, and they are typically complemented by visual inspection to increase the sample purity \citep[for instance, only $\simeq$1$-$3\% of candidates selected by arc-finders and CNNs are highly-probable lenses;][]{Sonnenfeld2018,Petrillo2019a}.

One distributed method that has been proven successful in the visual identification of strong lenses is citizen science. The general public has been involved in the classification of astronomical data through projects such as Galaxy Zoo \citep{Lintott2008}. The Space Warps project \citep{Marshall2016} pioneered the crowdsourced identification of strong gravitational lenses in surveys such as CFHTLS \citep{More2016} and HSC-SSP \citep{Sonnenfeld2020}. Using the Zooniverse framework, Space Warps display images on a web-based interface and ask volunteers whether a gravitational lens is present in the images. To assess the completeness and uncertainty of the classifications, Space Warps measures the performance of each user on a training set of simulated lenses, and weights the individual user contributions based on their `skill'.

These various lens search methods typically focus on specific lens configurations. On the one hand, galaxy-scale strong lens candidates identified from unresolved fibre spectra have massive foreground elliptical and Einstein radii limited by the fibre aperture sizes \citep[e.g.][]{Bolton2008,Brownstein2012}. On the other hand, searches using multiband images from wide-field, ground-based surveys are preferentially selecting candidates with wider image separations ($\geq$2\arcsec\ from the lens centre) ensuring robust lens and source deblending, and are therefore spanning both galaxy-scale and more massive group-, cluster-scale foreground halos \cite[e.g.][]{Belokurov2009,Sonnenfeld2018,Canameras2020}. Moreover, the visual grades assigned by strong lens expert and volunteers show large scatter due to the difficulty in distinguishing strong lenses from rings, spirals, mergers, and other contaminants in seeing-limited images \citep[e.g.][]{More2016,Rojas2021}.

In this study, we employed a crowdsourced approach to detect strong lenses in archival images from the {\it Hubble Space Telescope} (HST), in order to benefit from the higher angular resolutions, and to cover a broader range of lens and source galaxy types, lens potentials, and multiple image configurations. Instead of showing volunteers postage stamps of galaxies, we displayed cutouts of HST images sufficiently large to contain tens of objects. Although the original project was not designed for the detection of lenses but for the detection of asteroids, we asked the volunteers on Hubble Asteroid Hunter \footnote{\url{www.asteroidhunter.org}} (HAH) to tag possible lenses on the forum of the project, and the science team inspected all the tagged lenses. Several square degrees of HST imaging have been previously searched for the presence of lenses, for example in the HST AEGIS survey \citep{Moustakas2007}, the GEMS survey \citep{More2011}, the COSMOS survey \citep{Jackson2008,Faure2008,Pourrahmani2018}, and over 7 deg$^2$ of archival observations \citep{Pawase2014}. We extended the systematic morphological selection of strong lenses from the HST archive to a much larger area, without colour, brightness, and redshift cuts, in order to provide a statistically-significant and diverse sample of robust lens candidates for future spectroscopic follow-up and detailed modelling.

The structure of this paper is as follows. In Sect.~\ref{data}, we present the input data set, and in Sect.~\ref{method} we describe the inspection, classification, and light profile fitting. The results and properties of strong lens candidates are given in Sect.~\ref{results}. Finally, we compare with other search methods in Sect.~\ref{discussion}, and we summarise in Sect.~\ref{conclusion}. Throughout the paper we adopted the WMAP Seven-Year Cosmological parameters \citep{Jarosik2011} with ($\Omega_{\rm M},\Omega_{\Lambda},h) = (0.27,0.73,0.71)$.

\section{Data}
\label{data}

We analysed archival HST images from the Advanced Camera for Surveys Wide Field Channel (ACS/WFC) and Wide Field Camera 3 Ultraviolet and Visible (WFC3/UVIS) and Near-infrared (WFC3/IR) channels. These instruments and detectors have the largest field-of-views, thus the highest chance of containing serendipitously observed strong lenses. The observations were taken between 30 April 2002 (when the ACS camera was installed) and 24 April 2020 for ACS/WFC, and between 25 July 2009 (when the WFC3 obtained first light) and 24 April 2020 for WFC3/UVIS. This data is presented in the \citet{Kruk2022} paper. The WFC3/IR images were uploaded later to the project, after the analysis of ACS and WFC3/UVIS data was completed, and consisted of observations taken between 25 July 2009 and 1 June 2020. 

The analysis is complete with data taken and publicly available in the HST archives up to 1 June 2020. Observations based on general observer (GO) proposals are available in the HST archive one year after they were taken, therefore the last GO observation analysed were taken before June 2019. HST Snapshot observations are available in the archive immediately after they are acquired and they were analysed up to June 2020. One of the authors, citizen scientist Claude Cornen inspected the data released in the HST archives after June 2020 and found 12 new strong lens candidates. These were not included in the main analysis but presented separately in Appendix~\ref{recent}. 

In the Hubble Asteroid Hunter (HAH) project, volunteers inspected the single-band HST composite images in \textsc{PNG} format available from the European {\it Hubble Space Telescope} (eHST) archive (the same images are also available in MAST). These images were created from the \texttt{FITS} images by first applying autoscaling, which scales the image linearly from the 0.5\% pixel level to the 99.5\% pixel level (using autoscale=99.5, which is the default), and then applying a trigonometric asinh scaling.\footnote{More details on the parameters used can be found at \url{https://hla.stsci.edu/fitscutcgi_interface.html}}. 

The composite HST images were obtained by stacking individual HST dithered exposures, processed using the standard HST data processing pipeline by STScI: the exposures were aligned and processed with DrizzlePac\footnote{\url{https://hst-docs.stsci.edu/drizzpac}} \citep{Gonzaga2012} for geometric distortion corrections and cosmic ray removal.

In order to optimise the presentation of the images for the visual inspection of citizen scientists, we split the HST ACS/WFC and WFC3/UVIS composite images into four equal parts, of sizes 101\arcsec\,$\times$\,101\arcsec\ and 80\arcsec\,$\times$\,80\arcsec, respectively. For the WFC3/IR images we used the full frame with size 127\arcsec\,$\times$\,137\arcsec\ for the search. We selected all the composite HST images available in the archive based on the following criteria: an exposure time greater than 300 seconds, and a field-of-view greater than 7~arcmin$^2$ to exclude sub-frames (only for ACS/WFC and WFC3/UVIS). Occasionally, the same field was observed multiple times by HST. We showed all the images available, even though observations overlapped in the targeted region of the sky. 

In total, 145\,396 cutouts were inspected, corresponding to 45\,784 unique observations. The inspected images were in grey-scale, without colour information. The total area imaged by HST and inspected in this project is about 27~deg.$^2$. This includes the HST Cosmic Evolution Survey (COSMOS; \citealt{Koekemoer2007, Scoville2007}), as well as the CANDELS survey \citep{Grogin2011}. The estimate of the total area imaged takes into account the overlap between the HST pointings of the approximately same region of the sky with different instruments and filters. These overlapping observations were nonetheless inspected individually by the volunteers. Details on the images inspected are shown in Table \ref{data_table}.

\begin{table}
\caption{HST archival images inspected while searching for strong gravitational lenses in the Hubble Asteroid Hunter project.}
\renewcommand*{\arraystretch}{1.3}
\begin{tabular}{ccccc}
\hline
\hline
HST Instr. & FoV & Scale & Images & Cutouts \\
 & & (\arcsec/pix) & & \\
\hline
ACS/WFC & 202\arcsec\,$\times$\,202\arcsec & 0.05 & 22\,940 & 91\,760 \tabularnewline*
WFC3/UVIS & 160\arcsec\,$\times$\,160\arcsec & 0.04 & 10\,326 & 41\,118 \tabularnewline*
WFC3/IR & 123\arcsec\,$\times$\,136\arcsec & 0.13 & 12\,518 & 12\,518 \tabularnewline*
\hline
\end{tabular}
\newline
\newline
 {\footnotesize \setstretch{1} {\bf Notes.} The search contains HST observations taken and released between 2002 and June 2020, with exposure times larger than 300s. Images follow the standard HST processing pipeline and are single-band. ACS/WFC and WFC3/UVIS images were divided into 4 equal quadrants (cutouts) in order to maximise the detection of smaller arcs. WFC3/IR images were presented as they are.}
\label{data_table}
\end{table}

\section{Method}
\label{method}

\subsection{Visual identification by citizen scientists}

We identified gravitational lenses in the Hubble Asteroid Hunter citizen science project \citep{Kruk2022}, launched on 20 June 2019 and which ran until August 2020. The primary purpose of the project was to identify serendipitously observed asteroid trails in HST images while the telescope was observing targeted objects, proposed by astronomers.  In total, 11\,482 volunteers inspected the 145\,396 cutouts, each image being inspected by ten people. The project also had a dedicated forum, \textit{Talk}\footnote{\url{https://www.zooniverse.org/projects/sandorkruk/hubble-asteroid-hunter/talk}}, where the volunteers could tag cutouts containing interesting objects (with \#) and discuss about them. 

In contrast to other citizen science projects that are specifically designed to identify strong lenses (such as Space Warps) or projects that have the strong lens classification as part of their workflows (e.g. Galaxy Zoo), Hubble Asteroid Hunter did not provide a classification option for strong lenses. Instead, we asked the volunteers to specifically tag observations containing strong gravitational lenses on \textit{Talk}, including individual strong lenses (with \#gravitational\_lens) and clusters of galaxies with arcs corresponding to strong lensing (with \#cluster\_lens). For training, we provided several examples of strong gravitational lenses, based on previously identified lenses in HST images and targeted lenses, both in the Tutorial and the Field Guide of the project. The science team inspected and assessed all gravitational lenses as the project progressed, instead of analysing the classifications when the project was completed. Finally, it is worth mentioning that some of the citizen scientists on the project who tagged lenses participated in Space Warps and, therefore, have prior training and familiarity with the appearance of strong lenses. 

\subsection{Visual identification by authors}

In total, 2\,354 cutouts were tagged as containing individual strong lenses (with \#gravitational\_lens). Members of the science team (EOG, SK, CC) inspected all the cutouts, selected 417 unique gravitational lens candidates and added them to our catalogue.

\begin{figure}
    \centering
    \includegraphics[width=\columnwidth]{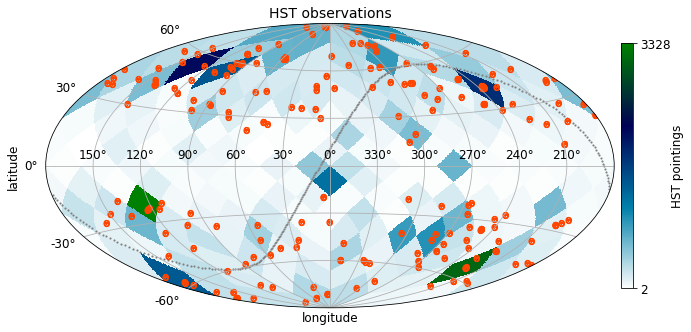} 
    \caption{Map in galactic coordinates showing the sky position of the newly discovered lenses serendipitously observed with HST (red points), over the densities of the parent sample of HST pointings analysed in the HAH project (in tiles of $\sim$215 deg.$^2$ each). The sky distribution of the HST observations with strong lens candidates is roughly isotropic, excluding the Galactic Plane. The grey dotted line represents the ecliptic.}
    \label{fig:map}
\end{figure}

The initial catalogue contained the position of the lens, the instrument and filter used. We used the ESASky\footnote{\url{https://sky.esa.int/}} portal \citep{Giordano2018} to retrace the sky coordinates of the lenses in the images. ESASky also reports the objects which appear in previous publications, matching their positions in the HST images with astronomical catalogues and papers from the CDS Simbad database. This eased our work to establish which lenses have been previously published. We also double-checked with other databases such as NED. We classified the objects as being `targeted' (in the case where the strong lenses were the target of the observations, judged based on the corresponding HST proposal), previously `published' (whether they appeared in other publications) or `unpublished'. A number of 165 strong lenses were the target of the HST observations, for example by the Sloan Lens ACS Survey \citep[SLACS,][]{Bolton2008}, the CFHTLS-Strong Lensing Legacy Survey \citep[SL2S,][]{More2012}, the CASSOWARY survey \citep{Stark2013}, or the BOSS Emission-Line Lens Survey \citep[BELLS,][]{Shu2016}. These targeted lenses were excluded from our analysis. The remaining 252 strong lenses were not the primary targets of the HST observations. We searched for these objects on ESASky, NED, and enquired with the 96 Principal Investigators of these HST observations whether they appear in previous publications. 54 have been previously identified and appear in other publications. Hence, we are left with 198 unknown, new lens candidates distributed on the sky according to Fig.~\ref{fig:map}.

Three members of the science team (EOG, SK, CC) assigned a grade to each lens, based on: the morphology, the shape of the source image and the quality and availability of a colour image. We discussed each object and agreed on a final grade. The criteria used for the classification are inspired by the convention of \cite{More2016}, and define the likelihood of each object to be a strong lens. {\tt Grades A} are almost certainly lenses. They are the most promising candidates based on the configuration of the lens, and the morphology in the colour- or grey-scale images. Their identification is based on the presence of an image and counter image, and on a clear colour separation between the lens and image (if colour information was available). {\tt Grades B} are highly probable lenses. These are also high-quality candidates showing probable lensing features, but with typically fewer clear and bright counter-images, and less colour information available than grades A. Their confirmation requires spectroscopic follow-up or lens modelling. {\tt Grades C} are maybe lenses. These possible strong lens systems mostly have single arcs without clear counter-images. They are more difficult to distinguish from non-lens contaminants, and some could be tidal tails, compact groups, or weakly lensed arcs.

We further classified the morphology of the source images into `Arc', which is the most common class, `Double' (two images), `Triple' (three images), `Quad' (four images), and `Ring' or `Cross' for the rarest Einstein ring and Einstein cross configurations. Additionally, we classified the morphology of the lenses into `elliptical', `disc' and `edge-on disc'.

\subsection{Measuring the arc radii}

A measurable parameter for the strong lenses in our sample is the arc radius, which can be used as a proxy for the Einstein radius. For the strong lenses made of a single arc, we measured the radius of the circle centered on the brightest pixel of the lensing galaxy and tangent to the arc, using both the DS9 and the Aladin `circle' tool. For the other configurations, we drew a circle, or the best matching ellipse, as a substitute for the outer tangential critical curve in the lens plane and measured its radius.

\subsection{Galfit modelling of the lenses}
\label{galfit}

In this subsection, we describe the fitting process using \texttt{GALFIT} \citep{Peng2002} to determine the parameters of the lenses: the magnitudes, effective radii, axis ratios, and position angles. These parameters are important for future tasks, such as lens modelling and source reconstruction, which are not covered in this work.

We fitted a total of 252 lenses with \texttt{GALFIT}, including the 198 previously unknown systems, and the 54 which were already identified and published, as discussed above. We used the sky position of the lenses to create cutouts with square sizes of 10\arcsec, centered on the objects, from the single-band composite HST images available in eHST. If multiple bands were available, we used the reddest band available, which was F814W in the majority of cases. The band used for fitting is shown in Table~\ref{tab:unpublished} and Table~\ref{tab:rediscoveries}.

To ensure that only the lenses are fitted, we masked out the other objects in the cutouts, including the arcs and any other background image, with \texttt{SExtractor} \citep{Bertin1996}. We created images of the HST Point Spread Function (PSF) using \texttt{TinyTim} \citep{TinyTim1995}, for the band corresponding to each image and the position of the object in the corresponding image, which we use for the fitting. We fitted the lenses with a single S\'ersic profile \citep{Sersic1968} described by the following equation:

\begin{equation}
\sum(r)=\sum_{\rm e}\exp\Bigg\{-k\left[(\frac{r}{r_{\rm e}})^{\frac{1}{n}}) - 1\right]\Bigg\},
\end{equation}
\noindent{where $\mathrm{\sum(r)}$ is the surface brightness at radius \textsl{r}, \textit{$r_{\rm e}$} is the half-light radius, $\sum_e$ is the effective surface brightness, \textit{k} is a normalisation coefficient and \textit{n} is the S\'ersic index. As starting parameters, we used a magnitude of 20, an axis ratio of 0.9, and an effective radius of 0.5\arcsec.

We used a fixed De Vaucouleurs profile ($n=4$) for the majority of the lenses (243 galaxies), which we classified as being elliptical galaxies. We fitted the remaining 23 galaxies which we classified as being discs with an exponential S\'ersic index ($n=1$). We note that in the case of group-scale lenses, we fitted multiple galaxies which we deemed to be the lenses, hence the total number of galaxies fitted is higher than the number of strong lens candidates. We show the \texttt{GALFIT} models and residuals of the A grade lenses in Fig.~\ref{fig:galfitB1} and Fig.~\ref{fig:galfitB2}.

To determine the magnitudes, we calculated the zero points based on the quantities provided in the \textsc{FITS} headers, using\footnote{\url{https://www.stsci.edu/hst/instrumentation/acs/data-analysis/zeropoints}}:
\begin{equation}
\begin{aligned}
m_{\rm AB} = & -2.5\log_{10}(F) -2.5\log_{10}({\tt PHOTFLAM}) \\
         & -2.5\log_{10}({\tt PHOTPLAM}) - 2.408,
\end{aligned}
\label{eq:mag}
\end{equation}
where {\tt PHOTFLAM} is the inverse sensitivity and represents the scaling factor necessary to transform an instrumental flux in units of electrons per second to a physical flux density, and the {\tt PHOTPLAM} is the pivot wavelength provided in the \textsc{FITS} header and used to derive the instrumental zero point magnitudes. For consistency, we also measured the magnitudes of the lenses with \texttt{SExtractor} ($\mathrm{mag}_{\rm AUTO}$ and $\mathrm{mag}_{\rm ISO}$). We compare the magnitudes estimated with \texttt{GALFIT} and \texttt{SExtractor} in Figure~\ref{fig:compare_magnitudes}} and find that the measurements are consistent between the different methods. 
 
\section{Results}
\label{results}

\begin{figure*}
\centering
 \includegraphics [width=1.03\textwidth]{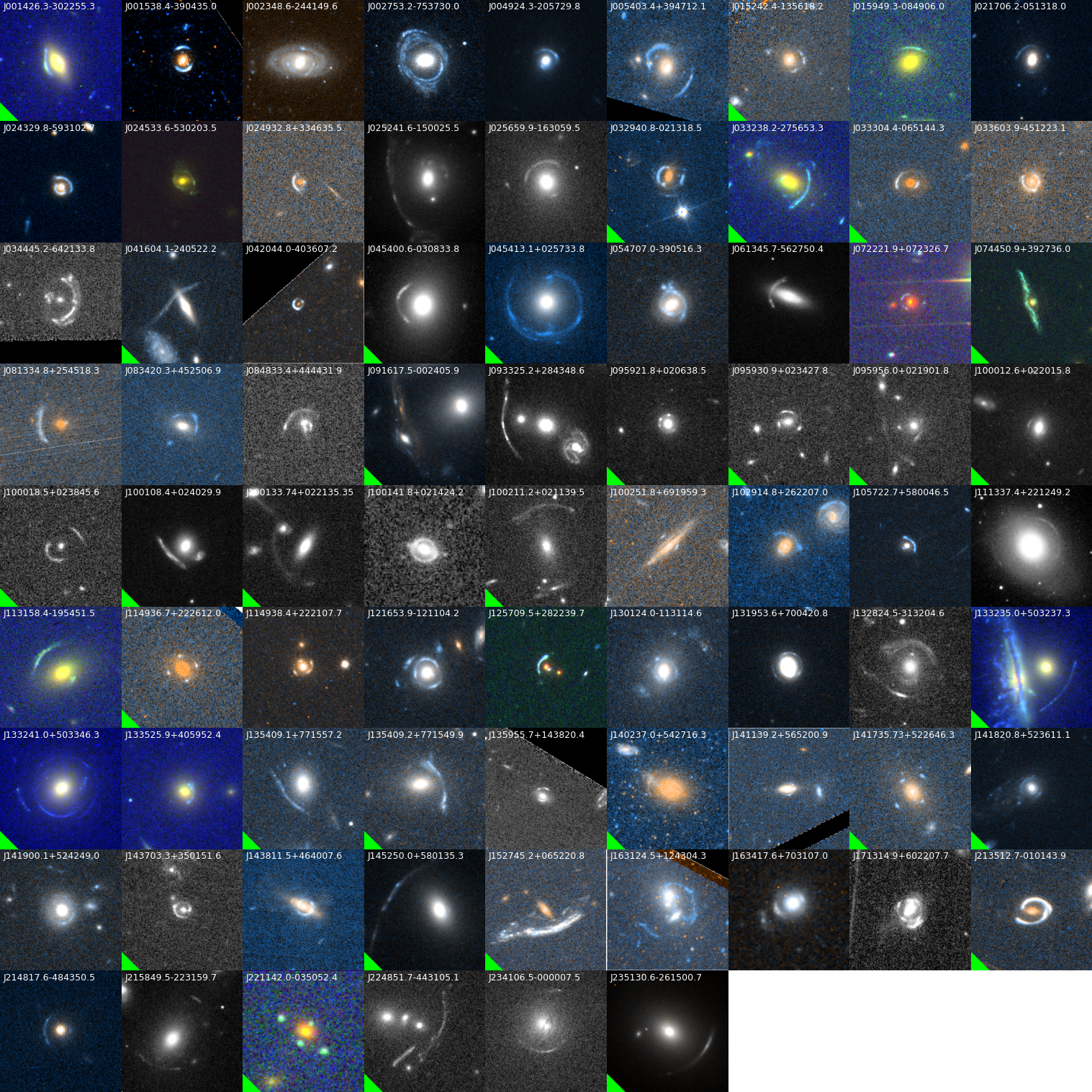} 
 \caption[Grade A lenses]{The 78 grade A HAH lenses identified in this study, with green triangles indicating the 33 ‘rediscovered’ lenses. The lenses are indicated using the sequence part of their name. Postage stamps are $10\arcsec\times10\arcsec$. The orientation of the images is North up and East is to the left.}
 \label{fig:grade_A}
\end{figure*}

\begin{figure*}
\centering
 \includegraphics [width=1.03\textwidth]{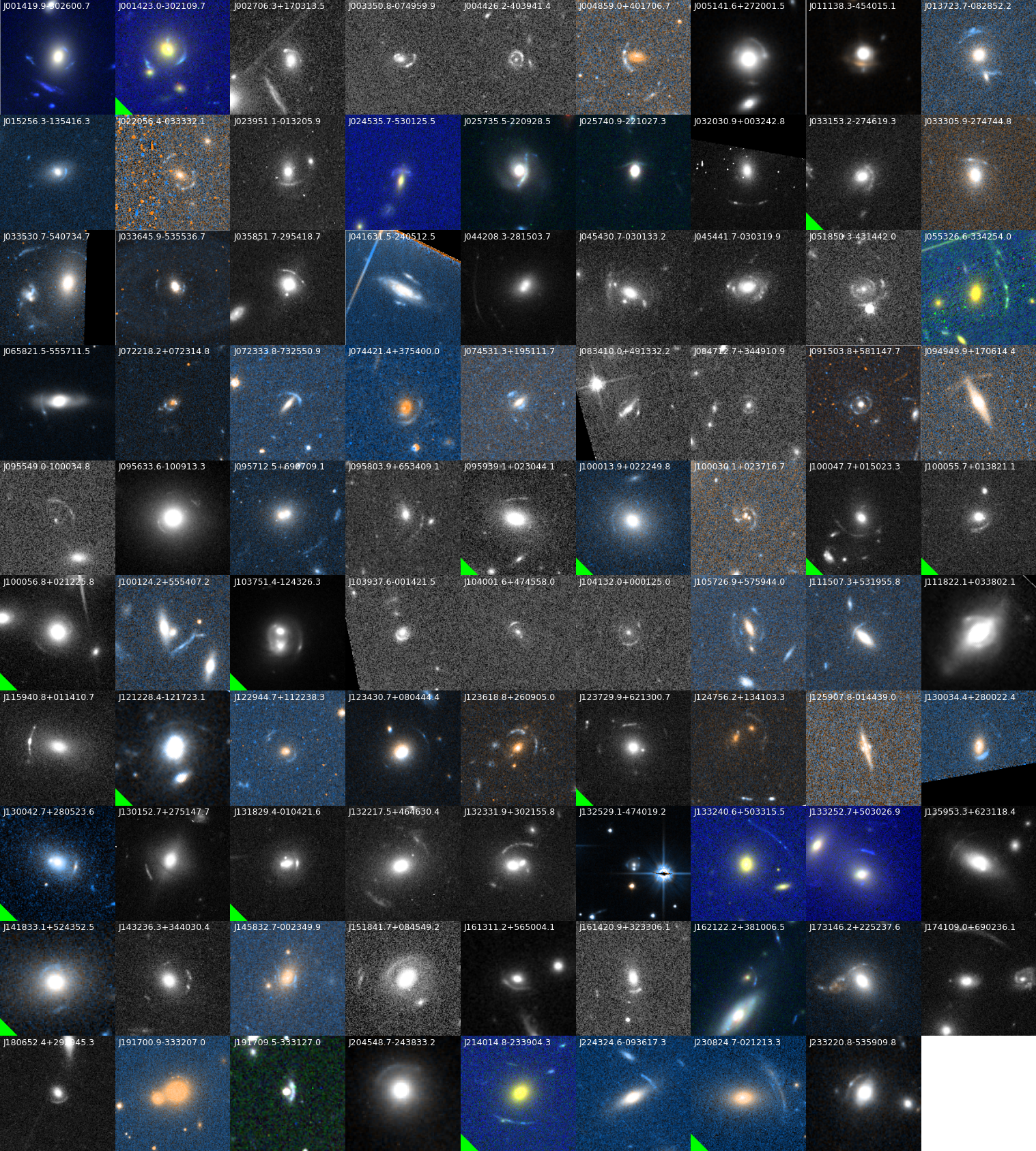} 
 \caption[Grade B lenses]{The 89 grade B HAH lenses identified in this study, with green triangles indicating the 15 ‘rediscovered’ lenses. The lenses are indicated using the sequence part of their name. Postage stamps are $10\arcsec\times10\arcsec$. The orientation of the images is North up and East is to the left.}
 \label{fig:grade_B}
\end{figure*}

\begin{figure*}
\centering
 \includegraphics [width=1.03\textwidth]{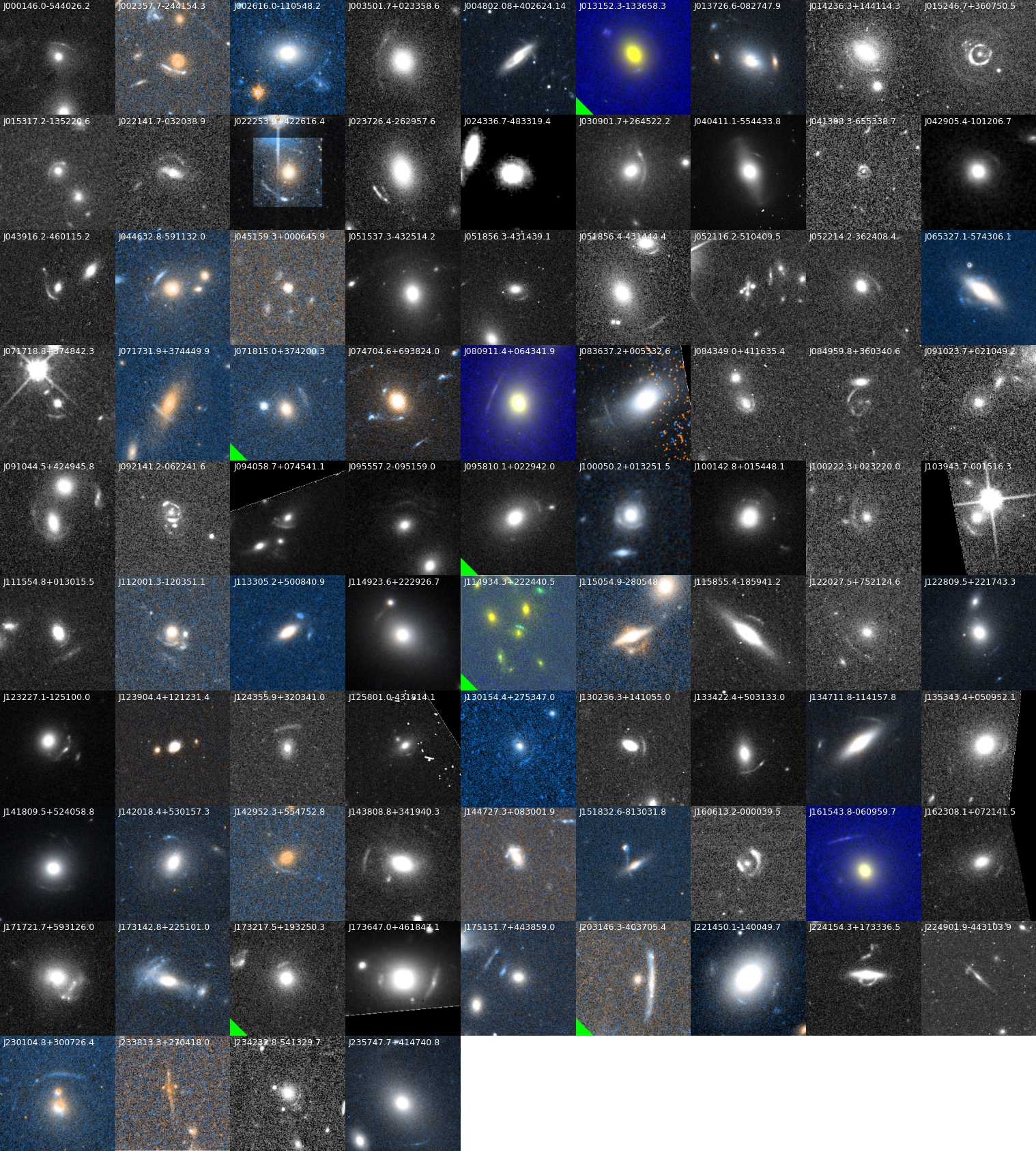} 
 \caption[Grade C lenses]{The 85 grade C HAH lenses identified in this study, with green triangles indicating the 6 ‘rediscovered’ lenses. The lenses are indicated using the sequence part of their name. Postage stamps are $10\arcsec\times10\arcsec$. The orientation of the images is North up and East is to the left.}
 \label{fig:grade_C}
\end{figure*}

We present the results on the 198 newly identified lens candidates (which we henceforth call `discovered') in Table~\ref{tab:unpublished}, and on the 54 previously published lenses (which we henceforth call `rediscovered') in Table~\ref{tab:rediscoveries}. The lenses in the tables are grouped by grades, and postage stamps of the \texttt{Grade A}, \texttt{Grade B}, and \texttt{Grade C} lenses are shown in Figures~\ref{fig:grade_A}, \ref{fig:grade_B}, and Figure~\ref{fig:grade_C}, respectively. In the tables we provide the IAU name, the $\mathrm{RA}$ and $\mathrm{Dec}$ coordinates of the candidate, the instrument corresponding to the HST observation in which the lens was found, and the filter used for light fitting. We also provide our measurements of the arc radii, $r_{\mathrm{arc}}$, as well as the magnitude, the effective radius, $r_{\rm e}$, the axis ratio, $q$, and the position angle, PA, measured using \texttt{GALFIT}. We also list our classification into Arc, Double, Ring, and Quad configurations.

While the sample mainly consists of galaxy-galaxy lens configurations, some of the lens galaxies may be part of a cluster, in which case the strong lensing effect can be enhanced by the mass distribution of the cluster. In addition, the presence of a structure along the line-of-sight, either in front or behind the main lens plane, will also perturb the light deflection. Firstly, we identify the lens candidates that are associated with known foreground galaxy clusters (from, e.g. MACS, Abell, RELICS catalogues) using the target name of the HST observations. Secondly, to get a more comprehensive overview of the lens candidates located in possible cluster fields, we also cross-match with the positions of cluster candidates selected (1) from SDSS DR8 data with the redMaPPer algorithm \citep[the SDSS catalogue from][]{Rykoff2016}, and (2) in the footprint of DESI legacy imaging surveys based on a photometric redshift clustering analysis \citep[DR8,][]{Zou2021}. Both surveys cover a significant fraction of the line-of-sights towards our HAH lenses. We use a maximal cross-match radius of 3\arcmin\ corresponding to $\simeq$1~Mpc at $z \sim 0.5$. Tables~\ref{tab:unpublished} and \ref{tab:rediscoveries} indicate whether each lens matches the position of a cluster, and Table~\ref{tab:clusters} gives additional information on the cluster names and angular separation. In total, 124 out of the 198 new lens discoveries and 32 out of the 54 rediscoveries are within 3\arcmin\ from the centroid position of a candidate or confirmed cluster.

Finally, we search for redshifts available for the lens galaxies. We provide either the photometric or spectroscopic redshift, if available, retrieved from NED or from the SDSS-IV DR16 \citep{Ahumada2020}. Additionally, the last column of Table~\ref{tab:rediscoveries} lists the reference papers for the candidates previously published.

\begin{figure*}
    \centering
    \includegraphics[width=\textwidth]{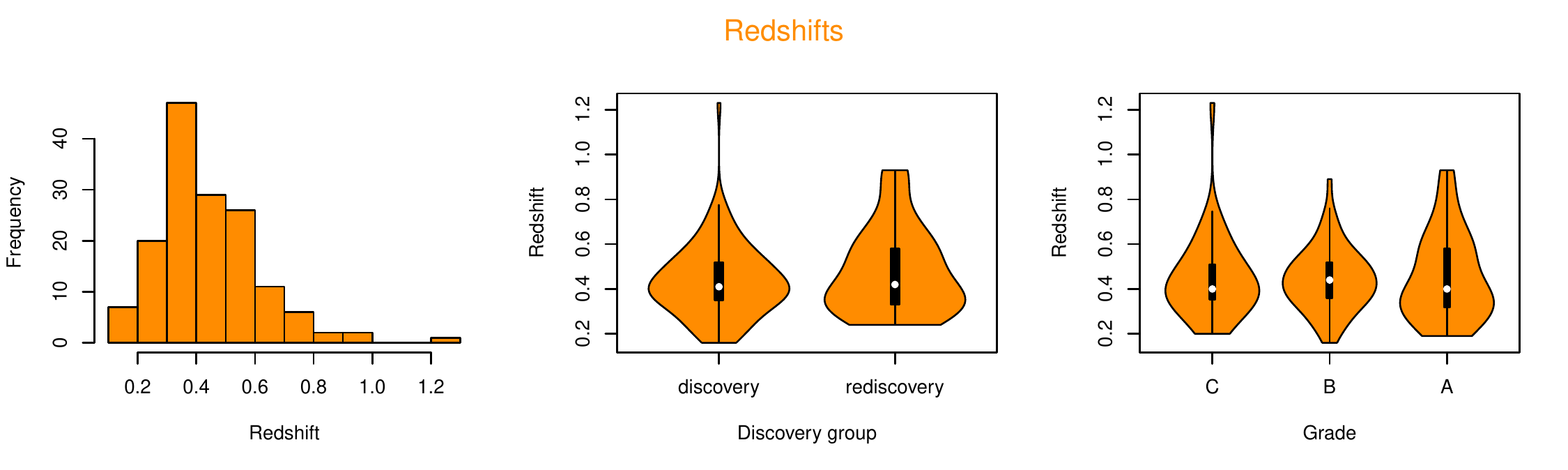}
    \caption{The first sub-figure on the left shows a histogram distribution of the redshifts retrieved with SDSS and NED (both photometric and spectroscopic). The middle sub-figure uses a violin plot to show the respective empirical distributions of discovered and rediscovered lenses. The right sub-figure shows violin plots per grade groups. The violin plots were fitted using a Gaussian kernel, with the software \texttt{R}.}
    \label{fig:redshifts}
\end{figure*}

\begin{figure*}
    \centering
    \includegraphics[width=\textwidth]{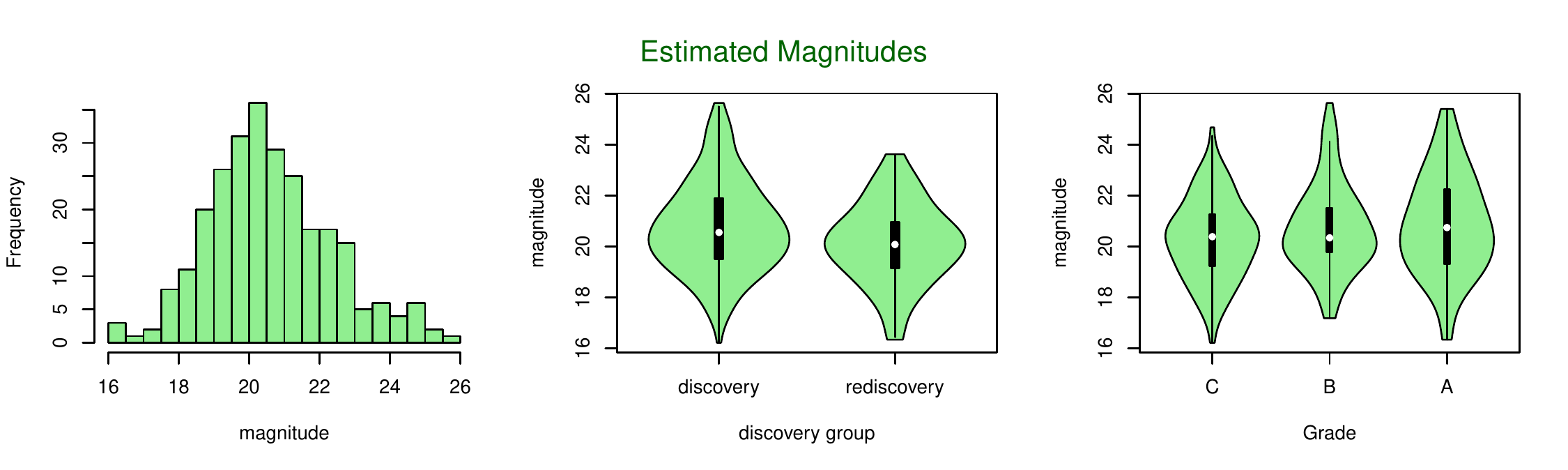}
    \caption{The left panel shows a histogram distribution of the apparent magnitudes as measured with \texttt{GALFIT} (in the filter mentioned in Table~\ref{tab:unpublished} and Table~\ref{tab:rediscoveries}). The middle panel shows violin plots as the respective empirical distributions of discovered and rediscovered lenses. The right panel shows empirical distributions of magnitudes for each grade group.}
    \label{fig:magnitudes}
\end{figure*}

\begin{figure*}
    \centering
    \includegraphics[width=\textwidth]{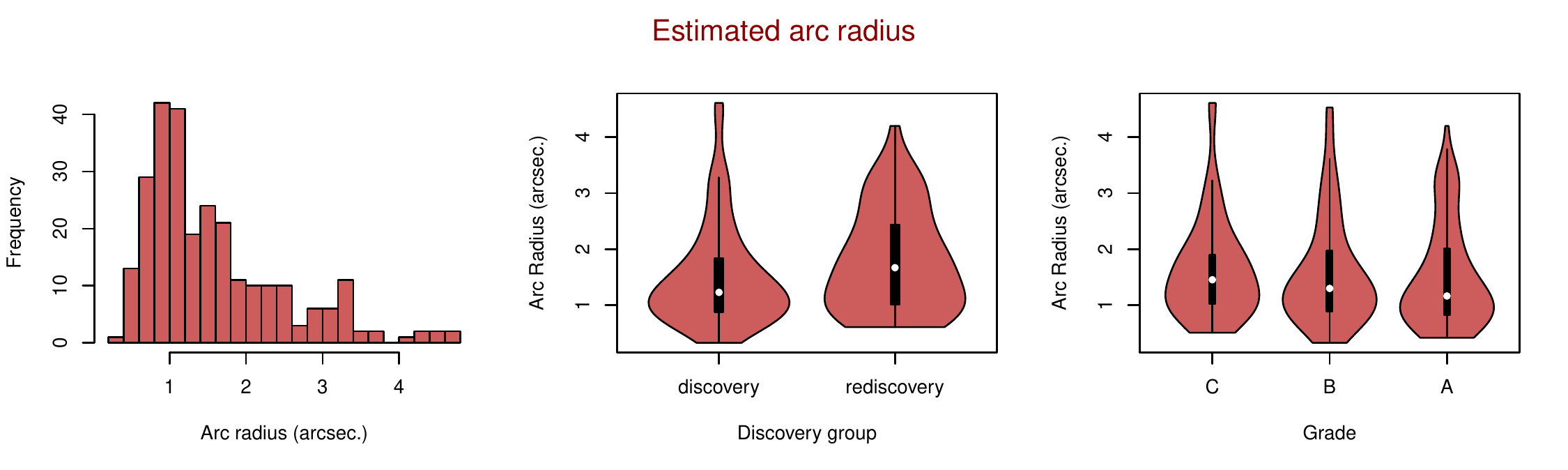}
    \caption{The first sub-figure on the left shows a histogram distribution of the measured arc radii. The middle sub-figure uses violin plots to show the respective empirical distributions of discovered and rediscovered lenses. The right sub-figure shows empirical distributions of arc radii per grade groups. The distributions are all skewed to the right.}
    \label{fig:arcradius}
\end{figure*}

\begin{figure*}
    \centering
    \includegraphics[width=\textwidth]{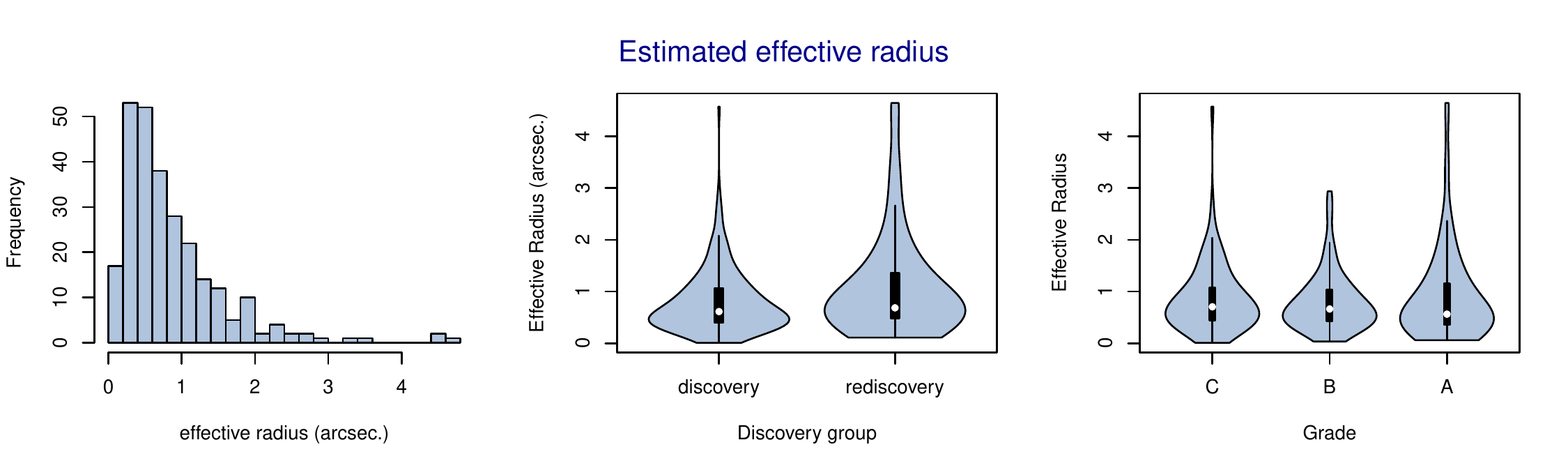}
    \caption{The first sub-figure on the left shows a histogram distribution of the effective radii measured with \texttt{GALFIT}. The middle and right sub-figure shows empirical Gaussian kernel distributions of lenses, separated per discovery (middle panel) and per grade (right panel).}
    \label{fig:re}
\end{figure*}

\begin{figure*}
    \centering
    \includegraphics[width=\textwidth]{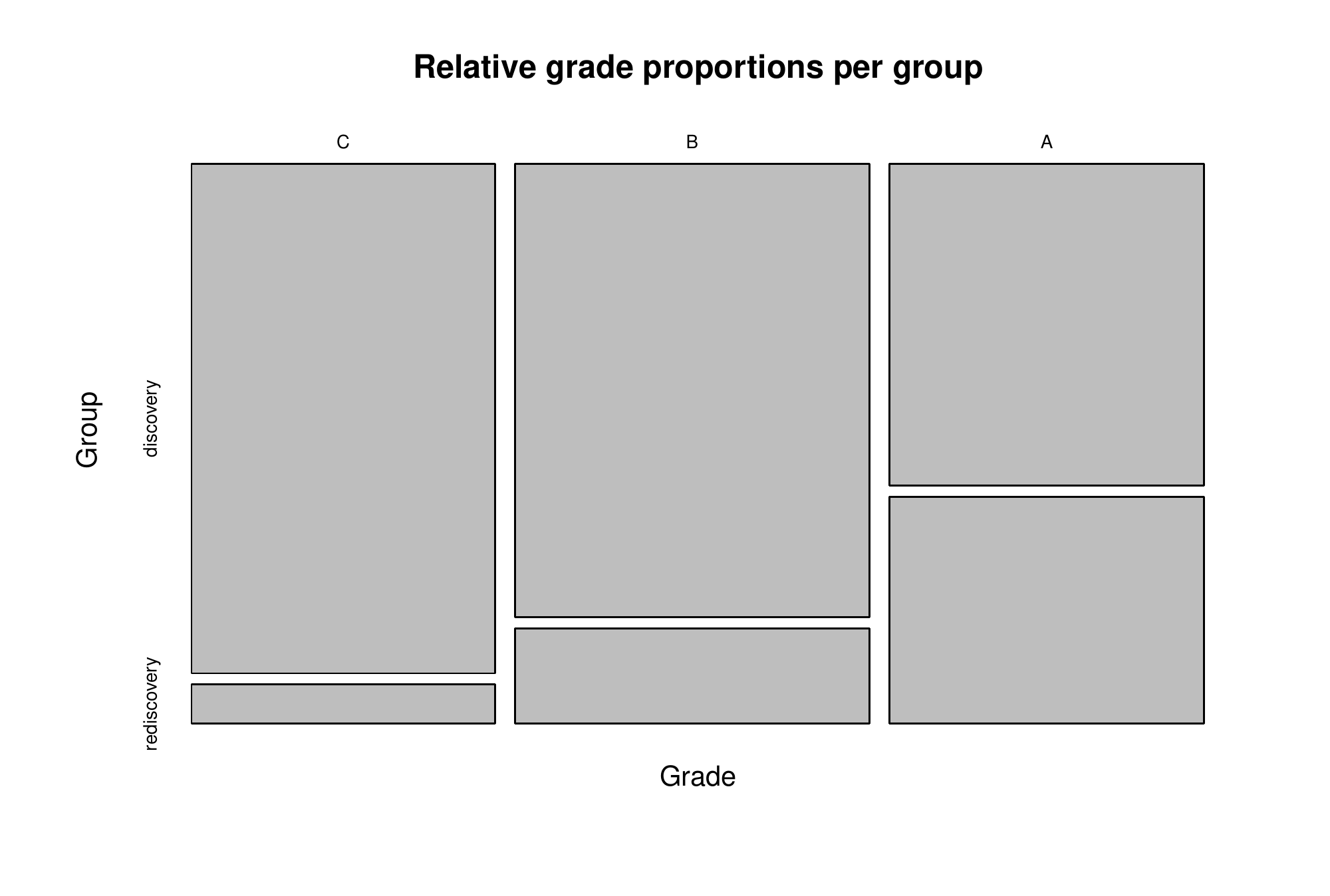}
    \caption{The mosaic plot shows the discovery groups and grades in a visual representation of a contingency table. The widths of the stacked boxes represent the fraction of each grade in the sample, while the heights of the boxes represent the proportion per discovery group. All the grades have been assigned by three members of the science team (EOG, SK, CC) after consensus following the individual evaluations. We observe that in the discovery group, we have a higher fraction of C grades compared to A grades. In the rediscovery group, there is a higher fraction of A grades compared to both B and C grades. Hence, the mosaic plot indicates that, overall, our grading scheme is rather conservative towards newly discovered lenses.
    }
    \label{fig:proportions}
\end{figure*}

\subsection{Distribution of lens properties}

In the previous sections, we described how the gravitational lenses were found and reported into discovered and rediscovered lenses and we explained how we fitted them in order to retrieve the lens parameters. In this subsection, we describe the results, namely the distributions of properties such as arc radii, and lens redshifts, magnitudes, and effective radii. The properties are grouped by either discovered or rediscovered lenses, and by visual grade.

In Figure~\ref{fig:redshifts}, we plot the distribution of photometric and spectroscopic redshifts from SDSS and NED for the lens galaxies for which this information was available. The median redshift of our sample is $z=0.41$, and the distribution is skewed to higher redshifts, up to $z=1.3$. The discovered and rediscovered lenses show similar distributions, while Grade A lenses are, on average, at lower redshifts compared to Grade B or C. The redshift distribution of the lenses in our sample is also broadly consistent with the lens galaxy redshifts from previous ground-based searches \citep[e.g.][]{Bolton2008,Stark2013}.

Figure ~\ref{fig:magnitudes} shows the distribution of magnitudes over all instruments and filters. We find that the magnitude of reported objects are distributed around an average value of $m=20.6$. Although our sample contains lenses observed with WFC3/IR, which are typically brighter, our average magnitude is 1.3 magnitudes fainter than the average magnitude $m=19.3$ of the sample of lenses found by \citet{Pawase2014} in a previous search in ACS images. The violin plots in Figure~\ref{fig:magnitudes} show that the magnitudes of the discovered lenses are fainter than those of rediscoveries ($m=20.8$ vs. $m=20.1$). Indeed, the faintest lens found has a measured magnitude of $m=26.4$ compared to $m=23.6$ in the rediscovered group. Moreover, a quarter of the rediscovered objects were detected above $m=21$, while a quarter of the new discoveries have a magnitude above $m=21.9$. Interestingly, Grade C lenses are brighter than Grade B or Grade A, showing that we did not assign the grade based on brightness, but on the morphology. It is important to note that the magnitudes are plotted together for both the ACS and WFC3 instruments and some are measured in different filters (see Table~\ref{tab:unpublished} for the filters used). 

As for the arc radii of the lenses, the distribution is right skewed, with a median of $r_{\mathrm{arc}}=1.58\arcsec$ and a mean of $r_{\mathrm{arc}}=1.94\arcsec$. Although it is consistent with the distribution of arc radii presented in \citet{Pawase2014}, we notice from the plot on the left in Figure~\ref{fig:arcradius} that our detected lenses include more extreme arc radii than their sample. Indeed, a few of the newly discovered arc radii range up to $r_{\mathrm{arc}}=4.5\arcsec$, which corresponds to group-scale lenses. In addition, for our newly discovered lenses, the median arc radius is $r_{\mathrm{arc}}=1.23\arcsec$ and the smaller arcs we find have a radius of only $r_{\mathrm{arc}}=0.33\arcsec$. This shows a relative improvement in finding smaller separation arcs in comparison to previously reported lenses (in the rediscovered group), for which median $r_{\mathrm{arc}}=1.67\arcsec$ and the smallest arcs have radii $r_{\mathrm{arc}}=0.61\arcsec$. Furthermore, a quarter of the rediscovered objects were found to have $r_{\mathrm{arc}}\leq0.88\arcsec$ while a quarter of our newly discovered objects have $r_{\mathrm{arc}}\leq0.82\arcsec$. This observed difference in the median arc radii and their distribution (shown in the central plot in Figure~\ref{fig:arcradius}) indicates that citizen science as a strong lens detection method has a potential to unveil arcs with smaller angular separations from the source. This potential deserves to be further investigated and quantified in future studies.

The distributions of lens galaxy effective radii (Figure~\ref{fig:re}) show a median $r_{\mathrm{e}}=0.66\arcsec$ and are consistent across discovered and rediscovered groups, as well as with the distribution in \citet{Pawase2014}.

Finally, the mosaic plot from Figure~\ref{fig:proportions} shows a visual representation of a proportion test on the contingency table of the grades assigned by members of the science team (EOG, SK, CC) to the discovered and rediscovered lenses. We can see that the C grades are relatively larger in proportion than the B grade and A grade lenses in the discovery group, while the A grades are more frequent than both the B and C grades in the rediscovery group.

Overall, the results we presented in this sub-section highlight the interest and importance of the contribution of citizen scientists and detection by human eye in order to continuously cross validate and improve the performance of classical algorithms.

\subsection{Properties of the high-quality candidates}
\label{individual_lenses}

The morphological search conducted in this paper results in strong lenses with high diversity of image configurations and lens potentials, and their future spectroscopic follow-up and detailed modelling will enable a range of valuable studies. In Sect.~\ref{studies}, we illustrate possible applications with the highest-quality, newly-discovered, grade A candidates in Fig.~\ref{fig:grade_A}. In Sect.~\ref{configurations}, we further describe the lensing configurations and we highlight the properties of few individual lenses.

\subsubsection{Prospects for scientific studies}
\label{studies}

Strong lenses with distant, isolated foreground galaxies are promising to improve our understanding of the total mass-density slope, $\gamma_{\rm tot}$, of early-type galaxies. While it is firmly established that $\gamma_{\rm tot}$ is nearly isothermal at low redshift \citep[e.g.][]{Treu2004,Koopmans2006,Cappellari2015}, its evolution at $z<1$ remains debated, with strong lensing studies suggesting a mild increase from $z \sim 1$ to $z \sim 0$ \citep[][]{Koopmans2006,Bolton2012,Sonnenfeld2013,Li2018}, and Jeans dynamical modelling and cosmological simulations favouring nearly constant $\gamma_{\rm tot}$ over this period \citep[e.g.][]{Wang2019,Derkenne2021}. Measurements are mainly restricted to early-type galaxies at $z \lesssim 0.6$ and need to be extended to higher redshifts in order to further test galaxy evolution models. Systems with the most distant, isolated deflectors (e.g. HAH J083420.3$+$452506.9 at $z_{\rm spec} = 0.65$) are very useful in that regard, as well as future spectroscopic follow-up of lenses with $z_{\rm phot} > 0.6$ (e.g. HAH J005403.4$+$394712.1, HAH J125709.5$+$282239.7).

Double source plane lenses are beneficial to infer tight constraints on the foreground total mass-density profiles \citep[e.g.][]{Tu2009}. Multiple images covering different angular separations from the lens centre, as for instance in HAH J121653.9$-$121104.2, and HAH J132824.5$-$313204.6, break parameter degeneracies in the lens models. Together with multiband photometry or stellar kinematic measurements, such lensing configurations enable to characterise the lens dark-matter distributions. Moreover, when the background sources lie at distinct redshifts, measuring the two Einstein radii can provide valuable and independent constraints on the equation of state of dark energy \citep{Collett2012}. Only a handful of double source plane lenses are known to date \citep[e.g.][]{Gavazzi2008,Tanaka2016}, and in our sample, HAH J002753.2$-$753730.0 and HAH J132824.5$-$313204.6 are good candidates. The former shows two opposite blue, thin arcs surrounding a foreground elliptical galaxy, and a broad fuzzy ring that could be a distinct background source. Possible applications to cosmology make these strong lenses well-suited for spectroscopic follow-up.

In addition, systems with isolated lens galaxies and bright, extended, and structured lensed arcs are typically prioritised to search for the presence of foreground dark-matter sub-halos and line-of-sight halos \citep[e.g.][]{Vegetti2009,Ritondale2019}. Due to their arc surface brightness distributions, HAH J025659.9$-$163059.5, HAH J105722.7$+$580046.5, and HAH J113158.4$-$195451.5 have the potential to put valuable constraints on foreground mass perturbations.

\subsubsection{Strong lensing configurations}
\label{configurations}

While Tables~\ref{tab:unpublished} and \ref{tab:clusters} indicates the presence of galaxy clusters along the lines-of-sight towards several candidates, the image separations and the morphologies in Fig.~\ref{fig:grade_A}, \ref{fig:grade_B} and \ref{fig:grade_C} show that in fact, most have galaxy-galaxy strong lens configurations. In particular, the light deflection is dominated by a single member or line-of-sight galaxy for 20 out of 21 newly-discovered grade A with foreground confirmed or candidate clusters. Only HAH J025241.6$-$150025.5 shows an extended arc and is located within 0.5\arcmin\ from the centroid position of a cluster candidate listed in \citet{Zou2021}. Three grade A have $r_{\rm arc} > 3$\arcsec, suggesting a major contribution from an extended group- or cluster-scale mass component to the external convergence and shear and, for instance, HAH J093325.2$+$284348.6 comprises a compact group of three foreground galaxies. These three systems are nonetheless absent from our compilation of cluster candidates selected from shallower ground-based imaging. In addition, Fig.~\ref{fig:grade_A} shows a compact group at $z_{\rm spec}=0.1895$ towards HAH J234106.5$-$000007.5, but this system has smaller image separation ($r_{\rm arc} \simeq 2.3$\arcsec), and the associated candidate from \citet{Zou2021} is $>2$\arcmin\ away. Towards the low-end of the image separation distribution, several grade A candidates have very compact configurations with sub-arcsec Einstein radii, as best shown by HAH J042044.0$-$403607.2 with $r_{\rm arc} \simeq 0.4$\arcsec. This extends current samples of compact, galaxy-scale strong lenses \citep[e.g. SLACS,][]{Bolton2008}, and illustrates the new discovery space that will be unlocked by the {\it Euclid} survey with comparable, high-resolution imaging over $\simeq$15,000\,deg$^2$. 

The sample contains a majority of extended, distorted arcs and rings, which are much less ambiguous than configurations with more compact lensed sources. HAH J001538.4$-$390435.0, HAH J005403.4$+$394712.1, and HAH J024329.8$-$593102.7 are examples of systems with pairs of blue lensed arcs while HAH J061345.7$-$562750.4 shows a single, but particularly elongated arc, and HAH J033603.9$-$451223.1 or HAH J100141.8$+$021424.2 exhibit near-complete Einstein rings. The difficulty in identifying doubles is coming from their similarity with compact groups and other non-lens contaminants in the grey-scale HST images seen by volunteers during the first stage of the classification. HAH J100251.8$+$691959.3 is the only new grade A candidate firmly identified as a blue, doubly-imaged source from the inspection of colour images by the science team. HAH J002348.6$-$244149.6 also shows compact images, but in a quadruple configuration.

In terms of morphological types, our sample contains interesting classes of foreground lens galaxies. While traditional selection methods are mainly restricted to massive elliptical lens galaxies, the angular resolution reached from space is key to extend to less common lens galaxy types, and to asymmetric, or more exotic configurations which are particularly difficult to identify from the ground. Firstly, the new A grade HAH J002348.6$-$244149.6 comprises a moderately inclined lens spiral galaxy with a bright, blue point-like image in the north and three images distributed evenly along a southern thin arc. This system extends the small number of known spiral deflectors with prominent discs \citep[see also, the SWELLS survey,][]{Treu2011} and, due to the relative positions of multiple images, its modelling has the potential to tightly constrain the galaxy bulge mass. Secondly, HAH J100251.8$+$691959.3 and HAH J143811.5$+$464007.6 show edge-on disc galaxies with $r_{\rm arc} \simeq 0.6$--0.7\arcsec. Thirdly, the flexibility of the crowdsourced classification is demonstrated by the complex structure of the previously published A grade lens HAH J133235.0$+$503237.3 \citep{Ragozzine2012}. A background galaxy is lensed into a thin elongated arc by a spectacular foreground environment, where an edge-on jellyfish galaxy and nearby ellipticals are embedded within the merging cluster A1758N $z=0.279$ \citep{Ebeling2019b}.

\begin{figure}
    \centering
    \includegraphics[width=\columnwidth]{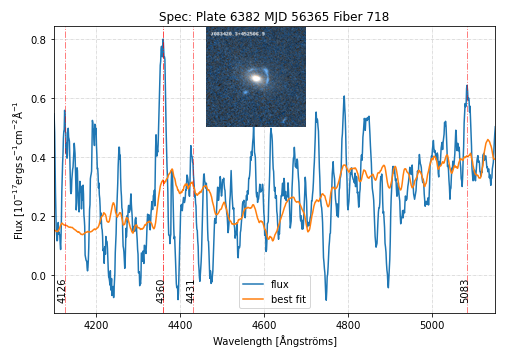}
    \caption{BOSS spectrum of the newly-discovered grade A candidate HAH J083420.3+452506.9 showing multiple emission lines inconsistent with the spectroscopic of the foreground early-type galaxy of 0.6495. The  blue and orange lines show the observed spectrum and the best-fit SDSS model for the central lens galaxy, respectively. The vertical lines mark the positions of the CIV-$\lambda1549$, HeII-$\lambda1640$, OIII-$\lambda1664$, and CIII]-$\lambda1908$ emission lines at $z \approx 1.667$, from left to right.}
    \label{fig:specJ083420}
\end{figure}

Most candidates with SDSS spectroscopic follow-up have large image separations (e.g. HAH J111337.4$+$221249.2), which limits the contribution from the lensed sources to the emission collected in the 2\arcsec\ diameter aperture fibres. This limits the number of candidates with spectroscopic redshift measurements for both the lens and source. We nonetheless inspected the SDSS spectra of all grades A, and identified multiple emission lines inconsistent with the lens redshift of 0.6495 for HAH J083420.3$+$452506.9 ($\sigma_* = 269 \pm 48$~km\,s$^{-1}$). Our estimate $r_{\rm arc} \simeq 0.6\arcsec$ suggests that the lens and source galaxies are blended in the SDSS spectrum, and that the background source is a star-forming galaxy with detections of CIV-$\lambda1549$, HeII-$\lambda1640$, OIII-$\lambda1664$, and CIII]-$\lambda1908$ emission lines (see Fig.~\ref{fig:specJ083420}). The [OII]$\lambda\lambda3727$\ doublet is also tentatively detected at 9942.6\,$\AA$\ and 9949.47\,$\AA$. By fitting the smoothed spectrum, we inferred a probable source redshift of $\approx$1.667. Higher signal-to-noise ratios are nonetheless needed to confirm this estimate.

\section{Discussion}
\label{discussion}

\subsection{Citizen science for lens finding}

Previous strong lens searches have relied on dedicated citizen science projects, the most well-known example being the Space Warps project \citep{Marshall2016}, which was also built within the Zooniverse framework. Space Warps has demonstrated that the visual inspection of volunteers is an efficient data mining tool for strong lenses. It has been successfully applied to the CFHT Legacy Survey finding 29 promising (and 59 total) new lens candidates \citep{More2016}. More recently, \cite{Sonnenfeld2020} used a dedicated algorithm, YATTALENS \citep{Sonnenfeld2018} and a citizen science approach on Space Warps to find lenses in the HSC survey, finding 14 definite lenses (grade A), 129 probable lenses (grade B), and 581 possible (grade C) lenses in total. Citizen science approaches have also been implemented for lens modelling \citep{Kung2015}. 

Our citizen science project and Space Warps employ different approaches for the inspection of images. Space Warps shows to the volunteers postage stamps of objects, generally pre-selected with some criteria (for example, objects within a redshift and mass range), in order to maximise the chances of detecting strong lenses and minimise the required volunteer effort. They also inserted simulated lenses in order to calibrate the response of the volunteers and measure the completeness. 

In contrast, in our project we show large fields-of-view to the volunteers and ask them to tag on `Talk' if they think the image contains a strong lens. While the Space Warps approach provides a measure of completeness, which is difficult to assess in our case, our project provides an unbiased search for strong lenses (since our images were not pre-selected with some criteria), being able to find more exotic/interesting lenses. \citet{Sonnenfeld2020} also mention in the Space Warps project that there were candidates flagged on the `Talk' section of the Space Warps project that would have otherwise been missed using the main workflow of the project. 

Additionally, using large fields-of-view (corresponding to 80\arcsec $\times$ 80\arcsec\ or larger) allowed us to explore the HST images much faster than showing individual postage stamps of galaxies. Finally, this is the first time the HST images are explored with a citizen science approach, showing the benefit of asking volunteers to inspect space-based images spanning decades in time.

\subsection{Comparison to previous $HST$ studies}
\label{comparison_HST} 

Previous visual searches for strong gravitational lenses in HST images has been undertaken by \citet{Faure2008} and \citet{Jackson2008} in the 1.64 deg$^2$ COSMOS field and \citet{Pawase2014} on 7 deg$^2$ of archival ACS/WFC I-band and WFC3/IR in the near-infrared channel F160W HST images taken until 31 August 2011. Additionally, \citet{Pourrahmani2018} trained a convolutional neural network on classifications by \citet{Faure2008} to identify strong lenses in COSMOS. 

In a first dedicated exploration of the COSMOS field, \citet{Faure2008} identified 67 low-grade strong galaxy-lens candidates with 20 of them displaying multiple images or obvious arcs. Their searches were limited to massive early-type lenses only with arcs at radii smaller than 5\arcsec. \cite{Jackson2008} visually explored all galaxies in COSMOS that are brighter than 25 mag finding two more certain and one probable lens, in addition to the lenses presented in \citet{Faure2008}. \citet{Pourrahmani2018} identified 11 more candidate strong lenses in COSMOS, using the automated LensFlow machine learning algorithm. Nine of the lenses we identified were reported in previously lens searches in COSMOS (shown in Table \ref{tab:rediscoveries}). We recover two more lenses in COSMOS (HAH J100142.8+015448.1, HAH J100222.3+023220.0) which were not previously reported. 

In the only other archival search for gravitational lenses in HST images, \citet{Pawase2014} explored 7 deg.$^2$ ACS/WFC I-band and WFC3/IR F160W HST images taken until 31 August 2011. They found 40 lens candidates in ACS and 9 in the WFC3/IR images. Even though the images we inspected overlap, we only recovered seven strong lenses (four secure cases) in common between the two studies (see Table~\ref{tab:rediscoveries}). There are 12 cases labelled by \citet{Pawase2014} as secure lenses not recovered in this study. In addition, eight of our grade A lenses (HAH J054707.0-390516.3, HAH J145250.0+580135.3, HAH J045413.1+025733.8, HAH J171314.9+602207.7, HAH J005403.4+394712.1, HAH J084833.4+444431.9, HAH J025241.6-150025.5, HAH J100108.4+024029.9) observed by HST before 31 August 2011 were not recovered by \citet{Pawase2014}. This relatively low overlap between the two samples is due to the different approaches: postage stamp inspection by experts (in the case of \citealt{Pawase2014}) and large cutout inspection by an army of citizen scientists (in our case). It also shows the difficulties in the visual inspection searches of strong lenses, as studies differ in how liberal or conservative are in the definition of lenses.

\subsection{Comparison to automated lens search}

Strong lens candidates in this paper were drawn from the largest and most extensive search of 27 deg$^2$ archival HST images taken until June 2020. The selection is purely based on morphological criteria, and relies on the systematic inspection of HST images without colour, brightness, or redshift pre-selections. This type of crowdsourced classification is expected to recover the majority of lensing configurations, as long as multiple images are spatially-resolved \citep[see, e.g.][]{Sonnenfeld2020}, and Sect.~\ref{results} illustrates that candidates indeed cover a large variety of lens and source galaxy types, image multiplicity and angular distributions, including exotic lenses.

This approach differs from automated searches in several aspects. Firstly, robots are generally run on wide-area, ground-based imaging and spectroscopic surveys, and focus on massive foreground early-type galaxies due to their high lensing cross-section \citep[e.g.][]{Oguri2010}, and to their smooth light distributions and typical rest-frame optical spectra facilitating the identification of signatures from background lensed images. This allows to deal with memory limitations but restricts automated classifications to relatively small pre-selected samples \citep[][]{Bolton2008,Belokurov2009,Shu2017,Sonnenfeld2018}. Similarly, despite the significant boost in classification accuracy offered by CNNs, recent lens searches on multiband imaging using machine learning have also focused on the population of elliptical lens galaxies \citep[e.g.][]{Jacobs2019,Petrillo2019a,Canameras2020}. Secondly, automated algorithms based on supervised neural networks reach the best classification performance \citep[][]{Metcalf2019}, but they are strongly dependent on the content of the ground truth data set, and require substantial examples of each morphological class and lens configuration for training. While deep residual networks are able to identify strong lenses with various image separations \citep[e.g.][]{Huang2020}, this poses challenges in developing machine learning pipelines reaching high completeness over broad ranges of lens halo masses ($\simeq$\,10$^{11}$--10$^{15}$\,M$_{\odot}$, from isolated galaxies to clusters).

\citet{Knabel2020} compare the citizen science classification of galaxy-scale lenses with automated deep learning and spectroscopic selections over a common footprint covered by the Galaxy and Mass Assembly and Kilo Degree surveys. Interestingly, the independent crowdsourced and machine learning searches result in essentially distinct sets of lens candidates. Despite substantial differences in the parent samples, these findings highlight the complementarity between the two approaches for ground-based searches.

Regarding space-based data, automated lens searches in optical and near-infrared imaging are currently also suffering from limitations. For instance, due to the lack of colour information, the deep learning search conducted in HST data by \citet{Pourrahmani2018} is restricted to the most massive lens ellipticals. The minor overlap with our present sample illustrates again that, even though machine learning reaches higher performance for specific classification tasks \citep[][]{Metcalf2019}, citizen science projects remain key in accessing the broad diversity of strong lenses.

\subsection{Future prospects}

Given the comparable angular resolution and sampling of {\it Euclid} and HST images, our results highlight the range of strong lensing studies that will become accessible with the wide {\it Euclid} survey. Given the large variety of lens configurations targeted in our archival search, and given the in-homogeneous selection process involving images with different depths and taken in different filters, we can not robustly measure the sample completeness. It is nonetheless interesting to note that our overall number of strong lens candidates is comparable to the re-scaled {\it Euclid} forecasts. \citet{Collett2015} predicted that 170\,000 galaxy-scale strong lenses will be discoverable from the final stack of the {\it Euclid} survey, for lens early-type galaxies and various assumptions on the source population and detectability criteria. This corresponds to $\simeq$300 strong lenses discoverable over our 27\,deg$^2$ footprint, while we found $\simeq$240 non-targeted candidates from the ACS and WFC3/UVIS classifications, including a few group-, cluster-scale and more exotic lenses which are not considered by \citet{Collett2015}. 

This adds further evidence that {\it Euclid} will be able to find large samples of $>$100\,000 strong lenses, while covering much broader ranges in lens redshifts and extending robust mass measurements to $z>1$. Strong lens searches in {\it Euclid} will likely benefit from joining CNN and crowdsourced classifications, for instance by involving citizens to assemble representative training sets, or to clean the contaminants from CNN selections in order to increase the sample purity.

\section{Conclusion}
\label{conclusion}

We present a sample of 198 new, high-quality strong lens candidates identified in 18 years of HST archival data. We performed a systematic search for non-targeted strong gravitational lenses in the entire HST archive, by inspecting observations taken with the ACS and WFC3 instruments between 2002 and 2020 with crowdsourcing.

We find 252 strong gravitational lens candidates appearing serendipitously in HST observations, after excluding targeted observations of known lenses. 54 of them have been previously mentioned in other studies, as presented in Table~\ref{tab:rediscoveries}. A total of 198 strong lens candidates were not previously reported in the literature. We categorised them into 45 grade A, 74 grade B, and 79 grade C based on their morphology, the shape of the source image, and the quality and availability of an HST colour image. Our selection purely based on morphological criteria results in strong lenses with various lens and source galaxy types, and various image multiplicities and angular distributions. This shows the potential of citizen science in accessing the broad diversity of strong lenses, and the complementarity with automated classification algorithms.

The sample contains a majority of extended, distorted arcs and rings, which are less ambiguous than compact lensed sources. While 124 (63\%) of the new strong lenses are associated with foreground spectroscopically-confirmed or candidate galaxy clusters, most of them have galaxy-scale configurations, with the light deflection dominated by a single member or line-of-sight galaxy. For instance, only three new grade A show a major contribution from the cluster-scale mass component. In addition, while we find a majority of lenses being ellipticals, $\lesssim$10\% of the foreground lens galaxies have light profiles consistent with exponential discs, corresponding either to edge-on or to moderately inclined spirals. Finally, a few systems have exotic configurations such as possible double source plane lenses.

In terms of distributions, the newly detected lenses are, on average, 1.3 magnitudes fainter than previous HST searches. The angular separations between multiple images are typically smaller than for lenses found in ground-based data. A quarter of our newly discovered systems have arc radii $\leq$0.82\arcsec, showing the advantage of combining high-resolution of HST imaging with crowdsourcing to select the most compact, galaxy-scale strong lenses for galaxy evolution studies.
The redshift distribution of the lens galaxies in our sample is consistent with the lens redshifts from previous ground-based searches.

Since we did not restrict our search to postage stamps of massive elliptical galaxies, as commonly done in strong lens searches, our study constitutes an unbiased search for lenses with the highest resolution currently possible with HST. This overall sample of 252 strong lenses is a useful benchmark for future lens searches in high-resolution images, such as those with {\it Euclid}, JWST or {\it Roman}.

This paper shows that crowdsourcing is a robust method to perform visual detection of strong lenses. Indeed, after receiving proper training and knowledge transfer, the volunteers demonstrated their ability to detect classical lenses as well as outliers such as exotic lens configurations. Although inspecting large field of view images might lead to a decrease in completeness, this approach has the advantage of considerably speeding up the detection process compared to inspecting postage stamps, where a pre-selection of the targets is often necessary. In addition, providing metadata for the images, such as the coordinates and links to ESASky, enabled the volunteers to perform further analysis and identify the objects in the archives. While using colour images for the inspection might have led to a larger number of lenses being identified, multiple bands were only available for a fraction of the Hubble images in the archives. Nevertheless, the volunteers demonstrated that they are able to identify strong lenses in grey-scale images, which is an important lesson for {\it Euclid}, where only a single optical band is available. It is important to note that, while it is significantly faster for thousands of citizen scientists to inspect the images than for a handful of professional astronomers, it still took one year of volunteer effort to explore more than 45\,000 HST observations. Given the scale of {\it Euclid}, a citizen science approach to explore the entire {\it Euclid} dataset of $\simeq$15\,000\,deg$^2$ is not feasible on its own. Combining artificial intelligence methods and crowdsourcing for the detection of strong lenses, for example through iterative training and validation, has thus a strong potential to produce reliable catalogues in shorter timescales and detect outliers such as exotic lenses. Our study also demonstrates the potential for using crowdsourcing to explore the growing JWST archives of images for strong gravitational lenses, as the high-resolution, infrared and multi-wavelength JWST observations will reveal even higher redshift lenses and with smaller Einstein radii.

To conclude, this series of Hubble Asteroid Hunter papers reaffirmed the importance of crowdsourcing in visually detecting complex objects such as strong gravitational lenses and Solar System objects. It also showed the benefits of exploring large archival datasets spanning decades in time to expand the horizons of future research.

\begin{acknowledgements}
We acknowledge the tremendous work made by the citizen scientist volunteers on the Hubble Asteroid Hunter project. Their contributions are individually acknowledged on \url{https://www.zooniverse.org/projects/sandorkruk/hubble-asteroid-hunter/about/results}. \\
\\
We thank the anonymous referee for their detailed and insightful comments, as they greatly contributed to enhance the quality of our paper. We thank Stella Seitz and Sherry Suyu for insightful discussions on the sample of strong gravitational lenses.\\
\\
This work has been conducted by EOG as part of the European Space Agency (ESA) student internship. SK and RB gratefully acknowledge support from the ESA Research Fellowship.\\
\\
This paper is based on observations made with the NASA/ESA {\it Hubble Space Telescope}, obtained from the data archive at the Space Telescope Science Institute. STScI is operated by the Association of Universities for Research in Astronomy, Inc. under NASA contract NAS 5-26555. Based on observations made with the NASA/ESA {\it Hubble Space Telescope}, and obtained from the Hubble Legacy Archive, which is a collaboration between the Space Telescope Science Institute (STScI/NASA), the Space Telescope European Coordinating Facility (ST-ECF/ESA) and the Canadian Astronomy Data Centre (CADC/NRC/CSA). \\
\\
The new strong lens candidates (Table~\ref{tab:unpublished}) were found in the following HST programs: 9405, 9414, 9427, 9458, 9476, 9483, 9500, 9722, 9753, 9770, 9822, 9836, 10096, 10134, 10152, 10200, 10207, 10325, 10326, 10334, 10395, 10420, 10491, 10496, 10503, 10504, 10505, 10521, 10523, 10569, 10626, 10635, 10816, 10825, 10861, 10875, 10880, 10881, 10997, 11142, 11588, 11597, 11613, 11697, 11734, 12063, 12064, 12104, 12166, 12195, 12209, 12238, 12253, 12286, 12313, 12319, 12362, 12476, 12477, 12515, 12546, 12549, 12555, 12575, 12591, 12756, 12884, 12898, 12937, 13023, 13024, 13307, 13352, 13364, 13393, 13412, 13442, 13495, 13496, 13514, 13641, 13657, 13695, 13698, 13711, 13750, 13845, 13942, 14096, 14098, 14118, 14165, 14199, 14594, 14662, 14766, 14808, 15063, 15117, 15121, 15183, 15212, 15230, 15275, 15287, 15307, 15320, 15378, 15446, 15495, 15608, 15642, 15644, 15654, 15696, 15843, 16025. We are extremely grateful to the PIs of the HST observations for their positive feedback and responses in including the objects in our catalogue and for the useful conversations.\\
\\
This publication uses data generated via the Zooniverse.org platform, development of which is funded by generous support, including a Global Impact Award from Google, and by a grant from the Alfred P. Sloan Foundation. Funding for the SDSS and SDSS-II has been provided by the Alfred P. Sloan Foundation, the Participating Institutions, the National Science Foundation, the U.S. Department of Energy, the National Aeronautics and Space Administration, the Japanese Monbukagakusho, the Max Planck Society, and the Higher Education Funding Council for England. The SDSS Web Site is http://www.sdss.org/. \\
\\
The SDSS is managed by the Astrophysical Research Consortium for the Participating Institutions. The Participating Institutions are the American Museum of Natural History, Astrophysical Institute Potsdam, University of Basel, University of Cambridge, Case Western Reserve University, University of Chicago, Drexel University, Fermilab, the Institute for Advanced Study, the Japan Participation Group, Johns Hopkins University, the Joint Institute for Nuclear Astrophysics, the Kavli Institute for Particle Astrophysics and Cosmology, the Korean Scientist Group, the Chinese Academy of Sciences (LAMOST), Los Alamos National Laboratory, the Max-Planck-Institute for Astronomy (MPIA), the Max-Planck-Institute for Astrophysics (MPA), New Mexico State University, Ohio State University, University of Pittsburgh, University of Portsmouth, Princeton University, the United States Naval Observatory and the University of Washington.\\
\\ 
This research made use of NASA's Astrophysics Data System Bibliographic Services. This work made extensive use of \textit{Astropy}\footnote{\url{http://www.astropy.org/}}, a community-developed core Python package for Astronomy \citep{Astropy} and of the Tool for Operations on Catalogues And Tables \citep[TOPCAT\footnote{\url{http://www.star.bris.ac.uk/~mbt/}};][]{Topcat}. This publication used \texttt{GALFIT} \citep{Peng2002} and \texttt{SExtractor} \citep{Bertin1996} for the lens fitting and parameter retrieval, as well as the software tool \texttt{R} \citep{R2020} and its packages \texttt{vioplot} \citep{Rvioplot} and \texttt{Hmisc} \citep{Rharrell2019package}  for preparation of the tables and data visualisation. This work also made use of Xmatch \citep{Budavari2013} and TOPCAT to perform cross-matching of the lenses with nearby clusters.

\end{acknowledgements}

\bibliographystyle{aa} 
\bibliography{references.bib}

\begin{thebibliography}{128}
\expandafter\ifx\csname natexlab\endcsname\relax\def\natexlab#1{#1}\fi

\bibitem[{Adler \& Kelly(2020)}]{Rvioplot}
Adler, D. \& Kelly, S.~T. 2020, vioplot: violin plot, r package version 0.3.7

\bibitem[{{Ahumada} {et~al.}(2020){Ahumada}, {Prieto}, {Almeida}, {Anders},
  {Anderson}, {Andrews}, {Anguiano}, {Arcodia}, {Armengaud}, {Aubert}, {Avila},
  {Avila-Reese}, {Badenes}, {Balland}, {Barger}, {Barrera-Ballesteros}, {Basu},
  {Bautista}, {Beaton}, {Beers}, {Benavides}, {Bender}, {Bernardi}, {Bershady},
  {Beutler}, {Bidin}, {Bird}, {Bizyaev}, {Blanc}, {Blanton}, {Boquien},
  {Borissova}, {Bovy}, {Brandt}, {Brinkmann}, {Brownstein}, {Bundy}, {Bureau},
  {Burgasser}, {Burtin}, {Cano-D{\'\i}az}, {Capasso}, {Cappellari}, {Carrera},
  {Chabanier}, {Chaplin}, {Chapman}, {Cherinka}, {Chiappini}, {Doohyun Choi},
  {Chojnowski}, {Chung}, {Clerc}, {Coffey}, {Comerford}, {Comparat}, {da
  Costa}, {Cousinou}, {Covey}, {Crane}, {Cunha}, {Ilha}, {Dai}, {Damsted},
  {Darling}, {Davidson}, {Davies}, {Dawson}, {De}, {de la Macorra}, {De Lee},
  {Queiroz}, {Deconto Machado}, {de la Torre}, {Dell'Agli}, {du Mas des
  Bourboux}, {Diamond-Stanic}, {Dillon}, {Donor}, {Drory}, {Duckworth},
  {Dwelly}, {Ebelke}, {Eftekharzadeh}, {Davis Eigenbrot}, {Elsworth},
  {Eracleous}, {Erfanianfar}, {Escoffier}, {Fan}, {Farr},
  {Fern{\'a}ndez-Trincado}, {Feuillet}, {Finoguenov}, {Fofie},
  {Fraser-McKelvie}, {Frinchaboy}, {Fromenteau}, {Fu}, {Galbany}, {Garcia},
  {Garc{\'\i}a-Hern{\'a}ndez}, {Oehmichen}, {Ge}, {Maia}, {Geisler}, {Gelfand},
  {Goddy}, {Gonzalez-Perez}, {Grabowski}, {Green}, {Grier}, {Guo}, {Guy},
  {Harding}, {Hasselquist}, {Hawken}, {Hayes}, {Hearty}, {Hekker}, {Hogg},
  {Holtzman}, {Horta}, {Hou}, {Hsieh}, {Huber}, {Hunt}, {Chitham}, {Imig},
  {Jaber}, {Angel}, {Johnson}, {Jones}, {J{\"o}nsson}, {Jullo}, {Kim},
  {Kinemuchi}, {Kirkpatrick}, {Kite}, {Klaene}, {Kneib}, {Kollmeier}, {Kong},
  {Kounkel}, {Krishnarao}, {Lacerna}, {Lan}, {Lane}, {Law}, {Le Goff}, {Leung},
  {Lewis}, {Li}, {Lian}, {Lin}, {Long}, {Longa-Pe{\~n}a}, {Lundgren}, {Lyke},
  {Ted Mackereth}, {MacLeod}, {Majewski}, {Manchado}, {Maraston}, {Martini},
  {Masseron}, {Masters}, {Mathur}, {McDermid}, {Merloni}, {Merrifield},
  {M{\'e}sz{\'a}ros}, {Miglio}, {Minniti}, {Minsley}, {Miyaji}, {Mohammad},
  {Mosser}, {Mueller}, {Muna}, {Mu{\~n}oz-Guti{\'e}rrez}, {Myers}, {Nadathur},
  {Nair}, {Nandra}, {do Nascimento}, {Nevin}, {Newman}, {Nidever}, {Nitschelm},
  {Noterdaeme}, {O'Connell}, {Olmstead}, {Oravetz}, {Oravetz}, {Osorio},
  {Pace}, {Padilla}, {Palanque-Delabrouille}, {Palicio}, {Pan}, {Pan},
  {Parker}, {Paviot}, {Peirani}, {Ram{\'r}ez}, {Penny}, {Percival},
  {Perez-Fournon}, {P{\'e}rez-R{\`a}fols}, {Petitjean}, {Pieri},
  {Pinsonneault}, {Poovelil}, {Povick}, {Prakash}, {Price-Whelan}, {Raddick},
  {Raichoor}, {Ray}, {Rembold}, {Rezaie}, {Riffel}, {Riffel}, {Rix}, {Robin},
  {Roman-Lopes}, {Rom{\'a}n-Z{\'u}{\~n}iga}, {Rose}, {Ross}, {Rossi},
  {Rowlands}, {Rubin}, {Salvato}, {S{\'a}nchez}, {S{\'a}nchez-Menguiano},
  {S{\'a}nchez-Gallego}, {Sayres}, {Schaefer}, {Schiavon}, {Schimoia},
  {Schlafly}, {Schlegel}, {Schneider}, {Schultheis}, {Schwope}, {Seo},
  {Serenelli}, {Shafieloo}, {Shamsi}, {Shao}, {Shen}, {Shetrone}, {Shirley},
  {Aguirre}, {Simon}, {Skrutskie}, {Slosar}, {Smethurst}, {Sobeck}, {Sodi},
  {Souto}, {Stark}, {Stassun}, {Steinmetz}, {Stello}, {Stermer},
  {Storchi-Bergmann}, {Streblyanska}, {Stringfellow}, {Stutz}, {Su{\'a}rez},
  {Sun}, {Taghizadeh-Popp}, {Talbot}, {Tayar}, {Thakar}, {Theriault}, {Thomas},
  {Thomas}, {Tinker}, {Tojeiro}, {Toledo}, {Tremonti}, {Troup}, {Tuttle},
  {Unda-Sanzana}, {Valentini}, {Vargas-Gonz{\'a}lez}, {Vargas-Maga{\~n}a},
  {V{\'a}zquez-Mata}, {Vivek}, {Wake}, {Wang}, {Weaver}, {Weijmans}, {Wild},
  {Wilson}, {Wilson}, {Wolthuis}, {Wood-Vasey}, {Yan}, {Yang}, {Y{\`e}che},
  {Zamora}, {Zarrouk}, {Zasowski}, {Zhang}, {Zhao}, {Zhao}, {Zheng}, {Zheng},
  {Zhu}, \& {Zou}}]{Ahumada2020}
{Ahumada}, R., {Prieto}, C.~A., {Almeida}, A., {et~al.} 2020, \apjs, 249, 3

\bibitem[{{Alard}(2006)}]{Alard2006}
{Alard}, C. 2006, arXiv e-prints, astro

\bibitem[{{Atek} {et~al.}(2014){Atek}, {Richard}, {Kneib}, {Clement}, {Egami},
  {Ebeling}, {Jauzac}, {Jullo}, {Laporte}, {Limousin}, \&
  {Natarajan}}]{Atek2014}
{Atek}, H., {Richard}, J., {Kneib}, J.-P., {et~al.} 2014, \apj, 786, 60

\bibitem[{{Barnab{\`e}} {et~al.}(2009){Barnab{\`e}}, {Czoske}, {Koopmans},
  {Treu}, {Bolton}, \& {Gavazzi}}]{Barnabe2009}
{Barnab{\`e}}, M., {Czoske}, O., {Koopmans}, L. V.~E., {et~al.} 2009, \mnras,
  399, 21

\bibitem[{{Belokurov} {et~al.}(2009){Belokurov}, {Evans}, {Hewett}, {Moiseev},
  {McMahon}, {Sanchez}, \& {King}}]{Belokurov2009}
{Belokurov}, V., {Evans}, N.~W., {Hewett}, P.~C., {et~al.} 2009, \mnras, 392,
  104

\bibitem[{{Bertin} \& {Arnouts}(1996)}]{Bertin1996}
{Bertin}, E. \& {Arnouts}, S. 1996, \aaps, 117, 393

\bibitem[{{Bettoni} {et~al.}(2019){Bettoni}, {Falomo}, {Scarpa}, {Negrello},
  {Omizzolo}, {Corradi}, {Reverte}, \& {Vulcani}}]{Bettoni2019}
{Bettoni}, D., {Falomo}, R., {Scarpa}, R., {et~al.} 2019, \apjl, 873, L14

\bibitem[{{Bhatawdekar} \& {Conselice}(2021)}]{Bhatawdekar2021}
{Bhatawdekar}, R. \& {Conselice}, C.~J. 2021, \apj, 909, 144

\bibitem[{{Bhatawdekar} {et~al.}(2019){Bhatawdekar}, {Conselice},
  {Margalef-Bentabol}, \& {Duncan}}]{Bhatawdekar2019}
{Bhatawdekar}, R., {Conselice}, C.~J., {Margalef-Bentabol}, B., \& {Duncan}, K.
  2019, \mnras, 486, 3805

\bibitem[{{Bolton} {et~al.}(2012){Bolton}, {Brownstein}, {Kochanek}, {Shu},
  {Schlegel}, {Eisenstein}, {Wake}, {Connolly}, {Maraston}, {Arneson}, \&
  {Weaver}}]{Bolton2012}
{Bolton}, A.~S., {Brownstein}, J.~R., {Kochanek}, C.~S., {et~al.} 2012, \apj,
  757, 82

\bibitem[{{Bolton} {et~al.}(2008){Bolton}, {Burles}, {Koopmans}, {Treu},
  {Gavazzi}, {Moustakas}, {Wayth}, \& {Schlegel}}]{Bolton2008}
{Bolton}, A.~S., {Burles}, S., {Koopmans}, L. V.~E., {et~al.} 2008, \apj, 682,
  964

\bibitem[{{Bom} {et~al.}(2017){Bom}, {Makler}, {Albuquerque}, \& {Brand
  t}}]{Bom2017}
{Bom}, C.~R., {Makler}, M., {Albuquerque}, M.~P., \& {Brand t}, C.~H. 2017,
  \aap, 597, A135

\bibitem[{{Brownstein} {et~al.}(2012){Brownstein}, {Bolton}, {Schlegel},
  {Eisenstein}, {Kochanek}, {Connolly}, {Maraston}, {Pandey}, {Seitz}, {Wake},
  {Wood-Vasey}, {Brinkmann}, {Schneider}, \& {Weaver}}]{Brownstein2012}
{Brownstein}, J.~R., {Bolton}, A.~S., {Schlegel}, D.~J., {et~al.} 2012, \apj,
  744, 41

\bibitem[{{Budavari} \& {Lee}(2013)}]{Budavari2013}
{Budavari}, T. \& {Lee}, M.~A. 2013, {Xmatch: GPU Enhanced Astronomic Catalog
  Cross-Matching}, Astrophysics Source Code Library, record ascl:1303.021

\bibitem[{{Ca{\~n}ameras} {et~al.}(2017){Ca{\~n}ameras}, {Nesvadba}, {Kneissl},
  {Frye}, {Gavazzi}, {Koenig}, {Le Floc'h}, {Limousin}, {Oteo}, \&
  {Scott}}]{Canameras2017}
{Ca{\~n}ameras}, R., {Nesvadba}, N., {Kneissl}, R., {et~al.} 2017, \aap, 604,
  A117

\bibitem[{{Ca{\~n}ameras} {et~al.}(2020){Ca{\~n}ameras}, {Schuldt}, {Suyu},
  {Taubenberger}, {Meinhardt}, {Leal-Taix{\'e}}, {Lemon}, {Rojas}, \&
  {Savary}}]{Canameras2020}
{Ca{\~n}ameras}, R., {Schuldt}, S., {Suyu}, S.~H., {et~al.} 2020, \aap, 644,
  A163

\bibitem[{{Cabanac} {et~al.}(2007){Cabanac}, {Alard}, {Dantel-Fort}, {Fort},
  {Gavazzi}, {Gomez}, {Kneib}, {Le F{\`e}vre}, {Mellier}, {Pello}, {Soucail},
  {Sygnet}, \& {Valls-Gabaud}}]{Cabanac2007}
{Cabanac}, R.~A., {Alard}, C., {Dantel-Fort}, M., {et~al.} 2007, \aap, 461, 813

\bibitem[{{Caminha} {et~al.}(2019){Caminha}, {Rosati}, {Grillo}, {Rosani},
  {Caputi}, {Meneghetti}, {Mercurio}, {Balestra}, {Bergamini}, {Biviano},
  {Nonino}, {Umetsu}, {Vanzella}, {Annunziatella}, {Broadhurst},
  {Delgado-Correal}, {Demarco}, {Koekemoer}, {Lombardi}, {Maier}, {Verdugo}, \&
  {Zitrin}}]{Caminha2019}
{Caminha}, G.~B., {Rosati}, P., {Grillo}, C., {et~al.} 2019, \aap, 632, A36

\bibitem[{{Cappellari} {et~al.}(2015){Cappellari}, {Romanowsky}, {Brodie},
  {Forbes}, {Strader}, {Foster}, {Kartha}, {Pastorello}, {Pota}, {Spitler},
  {Usher}, \& {Arnold}}]{Cappellari2015}
{Cappellari}, M., {Romanowsky}, A.~J., {Brodie}, J.~P., {et~al.} 2015, \apjl,
  804, L21

\bibitem[{Clowe {et~al.}(2012)Clowe, Markevitch, Brada{\v{c}}, Gonzalez, Chung,
  Massey, \& Zaritsky}]{Clowe2012}
Clowe, D., Markevitch, M., Brada{\v{c}}, M., {et~al.} 2012, The Astrophysical
  Journal, 758, 128

\bibitem[{{Collett}(2015)}]{Collett2015}
{Collett}, T.~E. 2015, \apj, 811, 20

\bibitem[{{Collett} {et~al.}(2012){Collett}, {Auger}, {Belokurov}, {Marshall},
  \& {Hall}}]{Collett2012}
{Collett}, T.~E., {Auger}, M.~W., {Belokurov}, V., {Marshall}, P.~J., \&
  {Hall}, A.~C. 2012, \mnras, 424, 2864

\bibitem[{{Courbin} {et~al.}(2018){Courbin}, {Bonvin}, {Buckley-Geer},
  {Fassnacht}, {Frieman}, {Lin}, {Marshall}, {Suyu}, {Treu}, {Anguita},
  {Motta}, {Meylan}, {Paic}, {Tewes}, {Agnello}, {Chao}, {Chijani}, {Gilman},
  {Rojas}, {Williams}, {Hempel}, {Kim}, {Lachaume}, {Rabus}, {Abbott}, {Allam},
  {Annis}, {Banerji}, {Bechtol}, {Benoit-L{\'e}vy}, {Brooks}, {Burke}, {Carnero
  Rosell}, {Carrasco Kind}, {Carretero}, {D'Andrea}, {da Costa}, {Davis},
  {DePoy}, {Desai}, {Flaugher}, {Fosalba}, {Garc{\'\i}a-Bellido}, {Gaztanaga},
  {Goldstein}, {Gruen}, {Gruendl}, {Gschwend}, {Gutierrez}, {Honscheid},
  {James}, {Kuehn}, {Kuhlmann}, {Kuropatkin}, {Lahav}, {Lima}, {Maia}, {March},
  {Marshall}, {McMahon}, {Menanteau}, {Miquel}, {Nord}, {Plazas}, {Sanchez},
  {Scarpine}, {Schindler}, {Schubnell}, {Sevilla-Noarbe}, {Smith},
  {Soares-Santos}, {Sobreira}, {Suchyta}, {Tarle}, {Tucker}, {Walker}, \&
  {Wester}}]{Courbin2018}
{Courbin}, F., {Bonvin}, V., {Buckley-Geer}, E., {et~al.} 2018, \aap, 609, A71

\bibitem[{{Derkenne} {et~al.}(2021){Derkenne}, {McDermid}, {Poci}, {Remus},
  {J{\o}rgensen}, \& {Emsellem}}]{Derkenne2021}
{Derkenne}, C., {McDermid}, R.~M., {Poci}, A., {et~al.} 2021, \mnras, 506, 3691

\bibitem[{{Desprez} {et~al.}(2018){Desprez}, {Richard}, {Jauzac}, {Martinez},
  {Siana}, \& {Cl{\'e}ment}}]{Desprez2018}
{Desprez}, G., {Richard}, J., {Jauzac}, M., {et~al.} 2018, \mnras, 479, 2630

\bibitem[{{Dessauges-Zavadsky} {et~al.}(2015){Dessauges-Zavadsky}, {Zamojski},
  {Schaerer}, {Combes}, {Egami}, {Swinbank}, {Richard}, {Sklias}, {Rawle},
  {Rex}, {Kneib}, {Boone}, \& {Blain}}]{Dessauges2015}
{Dessauges-Zavadsky}, M., {Zamojski}, M., {Schaerer}, D., {et~al.} 2015, \aap,
  577, A50

\bibitem[{{Diego} {et~al.}(2015){Diego}, {Broadhurst}, {Benitez}, {Lim}, \&
  {Lam}}]{Diego2015}
{Diego}, J.~M., {Broadhurst}, T., {Benitez}, N., {Lim}, J., \& {Lam}, D. 2015,
  \mnras, 449, 588

\bibitem[{{Dieleman} {et~al.}(2015){Dieleman}, {Willett}, \&
  {Dambre}}]{Dieleman2015}
{Dieleman}, S., {Willett}, K.~W., \& {Dambre}, J. 2015, \mnras, 450, 1441

\bibitem[{{D'Isanto} \& {Polsterer}(2018)}]{Disanto2018}
{D'Isanto}, A. \& {Polsterer}, K.~L. 2018, \aap, 609, A111

\bibitem[{{Ebeling} {et~al.}(2019){Ebeling}, {Atek}, {Edge}, {Kaiser}, {Kneib},
  {Limousin}, {McPartland}, {Repp}, {Richard}, \& {Toft}}]{Ebeling2019}
{Ebeling}, H., {Atek}, H., {Edge}, A.~C., {et~al.} 2019, {Beyond MACS: A
  Snapshot Survey of the Most Massive Clusters of Galaxies at z=0.5-1}, HST
  Proposal

\bibitem[{{Ebeling} \& {Kalita}(2019)}]{Ebeling2019b}
{Ebeling}, H. \& {Kalita}, B.~S. 2019, \apj, 882, 127

\bibitem[{{Ebeling} {et~al.}(2014){Ebeling}, {Ma}, \& {Barrett}}]{Ebeling2014}
{Ebeling}, H., {Ma}, C.-J., \& {Barrett}, E. 2014, \apjs, 211, 21

\bibitem[{{Ebeling} {et~al.}(2018){Ebeling}, {Stockmann}, {Richard}, {Zabl},
  {Brammer}, {Toft}, \& {Man}}]{Ebeling2018}
{Ebeling}, H., {Stockmann}, M., {Richard}, J., {et~al.} 2018, \apjl, 852, L7

\bibitem[{{Fassnacht} {et~al.}(2004){Fassnacht}, {Moustakas}, {Casertano},
  {Ferguson}, {Lucas}, \& {Park}}]{Fassnacht2004}
{Fassnacht}, C.~D., {Moustakas}, L.~A., {Casertano}, S., {et~al.} 2004, \apjl,
  600, L155

\bibitem[{{Faure} {et~al.}(2008){Faure}, {Kneib}, {Covone}, {Tasca},
  {Leauthaud}, {Capak}, {Jahnke}, {Smolcic}, {de la Torre}, {Ellis},
  {Finoguenov}, {Koekemoer}, {Le Fevre}, {Massey}, {Mellier}, {Refregier},
  {Rhodes}, {Scoville}, {Schinnerer}, {Taylor}, {Van Waerbeke}, \&
  {Walcher}}]{Faure2008}
{Faure}, C., {Kneib}, J.-P., {Covone}, G., {et~al.} 2008, \apjs, 176, 19

\bibitem[{{Fort} {et~al.}(1992){Fort}, {Le Fevre}, {Hammer}, \&
  {Cailloux}}]{Fort1992}
{Fort}, B., {Le Fevre}, O., {Hammer}, F., \& {Cailloux}, M. 1992, \apjl, 399,
  L125

\bibitem[{{Gavazzi} {et~al.}(2014){Gavazzi}, {Marshall}, {Treu}, \&
  {Sonnenfeld}}]{Gavazzi2014}
{Gavazzi}, R., {Marshall}, P.~J., {Treu}, T., \& {Sonnenfeld}, A. 2014, \apj,
  785, 144

\bibitem[{{Gavazzi} {et~al.}(2008){Gavazzi}, {Treu}, {Koopmans}, {Bolton},
  {Moustakas}, {Burles}, \& {Marshall}}]{Gavazzi2008}
{Gavazzi}, R., {Treu}, T., {Koopmans}, L. V.~E., {et~al.} 2008, \apj, 677, 1046

\bibitem[{{Giordano} {et~al.}(2018){Giordano}, {Racero}, {Norman},
  {Vall{\'e}s}, {Mer{\'\i}n}, {Baines}, {L{\'o}pez-Caniego}, {Mart{\'\i}}, {de
  Teodoro}, {Salgado}, {Sarmiento}, {Guti{\'e}rrez-S{\'a}nchez}, {Prieto},
  {Lorca}, {Alberola}, {Valtchanov}, {de Marchi}, {{\'A}lvarez}, \&
  {Arviset}}]{Giordano2018}
{Giordano}, F., {Racero}, E., {Norman}, H., {et~al.} 2018, Astronomy and
  Computing, 24, 97

\bibitem[{{Gonzaga} {et~al.}(2012){Gonzaga}, {Hack}, {Fruchter}, \&
  {Mack}}]{Gonzaga2012}
{Gonzaga}, S., {Hack}, W., {Fruchter}, A., \& {Mack}, J. 2012, {The DrizzlePac
  Handbook. (Baltimore, STScI)}

\bibitem[{{Grillo} {et~al.}(2008){Grillo}, {Lombardi}, {Rosati}, {Bertin},
  {Gobat}, {Demarco}, {Lidman}, {Motta}, \& {Nonino}}]{Grillo2008}
{Grillo}, C., {Lombardi}, M., {Rosati}, P., {et~al.} 2008, \aap, 486, 45

\bibitem[{{Grogin} {et~al.}(2011){Grogin}, {Kocevski}, {Faber}, {Ferguson},
  {Koekemoer}, {Riess}, {Acquaviva}, {Alexander}, {Almaini}, {Ashby}, {Barden},
  {Bell}, {Bournaud}, {Brown}, {Caputi}, {Casertano}, {Cassata}, {Castellano},
  {Challis}, {Chary}, {Cheung}, {Cirasuolo}, {Conselice}, {Roshan Cooray},
  {Croton}, {Daddi}, {Dahlen}, {Dav{\'e}}, {de Mello}, {Dekel}, {Dickinson},
  {Dolch}, {Donley}, {Dunlop}, {Dutton}, {Elbaz}, {Fazio}, {Filippenko},
  {Finkelstein}, {Fontana}, {Gardner}, {Garnavich}, {Gawiser}, {Giavalisco},
  {Grazian}, {Guo}, {Hathi}, {H{\"a}ussler}, {Hopkins}, {Huang}, {Huang},
  {Jha}, {Kartaltepe}, {Kirshner}, {Koo}, {Lai}, {Lee}, {Li}, {Lotz}, {Lucas},
  {Madau}, {McCarthy}, {McGrath}, {McIntosh}, {McLure}, {Mobasher},
  {Moustakas}, {Mozena}, {Nandra}, {Newman}, {Niemi}, {Noeske}, {Papovich},
  {Pentericci}, {Pope}, {Primack}, {Rajan}, {Ravindranath}, {Reddy}, {Renzini},
  {Rix}, {Robaina}, {Rodney}, {Rosario}, {Rosati}, {Salimbeni}, {Scarlata},
  {Siana}, {Simard}, {Smidt}, {Somerville}, {Spinrad}, {Straughn}, {Strolger},
  {Telford}, {Teplitz}, {Trump}, {van der Wel}, {Villforth}, {Wechsler},
  {Weiner}, {Wiklind}, {Wild}, {Wilson}, {Wuyts}, {Yan}, \& {Yun}}]{Grogin2011}
{Grogin}, N.~A., {Kocevski}, D.~D., {Faber}, S.~M., {et~al.} 2011, \apjs, 197,
  35

\bibitem[{{Guzzo} {et~al.}(2009){Guzzo}, {Schuecker}, {B{\"o}hringer},
  {Collins}, {Ortiz-Gil}, {de Grandi}, {Edge}, {Neumann}, {Schindler},
  {Altucci}, \& {Shaver}}]{Guzzo2009}
{Guzzo}, L., {Schuecker}, P., {B{\"o}hringer}, H., {et~al.} 2009, \aap, 499,
  357

\bibitem[{Harrell~Jr \& Harrell~Jr(2019)}]{Rharrell2019package}
Harrell~Jr, F.~E. \& Harrell~Jr, M. F.~E. 2019, CRAN2018, 2019, 235

\bibitem[{{Holwerda} {et~al.}(2015){Holwerda}, {Baldry}, {Alpaslan}, {Bauer},
  {Bland-Hawthorn}, {Brough}, {Brown}, {Cluver}, {Conselice}, {Driver},
  {Hopkins}, {Jones}, {L{\'o}pez-S{\'a}nchez}, {Loveday}, {Meyer}, \&
  {Moffett}}]{Holwerda2015}
{Holwerda}, B.~W., {Baldry}, I.~K., {Alpaslan}, M., {et~al.} 2015, \mnras, 449,
  4277

\bibitem[{{Horesh} {et~al.}(2010){Horesh}, {Maoz}, {Ebeling}, {Seidel}, \&
  {Bartelmann}}]{Horesh2010}
{Horesh}, A., {Maoz}, D., {Ebeling}, H., {Seidel}, G., \& {Bartelmann}, M.
  2010, \mnras, 406, 1318

\bibitem[{{Huang} {et~al.}(2020){Huang}, {Storfer}, {Ravi}, {Pilon}, {Domingo},
  {Schlegel}, {Bailey}, {Dey}, {Gupta}, {Herrera}, {Juneau}, {Landriau},
  {Lang}, {Meisner}, {Moustakas}, {Myers}, {Schlafly}, {Valdes}, {Weaver},
  {Yang}, \& {Y{\`e}che}}]{Huang2020}
{Huang}, X., {Storfer}, C., {Ravi}, V., {et~al.} 2020, \apj, 894, 78

\bibitem[{{Huertas-Company} {et~al.}(2015){Huertas-Company}, {Gravet},
  {Cabrera-Vives}, {P{\'e}rez-Gonz{\'a}lez}, {Kartaltepe}, {Barro}, {Bernardi},
  {Mei}, {Shankar}, {Dimauro}, {Bell}, {Kocevski}, {Koo}, {Faber}, \&
  {Mcintosh}}]{Huertas-Company2015}
{Huertas-Company}, M., {Gravet}, R., {Cabrera-Vives}, G., {et~al.} 2015, \apjs,
  221, 8

\bibitem[{{Jackson}(2008)}]{Jackson2008}
{Jackson}, N. 2008, \mnras, 389, 1311

\bibitem[{{Jacobs} {et~al.}(2019){Jacobs}, {Collett}, {Glazebrook},
  {Buckley-Geer}, {Diehl}, {Lin}, {McCarthy}, {Qin}, {Odden}, {Caso Escudero},
  {Dial}, {Yung}, {Gaitsch}, {Pellico}, {Lindgren}, {Abbott}, {Annis}, {Avila},
  {Brooks}, {Burke}, {Carnero Rosell}, {Carrasco Kind}, {Carretero}, {da
  Costa}, {De Vicente}, {Fosalba}, {Frieman}, {Garc{\'\i}a-Bellido},
  {Gaztanaga}, {Goldstein}, {Gruen}, {Gruendl}, {Gschwend}, {Hollowood},
  {Honscheid}, {Hoyle}, {James}, {Krause}, {Kuropatkin}, {Lahav}, {Lima},
  {Maia}, {Marshall}, {Miquel}, {Plazas}, {Roodman}, {Sanchez}, {Scarpine},
  {Serrano}, {Sevilla-Noarbe}, {Smith}, {Sobreira}, {Suchyta}, {Swanson},
  {Tarle}, {Vikram}, {Walker}, {Zhang}, \& {DES Collaboration}}]{Jacobs2019}
{Jacobs}, C., {Collett}, T., {Glazebrook}, K., {et~al.} 2019, \apjs, 243, 17

\bibitem[{{Jarosik} {et~al.}(2011){Jarosik}, {Bennett}, {Dunkley}, {Gold},
  {Greason}, {Halpern}, {Hill}, {Hinshaw}, {Kogut}, {Komatsu}, {Larson},
  {Limon}, {Meyer}, {Nolta}, {Odegard}, {Page}, {Smith}, {Spergel}, {Tucker},
  {Weiland}, {Wollack}, \& {Wright}}]{Jarosik2011}
{Jarosik}, N., {Bennett}, C.~L., {Dunkley}, J., {et~al.} 2011, \apjs, 192, 14

\bibitem[{{Jauzac} {et~al.}(2015){Jauzac}, {Richard}, {Jullo}, {Cl{\'e}ment},
  {Limousin}, {Kneib}, {Ebeling}, {Natarajan}, {Rodney}, {Atek}, {Massey},
  {Eckert}, {Egami}, \& {Rexroth}}]{Jauzac2015}
{Jauzac}, M., {Richard}, J., {Jullo}, E., {et~al.} 2015, \mnras, 452, 1437

\bibitem[{{Kelly} {et~al.}(2015){Kelly}, {Rodney}, {Treu}, {Foley}, {Brammer},
  {Schmidt}, {Zitrin}, {Sonnenfeld}, {Strolger}, {Graur}, {Filippenko}, {Jha},
  {Riess}, {Bradac}, {Weiner}, {Scolnic}, {Malkan}, {von der Linden}, {Trenti},
  {Hjorth}, {Gavazzi}, {Fontana}, {Merten}, {McCully}, {Jones}, {Postman},
  {Dressler}, {Patel}, {Cenko}, {Graham}, \& {Tucker}}]{kelly2015multiple}
{Kelly}, P.~L., {Rodney}, S.~A., {Treu}, T., {et~al.} 2015, Science, 347, 1123

\bibitem[{{Kikuchihara} {et~al.}(2020){Kikuchihara}, {Ouchi}, {Ono},
  {Mawatari}, {Chevallard}, {Harikane}, {Kojima}, {Oguri}, {Bruzual}, \&
  {Charlot}}]{Kikuchihara2020}
{Kikuchihara}, S., {Ouchi}, M., {Ono}, Y., {et~al.} 2020, \apj, 893, 60

\bibitem[{{Knabel} {et~al.}(2020){Knabel}, {Steele}, {Holwerda}, {Bridge},
  {Jacques}, {Hopkins}, {Bamford}, {Brown}, {Brough}, {Kelvin}, {Bilicki}, \&
  {Kielkopf}}]{Knabel2020}
{Knabel}, S., {Steele}, R.~L., {Holwerda}, B.~W., {et~al.} 2020, \aj, 160, 223

\bibitem[{{Koekemoer} {et~al.}(2007){Koekemoer}, {Aussel}, {Calzetti}, {Capak},
  {Giavalisco}, {Kneib}, {Leauthaud}, {Le F{\`e}vre}, {McCracken}, {Massey},
  {Mobasher}, {Rhodes}, {Scoville}, \& {Shopbell}}]{Koekemoer2007}
{Koekemoer}, A.~M., {Aussel}, H., {Calzetti}, D., {et~al.} 2007, \apjs, 172,
  196

\bibitem[{{Koester} {et~al.}(2010){Koester}, {Gladders}, {Hennawi}, {Sharon},
  {Wuyts}, {Rigby}, {Bayliss}, \& {Dahle}}]{Koester2010}
{Koester}, B.~P., {Gladders}, M.~D., {Hennawi}, J.~F., {et~al.} 2010, \apjl,
  723, L73

\bibitem[{{Koopmans} {et~al.}(2006){Koopmans}, {Treu}, {Bolton}, {Burles}, \&
  {Moustakas}}]{Koopmans2006}
{Koopmans}, L. V.~E., {Treu}, T., {Bolton}, A.~S., {Burles}, S., \&
  {Moustakas}, L.~A. 2006, \apj, 649, 599

\bibitem[{{Krist}(1995)}]{TinyTim1995}
{Krist}, J. 1995, in Astronomical Society of the Pacific Conference Series,
  Vol.~77, Astronomical Data Analysis Software and Systems IV, ed. R.~A.
  {Shaw}, H.~E. {Payne}, \& J.~J.~E. {Hayes}, 349

\bibitem[{{Kruk} {et~al.}(2022){Kruk}, {Garc{\'\i}a Mart{\'\i}n}, {Popescu},
  {Mer{\'\i}n}, {Mahlke}, {Carry}, {Thomson}, {Karadag}, {Dur{\'a}n}, {Racero},
  {Giordano}, {Baines}, {de Marchi}, \& {Laureijs}}]{Kruk2022}
{Kruk}, S., {Garc{\'\i}a Mart{\'\i}n}, P., {Popescu}, M., {et~al.} 2022, arXiv
  e-prints, arXiv:2202.00246

\bibitem[{Küng {et~al.}(2015)Küng, Saha, More, Baeten, Coles, Cornen,
  Macmillan, Marshall, More, Odermatt, Verma, \& Wilcox}]{Kung2015}
Küng, R., Saha, P., More, A., {et~al.} 2015, Monthly Notices of the Royal
  Astronomical Society, 447, 2170

\bibitem[{{Lagattuta} {et~al.}(2010){Lagattuta}, {Fassnacht}, {Auger},
  {Marshall}, {Brada{\v{c}}}, {Treu}, {Gavazzi}, {Schrabback}, {Faure}, \&
  {Anguita}}]{Lagattuta2010}
{Lagattuta}, D.~J., {Fassnacht}, C.~D., {Auger}, M.~W., {et~al.} 2010, \apj,
  716, 1579

\bibitem[{{Li} {et~al.}(2018){Li}, {Shu}, \& {Wang}}]{Li2018}
{Li}, R., {Shu}, Y., \& {Wang}, J. 2018, \mnras, 480, 431

\bibitem[{{Lintott} {et~al.}(2008){Lintott}, {Schawinski}, {Slosar}, {Land},
  {Bamford}, {Thomas}, {Raddick}, {Nichol}, {Szalay}, {Andreescu}, {Murray}, \&
  {Vandenberg}}]{Lintott2008}
{Lintott}, C.~J., {Schawinski}, K., {Slosar}, A., {et~al.} 2008, \mnras, 389,
  1179

\bibitem[{{Livermore} {et~al.}(2017){Livermore}, {Finkelstein}, \&
  {Lotz}}]{Livermore2017}
{Livermore}, R.~C., {Finkelstein}, S.~L., \& {Lotz}, J.~M. 2017, \apj, 835, 113

\bibitem[{{Marshall} {et~al.}(2009){Marshall}, {Hogg}, {Moustakas},
  {Fassnacht}, {Brada{\v{c}}}, {Schrabback}, \& {Blandford}}]{Marshall2009}
{Marshall}, P.~J., {Hogg}, D.~W., {Moustakas}, L.~A., {et~al.} 2009, \apj, 694,
  924

\bibitem[{{Marshall} {et~al.}(2007){Marshall}, {Treu}, {Melbourne}, {Gavazzi},
  {Bundy}, {Ammons}, {Bolton}, {Burles}, {Larkin}, {Le Mignant}, {Koo},
  {Koopmans}, {Max}, {Moustakas}, {Steinbring}, \& {Wright}}]{Marshall2007}
{Marshall}, P.~J., {Treu}, T., {Melbourne}, J., {et~al.} 2007, \apj, 671, 1196

\bibitem[{{Marshall} {et~al.}(2016){Marshall}, {Verma}, {More}, {Davis},
  {More}, {Kapadia}, {Parrish}, {Snyder}, {Wilcox}, {Baeten}, {Macmillan},
  {Cornen}, {Baumer}, {Simpson}, {Lintott}, {Miller}, {Paget}, {Simpson},
  {Smith}, {K{\"u}ng}, {Saha}, \& {Collett}}]{Marshall2016}
{Marshall}, P.~J., {Verma}, A., {More}, A., {et~al.} 2016, \mnras, 455, 1171

\bibitem[{{Metcalf} {et~al.}(2019){Metcalf}, {Meneghetti}, {Avestruz},
  {Bellagamba}, {Bom}, {Bertin}, {Cabanac}, {Courbin}, {Davies},
  {Decenci{\`e}re}, {Flamary}, {Gavazzi}, {Geiger}, {Hartley},
  {Huertas-Company}, {Jackson}, {Jacobs}, {Jullo}, {Kneib}, {Koopmans},
  {Lanusse}, {Li}, {Ma}, {Makler}, {Li}, {Lightman}, {Petrillo}, {Serjeant},
  {Sch{\"a}fer}, {Sonnenfeld}, {Tagore}, {Tortora}, {Tuccillo},
  {Valent{\'\i}n}, {Velasco-Forero}, {Verdoes Kleijn}, \&
  {Vernardos}}]{Metcalf2019}
{Metcalf}, R.~B., {Meneghetti}, M., {Avestruz}, C., {et~al.} 2019, \aap, 625,
  A119

\bibitem[{{Millon} {et~al.}(2020){Millon}, {Galan}, {Courbin}, {Treu}, {Suyu},
  {Ding}, {Birrer}, {Chen}, {Shajib}, {Sluse}, {Wong}, {Agnello}, {Auger},
  {Buckley-Geer}, {Chan}, {Collett}, {Fassnacht}, {Hilbert}, {Koopmans},
  {Motta}, {Mukherjee}, {Rusu}, {Sonnenfeld}, {Spiniello}, \& {Van de
  Vyvere}}]{Millon2020}
{Millon}, M., {Galan}, A., {Courbin}, F., {et~al.} 2020, \aap, 639, A101

\bibitem[{{More} {et~al.}(2012){More}, {Cabanac}, {More}, {Alard}, {Limousin},
  {Kneib}, {Gavazzi}, \& {Motta}}]{More2012}
{More}, A., {Cabanac}, R., {More}, S., {et~al.} 2012, \apj, 749, 38

\bibitem[{{More} {et~al.}(2011){More}, {Jahnke}, {More}, {Gallazzi}, {Bell},
  {Barden}, \& {H{\"a}u{\ss}ler}}]{More2011}
{More}, A., {Jahnke}, K., {More}, S., {et~al.} 2011, \apj, 734, 69

\bibitem[{{More} {et~al.}(2016){More}, {Verma}, {Marshall}, {More}, {Baeten},
  {Wilcox}, {Macmillan}, {Cornen}, {Kapadia}, {Parrish}, {Snyder}, {Davis},
  {Gavazzi}, {Lintott}, {Simpson}, {Miller}, {Smith}, {Paget}, {Saha},
  {K{\"u}ng}, \& {Collett}}]{More2016}
{More}, A., {Verma}, A., {Marshall}, P.~J., {et~al.} 2016, \mnras, 455, 1191

\bibitem[{Morishita {et~al.}(2017)Morishita, Abramson, Treu, Vulcani, Schmidt,
  Dressler, Poggianti, Malkan, Wang, Huang, {et~al.}}]{Morishita2017}
Morishita, T., Abramson, L.~E., Treu, T., {et~al.} 2017, The Astrophysical
  Journal, 835, 254

\bibitem[{{Moustakas} {et~al.}(2007){Moustakas}, {Marshall}, {Newman}, {Coil},
  {Cooper}, {Davis}, {Fassnacht}, {Guhathakurta}, {Hopkins}, {Koekemoer},
  {Konidaris}, {Lotz}, \& {Willmer}}]{Moustakas2007}
{Moustakas}, L.~A., {Marshall}, P., {Newman}, J.~A., {et~al.} 2007, \apjl, 660,
  L31

\bibitem[{{Oguri} \& {Marshall}(2010)}]{Oguri2010}
{Oguri}, M. \& {Marshall}, P.~J. 2010, \mnras, 405, 2579

\bibitem[{{Oldham} \& {Auger}(2018)}]{Oldham2018}
{Oldham}, L.~J. \& {Auger}, M.~W. 2018, \mnras, 476, 133

\bibitem[{{Pawase} {et~al.}(2014){Pawase}, {Courbin}, {Faure}, {Kokotanekova},
  \& {Meylan}}]{Pawase2014}
{Pawase}, R.~S., {Courbin}, F., {Faure}, C., {Kokotanekova}, R., \& {Meylan},
  G. 2014, \mnras, 439, 3392

\bibitem[{{Peng} {et~al.}(2002){Peng}, {Ho}, {Impey}, \& {Rix}}]{Peng2002}
{Peng}, C.~Y., {Ho}, L.~C., {Impey}, C.~D., \& {Rix}, H.-W. 2002, \aj, 124, 266

\bibitem[{{Petrillo} {et~al.}(2019){Petrillo}, {Tortora}, {Vernardos},
  {Koopmans}, {Verdoes Kleijn}, {Bilicki}, {Napolitano}, {Chatterjee},
  {Covone}, {Dvornik}, {Erben}, {Getman}, {Giblin}, {Heymans}, {de Jong},
  {Kuijken}, {Schneider}, {Shan}, {Spiniello}, \& {Wright}}]{Petrillo2019a}
{Petrillo}, C.~E., {Tortora}, C., {Vernardos}, G., {et~al.} 2019, \mnras, 484,
  3879

\bibitem[{{Pourrahmani} {et~al.}(2018){Pourrahmani}, {Nayyeri}, \&
  {Cooray}}]{Pourrahmani2018}
{Pourrahmani}, M., {Nayyeri}, H., \& {Cooray}, A. 2018, \apj, 856, 68

\bibitem[{Price-Whelan {et~al.}(2018)Price-Whelan, Sip{\H{o}}cz, G{\"u}nther,
  Lim, Crawford, Conseil, Shupe, Craig, Dencheva, Ginsburg, {et~al.}}]{Astropy}
Price-Whelan, A.~M., Sip{\H{o}}cz, B., G{\"u}nther, H., {et~al.} 2018, The
  Astronomical Journal, 156, 123

\bibitem[{{R Core Team}(2020)}]{R2020}
{R Core Team}. 2020, R: A Language and Environment for Statistical Computing, R
  Foundation for Statistical Computing, Vienna, Austria

\bibitem[{{Ragozzine} {et~al.}(2012){Ragozzine}, {Clowe}, {Markevitch},
  {Gonzalez}, \& {Brada{\v{c}}}}]{Ragozzine2012}
{Ragozzine}, B., {Clowe}, D., {Markevitch}, M., {Gonzalez}, A.~H., \&
  {Brada{\v{c}}}, M. 2012, \apj, 744, 94

\bibitem[{{Ratnatunga} {et~al.}(1995){Ratnatunga}, {Ostrander}, {Griffiths}, \&
  {Im}}]{Ratnatunga1995}
{Ratnatunga}, K.~U., {Ostrander}, E.~J., {Griffiths}, R.~E., \& {Im}, M. 1995,
  \apjl, 453, L5

\bibitem[{{Repp} \& {Ebeling}(2018)}]{Repp2018}
{Repp}, A. \& {Ebeling}, H. 2018, \mnras, 479, 844

\bibitem[{{Richard} {et~al.}(2015){Richard}, {Patricio}, {Martinez}, {Bacon},
  {Clement}, {Weilbacher}, {Soto}, {Wisotzki}, {Vernet}, {Pello}, {Schaye},
  {Turner}, \& {Martinsson}}]{Richard2015}
{Richard}, J., {Patricio}, V., {Martinez}, J., {et~al.} 2015, \mnras, 446, L16

\bibitem[{{Richard} {et~al.}(2010){Richard}, {Smith}, {Kneib}, {Ellis},
  {Sanderson}, {Pei}, {Targett}, {Sand}, {Swinbank}, {Dannerbauer}, {Mazzotta},
  {Limousin}, {Egami}, {Jullo}, {Hamilton-Morris}, \& {Moran}}]{Richard2010}
{Richard}, J., {Smith}, G.~P., {Kneib}, J.-P., {et~al.} 2010, \mnras, 404, 325

\bibitem[{{Ritondale} {et~al.}(2019){Ritondale}, {Vegetti}, {Despali}, {Auger},
  {Koopmans}, \& {McKean}}]{Ritondale2019}
{Ritondale}, E., {Vegetti}, S., {Despali}, G., {et~al.} 2019, \mnras, 485, 2179

\bibitem[{{Rojas} {et~al.}(2021){Rojas}, {Savary}, {Cl{\'e}ment}, {Maus},
  {Courbin}, {Lemon}, {Chan}, {Vernardos}, {Joseph}, {Ca{\~n}ameras}, \&
  {Galan}}]{Rojas2021}
{Rojas}, K., {Savary}, E., {Cl{\'e}ment}, B., {et~al.} 2021, arXiv e-prints,
  arXiv:2109.00014

\bibitem[{{Rykoff} {et~al.}(2014){Rykoff}, {Rozo}, {Busha}, {Cunha},
  {Finoguenov}, {Evrard}, {Hao}, {Koester}, {Leauthaud}, {Nord}, {Pierre},
  {Reddick}, {Sadibekova}, {Sheldon}, \& {Wechsler}}]{Rykoff2014}
{Rykoff}, E.~S., {Rozo}, E., {Busha}, M.~T., {et~al.} 2014, \apj, 785, 104

\bibitem[{{Rykoff} {et~al.}(2016){Rykoff}, {Rozo}, {Busha}, {Cunha},
  {Finoguenov}, {Evrard}, {Hao}, {Koester}, {Leauthaud}, {Nord}, {Pierre},
  {Reddick}, {Sadibekova}, {Sheldon}, \& {Wechsler}}]{Rykoff2016}
{Rykoff}, E.~S., {Rozo}, E., {Busha}, M.~T., {et~al.} 2016, VizieR Online Data
  Catalog, J/ApJ/785/104

\bibitem[{{Samui} \& {Samui Pal}(2017)}]{Samui2017}
{Samui}, S. \& {Samui Pal}, S. 2017, \na, 51, 169

\bibitem[{{Schaefer} {et~al.}(2018){Schaefer}, {Geiger}, {Kuntzer}, \&
  {Kneib}}]{Schaefer2018}
{Schaefer}, C., {Geiger}, M., {Kuntzer}, T., \& {Kneib}, J.~P. 2018, \aap, 611,
  A2

\bibitem[{{Schirmer} {et~al.}(2010){Schirmer}, {Suyu}, {Schrabback},
  {Hildebrandt}, {Erben}, \& {Halkola}}]{Schirmer2010}
{Schirmer}, M., {Suyu}, S., {Schrabback}, T., {et~al.} 2010, \aap, 514, A60

\bibitem[{{Schuldt} {et~al.}(2021){Schuldt}, {Suyu}, {Ca{\~n}ameras},
  {Taubenberger}, {Meinhardt}, {Leal-Taix{\'e}}, \& {Hsieh}}]{Schuldt2021}
{Schuldt}, S., {Suyu}, S.~H., {Ca{\~n}ameras}, R., {et~al.} 2021, \aap, 651,
  A55

\bibitem[{{Scoville} {et~al.}(2007){Scoville}, {Aussel}, {Brusa}, {Capak},
  {Carollo}, {Elvis}, {Giavalisco}, {Guzzo}, {Hasinger}, {Impey}, {Kneib},
  {LeFevre}, {Lilly}, {Mobasher}, {Renzini}, {Rich}, {Sanders}, {Schinnerer},
  {Schminovich}, {Shopbell}, {Taniguchi}, \& {Tyson}}]{Scoville2007}
{Scoville}, N., {Aussel}, H., {Brusa}, M., {et~al.} 2007, \apjs, 172, 1

\bibitem[{{Sereno} \& {Paraficz}(2014)}]{Sereno2014}
{Sereno}, M. \& {Paraficz}, D. 2014, \mnras, 437, 600

\bibitem[{{Sersic}(1968)}]{Sersic1968}
{Sersic}, J.~L. 1968, {Atlas de Galaxias Australes}

\bibitem[{{Sharon} {et~al.}(2020){Sharon}, {Bayliss}, {Dahle}, {Dunham},
  {Florian}, {Gladders}, {Johnson}, {Mahler}, {Paterno-Mahler}, {Rigby},
  {Whitaker}, {Akhshik}, {Koester}, {Murray}, {Remolina Gonz{\'a}lez}, \&
  {Wuyts}}]{Sharon2020}
{Sharon}, K., {Bayliss}, M.~B., {Dahle}, H., {et~al.} 2020, \apjs, 247, 12

\bibitem[{{Shu} {et~al.}(2016){Shu}, {Bolton}, {Kochanek}, {Oguri},
  {P{\'e}rez-Fournon}, {Zheng}, {Mao}, {Montero-Dorta}, {Brownstein},
  {Marques-Chaves}, \& {M{\'e}nard}}]{Shu2016}
{Shu}, Y., {Bolton}, A.~S., {Kochanek}, C.~S., {et~al.} 2016, \apj, 824, 86

\bibitem[{{Shu} {et~al.}(2017){Shu}, {Brownstein}, {Bolton}, {Koopmans},
  {Treu}, {Montero-Dorta}, {Auger}, {Czoske}, {Gavazzi}, {Marshall}, \&
  {Moustakas}}]{Shu2017}
{Shu}, Y., {Brownstein}, J.~R., {Bolton}, A.~S., {et~al.} 2017, \apj, 851, 48

\bibitem[{{Smail} {et~al.}(2007){Smail}, {Swinbank}, {Richard}, {Ebeling},
  {Kneib}, {Edge}, {Stark}, {Ellis}, {Dye}, {Smith}, \& {Mullis}}]{Smail2007}
{Smail}, I., {Swinbank}, A.~M., {Richard}, J., {et~al.} 2007, \apjl, 654, L33

\bibitem[{{Sonnenfeld} {et~al.}(2018){Sonnenfeld}, {Chan}, {Shu}, {More},
  {Oguri}, {Suyu}, {Wong}, {Lee}, {Coupon}, {Yonehara}, {Bolton}, {Jaelani},
  {Tanaka}, {Miyazaki}, \& {Komiyama}}]{Sonnenfeld2018}
{Sonnenfeld}, A., {Chan}, J. H.~H., {Shu}, Y., {et~al.} 2018, \pasj, 70, S29

\bibitem[{{Sonnenfeld} {et~al.}(2013){Sonnenfeld}, {Treu}, {Gavazzi}, {Suyu},
  {Marshall}, {Auger}, \& {Nipoti}}]{Sonnenfeld2013}
{Sonnenfeld}, A., {Treu}, T., {Gavazzi}, R., {et~al.} 2013, \apj, 777, 98

\bibitem[{{Sonnenfeld} {et~al.}(2015){Sonnenfeld}, {Treu}, {Marshall}, {Suyu},
  {Gavazzi}, {Auger}, \& {Nipoti}}]{Sonnenfeld2015}
{Sonnenfeld}, A., {Treu}, T., {Marshall}, P.~J., {et~al.} 2015, \apj, 800, 94

\bibitem[{{Sonnenfeld} {et~al.}(2020){Sonnenfeld}, {Verma}, {More}, {Baeten},
  {Macmillan}, {Wong}, {Chan}, {Jaelani}, {Lee}, {Oguri}, {Rusu}, {Veldthuis},
  {Trouille}, {Marshall}, {Hutchings}, {Allen}, {O'Donnell}, {Cornen}, {Davis},
  {McMaster}, {Lintott}, \& {Miller}}]{Sonnenfeld2020}
{Sonnenfeld}, A., {Verma}, A., {More}, A., {et~al.} 2020, \aap, 642, A148

\bibitem[{Stark {et~al.}(2013)Stark, Auger, Belokurov, Jones, Robertson, Ellis,
  Sand, Moiseev, Eagle, \& Myers}]{Stark2013}
Stark, D.~P., Auger, M., Belokurov, V., {et~al.} 2013, Monthly Notices of the
  Royal Astronomical Society, 436, 1040

\bibitem[{{Suyu} {et~al.}(2010){Suyu}, {Marshall}, {Auger}, {Hilbert},
  {Blandford}, {Koopmans}, {Fassnacht}, \& {Treu}}]{Suyu2010}
{Suyu}, S.~H., {Marshall}, P.~J., {Auger}, M.~W., {et~al.} 2010, \apj, 711, 201

\bibitem[{{Suyu} {et~al.}(2014){Suyu}, {Treu}, {Hilbert}, {Sonnenfeld},
  {Auger}, {Blandford}, {Collett}, {Courbin}, {Fassnacht}, {Koopmans},
  {Marshall}, {Meylan}, {Spiniello}, \& {Tewes}}]{Suyu2014}
{Suyu}, S.~H., {Treu}, T., {Hilbert}, S., {et~al.} 2014, \apjl, 788, L35

\bibitem[{{Swinbank} {et~al.}(2015){Swinbank}, {Dye}, {Nightingale},
  {Furlanetto}, {Smail}, {Cooray}, {Dannerbauer}, {Dunne}, {Eales}, {Gavazzi},
  {Hunter}, {Ivison}, {Negrello}, {Oteo-Gomez}, {Smit}, {van der Werf}, \&
  {Vlahakis}}]{Swinbank2015}
{Swinbank}, A.~M., {Dye}, S., {Nightingale}, J.~W., {et~al.} 2015, \apjl, 806,
  L17

\bibitem[{{Talbot} {et~al.}(2021){Talbot}, {Brownstein}, {Dawson}, {Kneib}, \&
  {Bautista}}]{Talbot2021}
{Talbot}, M.~S., {Brownstein}, J.~R., {Dawson}, K.~S., {Kneib}, J.-P., \&
  {Bautista}, J. 2021, \mnras, 502, 4617

\bibitem[{{Tanaka} {et~al.}(2016){Tanaka}, {Wong}, {More}, {Dezuka}, {Egami},
  {Oguri}, {Suyu}, {Sonnenfeld}, {Higuchi}, {Komiyama}, {Miyazaki}, {Onoue},
  {Oyamada}, \& {Utsumi}}]{Tanaka2016}
{Tanaka}, M., {Wong}, K.~C., {More}, A., {et~al.} 2016, \apjl, 826, L19

\bibitem[{{Taylor}(2005)}]{Topcat}
{Taylor}, M.~B. 2005, in Astronomical Society of the Pacific Conference Series,
  Vol. 347, Astronomical Data Analysis Software and Systems XIV, ed.
  P.~{Shopbell}, M.~{Britton}, \& R.~{Ebert}, 29

\bibitem[{{Treu} {et~al.}(2011){Treu}, {Dutton}, {Auger}, {Marshall}, {Bolton},
  {Brewer}, {Koo}, \& {Koopmans}}]{Treu2011}
{Treu}, T., {Dutton}, A.~A., {Auger}, M.~W., {et~al.} 2011, \mnras, 417, 1601

\bibitem[{{Treu} \& {Koopmans}(2004)}]{Treu2004}
{Treu}, T. \& {Koopmans}, L. V.~E. 2004, \apj, 611, 739

\bibitem[{{Tu} {et~al.}(2009){Tu}, {Gavazzi}, {Limousin}, {Cabanac},
  {Marshall}, {Fort}, {Treu}, {P{\'e}llo}, {Jullo}, {Kneib}, \&
  {Sygnet}}]{Tu2009}
{Tu}, H., {Gavazzi}, R., {Limousin}, M., {et~al.} 2009, \aap, 501, 475

\bibitem[{{Vanzella} {et~al.}(2021){Vanzella}, {Caminha}, {Rosati}, {Mercurio},
  {Castellano}, {Meneghetti}, {Grillo}, {Sani}, {Bergamini}, {Calura},
  {Caputi}, {Cristiani}, {Cupani}, {Fontana}, {Gilli}, {Grazian}, {Gronke},
  {Mignoli}, {Nonino}, {Pentericci}, {Tozzi}, {Treu}, {Balestra}, \&
  {Dijkstra}}]{vanzella2021muse}
{Vanzella}, E., {Caminha}, G.~B., {Rosati}, P., {et~al.} 2021, \aap, 646, A57

\bibitem[{{Vegetti} \& {Koopmans}(2009)}]{Vegetti2009}
{Vegetti}, S. \& {Koopmans}, L.~V.~E. 2009, \mnras, 392, 945

\bibitem[{{Walmsley} {et~al.}(2022){Walmsley}, {Lintott}, {G{\'e}ron}, {Kruk},
  {Krawczyk}, {Willett}, {Bamford}, {Kelvin}, {Fortson}, {Gal}, {Keel},
  {Masters}, {Mehta}, {Simmons}, {Smethurst}, {Smith}, {Baeten}, \&
  {Macmillan}}]{Walmsley2022}
{Walmsley}, M., {Lintott}, C., {G{\'e}ron}, T., {et~al.} 2022, \mnras, 509,
  3966

\bibitem[{{Walmsley} {et~al.}(2020){Walmsley}, {Smith}, {Lintott}, {Gal},
  {Bamford}, {Dickinson}, {Fortson}, {Kruk}, {Masters}, {Scarlata}, {Simmons},
  {Smethurst}, \& {Wright}}]{Walmsley2020}
{Walmsley}, M., {Smith}, L., {Lintott}, C., {et~al.} 2020, \mnras, 491, 1554

\bibitem[{{Walsh} {et~al.}(1979){Walsh}, {Carswell}, \& {Weymann}}]{Walsh1979}
{Walsh}, D., {Carswell}, R.~F., \& {Weymann}, R.~J. 1979, \nat, 279, 381

\bibitem[{{Wang} {et~al.}(2019){Wang}, {Vogelsberger}, {Xu}, {Shen}, {Mao},
  {Barnes}, {Li}, {Marinacci}, {Torrey}, {Springel}, \& {Hernquist}}]{Wang2019}
{Wang}, Y., {Vogelsberger}, M., {Xu}, D., {et~al.} 2019, \mnras, 490, 5722

\bibitem[{{Welch} {et~al.}(2022){Welch}, {Coe}, {Diego}, {Zitrin},
  {Zackrisson}, {Dimauro}, {Jim{\'e}nez-Teja}, {Kelly}, {Mahler}, {Oguri},
  {Timmes}, {Windhorst}, {Florian}, {de Mink}, {Avila}, {Anderson}, {Bradley},
  {Sharon}, {Vikaeus}, {McCandliss}, {Brada{\v{c}}}, {Rigby}, {Frye}, {Toft},
  {Strait}, {Trenti}, {Sharma}, {Andrade-Santos}, \&
  {Broadhurst}}]{welch2022star}
{Welch}, B., {Coe}, D., {Diego}, J.~M., {et~al.} 2022, \nat, 603, 815

\bibitem[{{Wong} {et~al.}(2020){Wong}, {Suyu}, {Chen}, {Rusu}, {Millon},
  {Sluse}, {Bonvin}, {Fassnacht}, {Taubenberger}, {Auger}, {Birrer}, {Chan},
  {Courbin}, {Hilbert}, {Tihhonova}, {Treu}, {Agnello}, {Ding}, {Jee},
  {Komatsu}, {Shajib}, {Sonnenfeld}, {Blandford}, {Koopmans}, {Marshall}, \&
  {Meylan}}]{Wong2020}
{Wong}, K.~C., {Suyu}, S.~H., {Chen}, G. C.~F., {et~al.} 2020, \mnras, 498,
  1420

\bibitem[{Zitrin {et~al.}(2015)Zitrin, Fabris, Merten, Melchior, Meneghetti,
  Koekemoer, Coe, Maturi, Bartelmann, Postman, {et~al.}}]{Zitrin2015}
Zitrin, A., Fabris, A., Merten, J., {et~al.} 2015, The Astrophysical Journal,
  801, 44

\bibitem[{{Zou} {et~al.}(2021){Zou}, {Gao}, {Xu}, {Zhou}, {Ma}, {Zhou},
  {Zhang}, {Nie}, {Wang}, \& {Xue}}]{Zou2021}
{Zou}, H., {Gao}, J., {Xu}, X., {et~al.} 2021, \apjs, 253, 56

\end{thebibliography}

\clearpage


\appendix
\onecolumn
\section{List of strong lens candidates}
\vspace{0.2cm}
\setlongtables{\scriptsize
\setlength\LTcapwidth{\textwidth}
\begin{longtable}{p{90pt}rrrlllrrrrlll}
\caption{Newly discovered lenses.}
\tabularnewline 
\hline
\multicolumn{1}{l}{IAU Name}&\multicolumn{1}{c}{ID}&\multicolumn{1}{c}{RA}&\multicolumn{1}{c}{Dec}&\multicolumn{1}{c}{$r_{\rm arc}\,(\arcsec)$}&\multicolumn{1}{c}{instru.}&\multicolumn{1}{c}{filter}&\multicolumn{1}{c}{$mag$}&\multicolumn{1}{c}{$r_{\rm e}\,(\arcsec)$}&\multicolumn{1}{c}{$q$}&\multicolumn{1}{c}{$PA$}&\multicolumn{1}{c}{class}&\multicolumn{1}{c}{clust.}&\multicolumn{1}{c}{$z$}
\tabularnewline 
\hline
\endfirsthead
\caption[]{\em continued.} 
\tabularnewline
\hline
\multicolumn{1}{l}{IAU Name}&\multicolumn{1}{c}{ID}&\multicolumn{1}{c}{RA}&\multicolumn{1}{c}{Dec}&\multicolumn{1}{c}{$r_{\rm arc}$\,(\arcsec)}&\multicolumn{1}{c}{instru.}&\multicolumn{1}{c}{filter}&\multicolumn{1}{c}{$mag$}&\multicolumn{1}{c}{$r_{\rm e}$\,(\arcsec)}&\multicolumn{1}{c}{$q$}&\multicolumn{1}{c}{$PA$}&\multicolumn{1}{c}{class}&\multicolumn{1}{c}{clust.}&\multicolumn{1}{c}{$z$}
\tabularnewline*
\hline
\tabularnewline*
\endhead
\tabularnewline*
\hline
\endfoot
\label{tab:unpublished}
{\bfseries A grade Lenses}&&&&&&&&&&&&&\tabularnewline*
\tabularnewline*
~~HAH $J001538.4-390435.0$&$  1$&$  3.91007$&$ 39.07643$&0.79&UVIS&f475w&$25.4$&$ 0.06$&$0.93$&$ -81.9$&Arc& & \tabularnewline
~~HAH $J002348.6-244149.6$&$  2$&$  5.95275$&$-24.69702$&1.07&ACS&f814w&$20.1$&$ 1.45$&$0.56$&$  89.1$&Quad&DClust1& \tabularnewline
~~HAH $J002753.2-753730.0$&$  3$&$  6.97150$&$-75.62500$&1.71&ACS&f606w&$19.6$&$ 1.10$&$0.93$&$-112.6$&Ring&DClust2& \tabularnewline
~~HAH $J004924.3-205729.8$&$  4$&$ 12.35113$&$-20.95829$&0.78&ACS&f606w&$20.3$&$ 0.80$&$0.77$&$ -19.7$&Arc&DClust3& \tabularnewline
~~HAH $J005403.4+394712.1$&$  5$&$ 13.51445$&$ 39.78661$&2.05&ACS&f814w&$20.2$&$ 0.85$&$0.81$&$ 155.8$&Arc& &0.73 P\tabularnewline
~~HAH $J015949.3-084906.0$&$  6$&$ 29.95549$&$ -8.81828$&1.17&ACS&f606w&$21.1$&$ 0.35$&$0.92$&$ 128.6$&Arc&DClust4&0.39 P\tabularnewline
~~HAH $J021706.2-051318.0$&$  7$&$ 34.27592$&$ -5.22168$&1.14&ACS&f814w&$21.7$&$ 0.54$&$0.68$&$  -9.8$&Arc& &0.59 P\tabularnewline
~~HAH $J024329.8-593102.7$&$  8$&$ 40.87441$&$-59.51759$&0.63&UVIS&f200lp&$23.1$&$ 0.35$&$0.81$&$  95.4$&Arc&DClust5&\tabularnewline
~~HAH $J024533.6-530203.5$&$  9$&$ 41.39015$&$-53.03433$&0.73&ACS&f814w&$20.1$&$ 0.39$&$0.88$&$ -24.2$&Ring&DClust6& \tabularnewline
~~HAH $J024932.8+334635.5$&$ 10$&$ 42.38714$&$ 33.77651$&0.55&ACS&f775w&$25.1$&$ 0.32$&$0.39$&$  89.7$&Arc& & \tabularnewline
~~HAH $J025241.6-150025.5$&$ 11$&$ 43.17305$&$-15.00712$&3.23&ACS&f814w&$21.3$&$ 0.36$&$0.90$&$-111.1$&Arc&DClust7& \tabularnewline
~~HAH $J025659.9-163059.5$&$ 12$&$ 44.24984$&$-16.51655$&1.48&ACS&f606w&$18.8$&$ 2.08$&$0.71$&$-164.4$&Arc&DClust8&0.31 P\tabularnewline
~~HAH $J033603.9-451223.1$&$ 13$&$ 54.01630$&$-45.20642$&0.63&ACS&f814w&$22.6$&$ 0.31$&$0.79$&$-176.5$&Ring& & \tabularnewline
~~HAH $J034445.2-642133.8$&$ 14$&$ 56.18856$&$-64.35941$&1.42&ACS&f606w&$24.2$&$ 0.10$&$1.00$&$ 153.0$&Arc&DClust9& \tabularnewline
~~HAH $J042044.0-403607.2$&$ 15$&$ 65.18326$&$-40.60200$&0.42&UVIS&f814w&$22.7$&$ 0.48$&$0.84$&$ 152.3$&Ring& & \tabularnewline
~~HAH $J054707.0-390516.3$&$ 16$&$ 86.77923$&$-39.08786$&0.74&ACS&f814w&$19.7$&$ 0.40$&$0.80$&$ 139.2$&Ring&DClust10& \tabularnewline
~~HAH $J061345.7-562750.4$&$ 17$&$ 93.44050$&$-56.46400$&1.2&ACS&f606w&$18.9$&$ 1.08$&$0.42$&$  68.7$&Arc&DClust11& \tabularnewline
~~HAH $J072221.9+072326.7$&$ 18$&$110.59159$&$  7.39076$&0.59&ACS&f814w&$20.3$&$ 0.74$&$0.91$&$ -99.5$&Arc&DClust12& \tabularnewline
~~HAH $J081334.8+254518.3$&$ 19$&$123.39489$&$ 25.75509$&1.6&IR&f140w&$19.1$&$ 1.04$&$0.83$&$  52.6$&Arc& & \tabularnewline
~~HAH $J083420.3+452506.9$&$ 20$&$128.58462$&$ 45.41860$&1.04&ACS&f814w&$21.2$&$ 0.65$&$0.61$&$-107.9$&Arc&DClust13&0.65 S\tabularnewline
~~HAH $J084833.4+444431.9$&$ 21$&$132.13930$&$ 44.74220$&1.2&ACS&f814w&$22.5$&$ 0.27$&$0.81$&$ 120.9$&Arc& &0.19 P\tabularnewline
~~ $ $&$ $&$ $&$ $&1.2&ACS&f814w&$24.2$&$ 0.22$&$0.13$&$-174.1$&Arc& &0.19 P\tabularnewline
~~ $ $&$ $&$ $&$ $&1.2&ACS&f814w&$23.6$&$ 0.20$&$0.46$&$-158.5$&Arc& &0.19 P\tabularnewline
~~HAH $J093325.2+284348.6$&$ 22$&$143.35520$&$ 28.73018$&3.27&ACS&f606w&$20.6$&$ 0.10$&$0.77$&$-130.3$&Arc& &0.4 P\tabularnewline
~~ $ $&$ $&$ $&$ $&3.27&ACS&f606w&$19.7$&$ 1.38$&$0.84$&$-103.8$&Arc& &0.4 P\tabularnewline
~~HAH $J100141.8+021424.2$&$ 23$&$150.42410$&$  2.24010$&1.46&ACS&f814w&$23.7$&$ 0.26$&$0.80$&$-104.2$&Ring& &0.34 P\tabularnewline
~~HAH $J100251.8+691959.3$&$ 24$&$150.71567$&$ 69.33316$&0.7&ACS&f606w&$22.0$&$ 1.21$&$0.22$&$ 133.9$&Double& &0.56 P\tabularnewline
~~HAH $J102914.8+262207.0$&$ 25$&$157.31170$&$ 26.36860$&0.64&ACS&f475w&$23.6$&$ 0.36$&$0.69$&$ 108.2$&Arc&DClust14&0.57 P\tabularnewline
~~HAH $J105722.7+580046.5$&$ 26$&$164.34460$&$ 58.01290$&0.46&ACS&f814w&$22.7$&$ 0.32$&$0.75$&$-105.9$&Arc&DClust15&0.54 P\tabularnewline
~~HAH $J111337.4+221249.2$&$ 27$&$168.40564$&$ 22.21368$&2.11&UVIS&f350lp&$18.8$&$ 1.94$&$0.74$&$-135.1$&Arc&DClust16&0.41 S\tabularnewline
~~HAH $J113158.4-195451.5$&$ 28$&$172.99340$&$-19.91430$&1.66&ACS&f814w&$18.8$&$ 1.25$&$0.65$&$ 114.5$&Arc& & \tabularnewline
~~HAH $J114938.4+222107.7$&$ 29$&$177.41020$&$ 22.35200$&0.8&ACS&f814w&$20.8$&$ 0.38$&$0.66$&$  58.8$&Arc&DClust17&0.42 P\tabularnewline
~~HAH $J121653.9-121104.2$&$ 30$&$184.22457$&$-12.18464$&1.12&ACS&f814w&$19.2$&$ 0.54$&$0.97$&$ -57.9$&Ring& & \tabularnewline
~~HAH $J125709.5+282239.7$&$ 31$&$194.28970$&$ 28.37770$&0.88&ACS&f475w&$26.4$&$ 1.48$&$0.05$&$ 140.0$&Arc& &0.78 P\tabularnewline
~~ $ $&$ $&$ $&$ $&0.88&ACS&f475w&$24.6$&$ 0.17$&$0.55$&$ -93.9$&Arc& &0.78 P\tabularnewline
~~HAH $J130124.0-113114.6$&$ 32$&$195.34996$&$-11.52072$&0.91&ACS&f814w&$19.1$&$ 1.13$&$0.67$&$ -29.2$&Ring& & \tabularnewline
~~HAH $J131953.6+700420.8$&$ 33$&$199.97339$&$ 70.07245$&1.16&ACS&f606w&$20.0$&$ 0.46$&$0.85$&$-157.0$&Ring&DClust18& \tabularnewline
~~HAH $J132824.5-313204.6$&$ 34$&$202.10207$&$-31.53468$&1.4&UVIS&f475w&$22.0$&$ 0.94$&$0.69$&$ 179.8$&Arc& & \tabularnewline
~~HAH $J133525.9+405952.4$&$ 35$&$203.85800$&$ 40.99790$&0.95&ACS&f814w&$18.9$&$ 1.49$&$0.74$&$ 102.6$&Arc&DClust19&0.29 P\tabularnewline
~~HAH $J135955.7+143820.4$&$ 36$&$209.98235$&$ 14.63882$&0.56&UVIS&f390w&$22.9$&$ 0.55$&$0.94$&$  44.7$&Arc& & \tabularnewline
~~HAH $J141139.2+565200.9$&$ 37$&$212.91351$&$ 56.86694$&0.51&UVIS&f475x&$21.2$&$ 0.74$&$0.45$&$  92.6$&Ring&DClust20& \tabularnewline
~~HAH $J141900.1+524249.0$&$ 38$&$214.75050$&$ 52.71360$&1.12&ACS&f814w&$20.9$&$ 0.44$&$0.94$&$-136.8$&Arc& &0.46 P\tabularnewline
~~HAH $J143811.5+464007.6$&$ 39$&$219.54794$&$ 46.66877$&0.57&ACS&f435w&$22.1$&$ 0.46$&$0.46$&$-118.5$&Ring& & \tabularnewline
~~HAH $J163124.5+124304.3$&$ 40$&$247.85224$&$ 12.71771$&2.31&UVIS&f814w&$22.5$&$ 0.52$&$0.92$&$ -45.1$&Arc& &0.64 S\tabularnewline
~~HAH $J163417.6+703107.0$&$ 41$&$248.57350$&$ 70.51860$&1.13&IR&f140w&$19.1$&$ 0.46$&$0.93$&$ 138.3$&Arc& & \tabularnewline
~~HAH $J171314.9+602207.7$&$ 42$&$258.31200$&$ 60.36890$&0.94&ACS&f814w&$21.2$&$ 0.27$&$0.62$&$ -17.9$&Ring& &0.26 P\tabularnewline
~~ $ $&$ $&$ $&$ $&0.94&ACS&f814w&$24.6$&$ 1.95$&$0.00$&$ -28.7$&Ring& &0.26 P\tabularnewline
~~HAH $J214817.6-484350.5$&$ 43$&$327.07301$&$-48.73045$&1.12&UVIS&f200lp&$23.1$&$ 0.39$&$0.89$&$  98.0$&Arc&DClust21& \tabularnewline
~~HAH $J215849.5-223159.7$&$ 44$&$329.70656$&$-22.53325$&3.27&ACS&f775w&$19.2$&$ 2.38$&$0.68$&$ -23.8$&Arc& & \tabularnewline
~~HAH $J234106.5-000007.5$&$ 45$&$355.27743$&$ -0.00210$&2.31&UVIS&f390w&$19.9$&$ 1.43$&$0.92$&$ -50.2$&Arc&DClust22&0.19 S\tabularnewline*
\tabularnewline
\hline
{\bfseries B grade Lenses}&&&&&&&&&&&&&\tabularnewline*
\tabularnewline
~~HAH $J001419.9-302600.7$&$ 46$&$  3.58309$&$-30.43355$&1.3&ACS&f814w&$18.6$&$ 0.93$&$0.67$&$ 177.1$&Arc&DClust23&0.29 $\star$\tabularnewline*
~~HAH $J002706.3+170313.5$&$ 47$&$  6.77622$&$ 17.05375$&0.8&ACS&f814w&$21.5$&$ 0.72$&$0.67$&$-174.9$&Arc&DClust24& \tabularnewline
~~HAH $J003350.8-074959.9$&$ 48$&$  8.46170$&$ -7.83330$&0.88&ACS&f814w&$22.1$&$ 0.45$&$0.65$&$ -95.4$&Arc&DClust25&0.16 P\tabularnewline
~~HAH $J004426.2-403941.4$&$ 49$&$ 11.10937$&$-40.66150$&0.55&ACS&f606w&$23.9$&$ 0.44$&$0.71$&$-120.6$&Ring&DClust26& \tabularnewline
~~HAH $J004859.0+401706.7$&$ 50$&$ 12.24571$&$ 40.28517$&1.05&ACS&f606w&$24.6$&$ 0.30$&$0.29$&$  69.4$&Arc& & \tabularnewline
~~HAH $J005141.6+272001.5$&$ 51$&$ 12.92343$&$ 27.33376$&1.3&IR&f110w&$17.6$&$ 0.81$&$0.97$&$  -1.4$&Arc&DClust27&0.36 P\tabularnewline
~~HAH $J011138.3-454015.1$&$ 52$&$ 17.90972$&$-45.67088$&0.93&IR&f105w&$19.4$&$ 0.29$&$0.82$&$ 146.2$&Arc&DClust28& \tabularnewline
~~HAH $J013723.7-082852.2$&$ 53$&$ 24.34861$&$ -8.48115$&1.87&ACS&f606w&$20.1$&$ 0.43$&$0.88$&$-178.1$&Arc&DClust29&0.52 P\tabularnewline
~~HAH $J015256.3-135416.3$&$ 54$&$ 28.23476$&$-13.90454$&0.59&ACS&f814w&$21.9$&$ 0.55$&$0.73$&$ -86.4$&Arc& & \tabularnewline
~~HAH $J022056.4-033332.1$&$ 55$&$ 35.23505$&$ -3.55899$&1.46&ACS&f850lp&$21.5$&$ 0.33$&$0.56$&$-130.1$&Arc& & \tabularnewline
~~HAH $J023951.1-013205.9$&$ 56$&$ 39.96294$&$ -1.53500$&1.1&ACS&f814w&$19.8$&$ 0.45$&$0.68$&$   0.2$&Arc&DClust30&0.48 P\tabularnewline
~~HAH $J024535.7-530125.5$&$ 57$&$ 41.39911$&$-53.02377$&0.86&ACS&f814w&$20.0$&$ 0.61$&$0.48$&$ -11.2$&Arc&DClust31& \tabularnewline
~~HAH $J025735.5-220928.5$&$ 58$&$ 44.39802$&$-22.15791$&0.84&ACS&f814w&$18.8$&$ 1.84$&$0.74$&$-161.8$&Arc&DClust32& \tabularnewline
~~HAH $J025740.9-221027.3$&$ 59$&$ 44.42063$&$-22.17426$&0.85&ACS&f814w&$20.0$&$ 0.29$&$0.59$&$ 168.8$&Arc&DClust33& \tabularnewline
~~HAH $J032030.9+003242.8$&$ 60$&$ 50.12858$&$  0.54522$&2.91&ACS&f775w&$19.4$&$ 0.69$&$0.53$&$-102.5$&Arc&DClust34&0.39 S\tabularnewline
~~HAH $J033305.9-274744.8$&$ 61$&$ 53.27467$&$-27.79577$&1.3&ACS&f606w&$21.1$&$ 0.37$&$0.74$&$-174.9$&Arc& &0.35 P$\star$\tabularnewline
~~HAH $J033530.7-540734.7$&$ 62$&$ 53.87735$&$-54.12630$&3.23&ACS&f814w&$20.2$&$ 1.17$&$0.34$&$  -4.0$&Arc&DClust35& \tabularnewline
~~ $ $&$ $&$ $&$ $&3.23&ACS&f814w&$20.0$&$15.49$&$0.52$&$  -0.1$&Arc& & \tabularnewline
~~HAH $J033645.9-535536.7$&$ 63$&$ 54.18759$&$-53.92647$&1.21&UVIS&f814w&$20.4$&$ 0.33$&$0.77$&$-149.5$&Arc&DClust36& \tabularnewline
~~HAH $J035851.7-295418.7$&$ 64$&$ 59.71560$&$-29.90520$&1&ACS&f814w&$20.1$&$ 0.50$&$0.76$&$ -28.2$&Arc&DClust37&0.42 $\star$\tabularnewline
~~HAH $J041631.5-240512.5$&$ 65$&$ 64.13140$&$-24.08680$&1.23&ACS&f814w&$21.1$&$ 1.06$&$0.34$&$-116.5$&Double&DClust38& \tabularnewline
~~HAH $J044208.3-281503.7$&$ 66$&$ 70.53440$&$-28.25102$&4.4&ACS&f814w&$19.9$&$ 1.16$&$0.58$&$-103.0$&Arc&DClust39& \tabularnewline
~~ $ $&$ $&$ $&$ $&4.4&ACS&f814w&$19.5$&$ 1.62$&$0.66$&$ -46.3$&Arc& & \tabularnewline
~~HAH $J045430.7-030133.2$&$ 67$&$ 73.62796$&$ -3.02590$&1.47&ACS&f814w&$20.3$&$ 0.66$&$0.77$&$ -97.0$&Arc&DClust40& \tabularnewline
~~HAH $J045441.7-030319.9$&$ 68$&$ 73.67378$&$ -3.05553$&1.18&ACS&f814w&$20.1$&$ 0.51$&$0.76$&$ -82.1$&Arc&DClust41&0.61 P\tabularnewline
~~HAH $J051850.3-431442.0$&$ 69$&$ 79.70954$&$-43.24500$&1.56&ACS&f814w&$21.4$&$ 1.10$&$0.80$&$  99.1$&Arc&DClust42& \tabularnewline
~~HAH $J055326.6-334254.0$&$ 70$&$ 88.36070$&$-33.71500$&2.65&ACS&f814w&$20.0$&$ 0.36$&$0.77$&$ -18.5$&Arc&DClust43& \tabularnewline
~~HAH $J065821.5-555711.5$&$ 71$&$104.58970$&$-55.95316$&1.88&ACS&f606w&$19.9$&$ 0.87$&$0.43$&$ -88.6$&Arc& & \tabularnewline
~~HAH $J072218.2+072314.8$&$ 72$&$110.57599$&$  7.38745$&0.6&ACS&f814w&$21.4$&$ 0.44$&$0.60$&$ 146.3$&Arc&DClust44& \tabularnewline
~~HAH $J072333.8-732550.9$&$ 73$&$110.89080$&$-73.43080$&0.97&ACS&f814w&$21.3$&$ 0.42$&$0.31$&$ 139.8$&Arc&DClust45& \tabularnewline
~~HAH $J074421.4+375400.0$&$ 74$&$116.08890$&$ 37.89990$&1.27&UVIS&f606w&$24.3$&$ 0.40$&$0.28$&$-159.4$&Arc& & \tabularnewline
~~HAH $J074531.3+195111.7$&$ 75$&$116.38083$&$ 19.85327$&0.73&UVIS&f606w&$22.4$&$ 0.45$&$0.79$&$ 112.4$&Double& &0.44 P\tabularnewline
~~HAH $J083410.0+491332.2$&$ 76$&$128.54159$&$ 49.22560$&0.85&ACS&f606w&$23.0$&$ 0.83$&$0.46$&$  -8.7$&Arc& & \tabularnewline
~~HAH $J084712.7+344910.9$&$ 77$&$131.80310$&$ 34.81970$&0.69&ACS&f814w&$22.9$&$ 0.13$&$0.84$&$  26.1$&Arc&DClust46& \tabularnewline
~~HAH $J091503.8+581147.7$&$ 78$&$138.76580$&$ 58.19660$&0.85&UVIS&f350lp&$22.8$&$ 0.50$&$0.89$&$ 134.8$&Ring& & \tabularnewline
~~HAH $J094949.9+170614.4$&$ 79$&$147.45790$&$ 17.10399$&2.3&ACS&f814w&$19.8$&$ 0.57$&$0.40$&$  30.4$&Arc&DClust47&0.39 P\tabularnewline
~~HAH $J095549.0-100034.8$&$ 80$&$148.95441$&$-10.00973$&1.16&ACS&f606w&$24.5$&$ 0.47$&$0.34$&$  36.0$&Arc& & \tabularnewline
~~HAH $J095633.6-100913.3$&$ 81$&$149.13990$&$-10.15370$&1.31&ACS&f606w&$17.6$&$ 2.00$&$0.86$&$  19.2$&Arc& & \tabularnewline
~~HAH $J095712.5+690709.1$&$ 82$&$149.30206$&$ 69.11924$&3.79&ACS&f814w&$19.7$&$ 0.96$&$0.81$&$ 145.4$&Arc& &0.52 P\tabularnewline
~~ $ $&$ $&$ $&$ $&3.79&ACS&f814w&$20.4$&$ 1.01$&$0.70$&$-139.8$&Arc& &0.52 P\tabularnewline
~~HAH $J095803.9+653409.1$&$ 83$&$149.51620$&$ 65.56920$&1.44&ACS&f814w&$21.0$&$ 1.66$&$0.77$&$   4.4$&Arc&DClust48& \tabularnewline
~~HAH $J100030.1+023716.7$&$ 84$&$150.12550$&$  2.62130$&0.84&ACS&f814w&$25.6$&$ 0.03$&$0.41$&$  39.6$&Arc&DClust49&0.52 P\tabularnewline
~~ $ $&$ $&$ $&$ $&0.84&ACS&f814w&$24.8$&$ 0.36$&$0.27$&$-103.8$&Arc& &0.52 P\tabularnewline
~~ $ $&$ $&$ $&$ $&0.84&ACS&f814w&$24.5$&$ 2.60$&$0.42$&$ -93.1$&Arc& &0.52 P\tabularnewline
~~HAH $J100124.2+555407.2$&$ 85$&$150.35080$&$ 55.90200$&1.45&ACS&f814w&$21.2$&$ 0.13$&$0.62$&$-165.7$&Arc&DClust50&0.36 S\tabularnewline
~~ $ $&$ $&$ $&$ $&1.45&ACS&f814w&$20.8$&$ 0.70$&$0.35$&$-169.3$&Arc& &0.36 S\tabularnewline
~~HAH $J103937.6-001421.5$&$ 86$&$159.92930$&$ -0.26555$&0.51&ACS&f814w&$21.4$&$ 1.10$&$0.90$&$ -24.6$&Arc& &0.23 P\tabularnewline
~~HAH $J104001.6+474558.0$&$ 87$&$160.00670$&$ 47.76612$&0.66&ACS&f435w&$22.8$&$ 0.53$&$0.50$&$  60.0$&Arc& &0.35 P\tabularnewline
~~HAH $J104132.0+000125.0$&$ 88$&$160.38332$&$ -0.02360$&0.99&ACS&f606w&$23.7$&$ 0.62$&$0.88$&$  45.9$&Ring&DClust51&0.67 P\tabularnewline
~~HAH $J105726.9+575944.0$&$ 89$&$164.36207$&$ 57.99555$&1.47&ACS&f814w&$22.1$&$ 0.43$&$0.43$&$-157.7$&Arc&DClust52&0.24 P\tabularnewline
~~HAH $J111507.3+531955.8$&$ 90$&$168.78051$&$ 53.33217$&1.12&ACS&f606w&$21.2$&$ 0.33$&$0.62$&$-136.6$&Arc&DClust53&0.44 P\tabularnewline
~~HAH $J111822.1+033802.1$&$ 91$&$169.59242$&$  3.63393$&1.33&IR&f110w&$17.4$&$ 1.10$&$0.60$&$ -44.4$&Ring&DClust54&0.42 S\tabularnewline
~~HAH $J115940.8+011410.7$&$ 92$&$179.92011$&$  1.23632$&2.5&ACS&f606w&$19.3$&$ 2.94$&$0.56$&$-102.7$&Arc& &0.44 S\tabularnewline
~~HAH $J122944.7+112238.3$&$ 93$&$187.43610$&$ 11.37730$&1.3&ACS&f814w&$20.2$&$ 0.26$&$0.80$&$  84.3$&Arc&DClust55&0.41 P\tabularnewline
~~HAH $J123430.7+080444.4$&$ 94$&$188.62785$&$  8.07903$&2.1&ACS&f814w&$20.3$&$ 1.22$&$0.85$&$ 130.0$&Arc& &0.5 S\tabularnewline
~~HAH $J123618.8+260905.0$&$ 95$&$189.07830$&$ 26.15140$&1.54&ACS&f814w&$21.3$&$ 0.66$&$0.68$&$ -50.9$&Arc&DClust56& \tabularnewline
~~HAH $J124756.2+134103.3$&$ 96$&$191.98435$&$ 13.68426$&1.65&ACS&f814w&$22.0$&$ 1.34$&$0.86$&$ 105.2$&Arc& & \tabularnewline
~~HAH $J125907.8-014439.0$&$ 97$&$194.78235$&$ -1.74418$&0.33&ACS&f814w&$22.2$&$ 0.68$&$0.26$&$-164.8$&Double&DClust57& \tabularnewline
~~HAH $J130034.4+280022.4$&$ 98$&$195.14321$&$ 28.00622$&0.7&ACS&f814w&$20.7$&$ 0.20$&$0.61$&$ 174.3$&Arc& &0.49 P\tabularnewline
~~HAH $J130152.7+275147.7$&$ 99$&$195.46939$&$ 27.86325$&2.17&ACS&f475w&$20.2$&$ 1.15$&$0.84$&$ 157.9$&Arc&DClust58&0.27 S\tabularnewline
~~HAH $J132217.5+464630.4$&$100$&$200.57300$&$ 46.77510$&3.19&ACS&f814w&$18.4$&$ 1.49$&$0.66$&$ -60.9$&Arc&DClust59&0.37 S\tabularnewline
~~HAH $J132331.9+302155.8$&$101$&$200.88300$&$ 30.36550$&2.5&ACS&f814w&$21.1$&$ 0.36$&$0.87$&$  64.7$&Arc&DClust60&0.46 S\tabularnewline
~~ $ $&$ $&$ $&$ $&2.5&ACS&f814w&$19.3$&$ 0.82$&$0.73$&$  70.1$&Arc& &0.46 S\tabularnewline
~~HAH $J132529.1-474019.2$&$102$&$201.37147$&$-47.67202$&0.71&UVIS&f814w&$21.4$&$ 0.40$&$0.95$&$ -65.8$&Ring& & \tabularnewline
~~HAH $J133240.6+503315.5$&$103$&$203.16900$&$ 50.55430$&3.3&ACS&f814w&$19.8$&$ 0.46$&$0.93$&$-142.5$&Arc&DClust61&0.33 P\tabularnewline
~~HAH $J133252.7+503026.9$&$104$&$203.21991$&$ 50.50747$&1.45&ACS&f814w&$18.7$&$ 2.39$&$0.80$&$-150.4$&Arc&DClust62&0.29 P\tabularnewline
~~HAH $J135953.3+623118.4$&$105$&$209.97199$&$ 62.52180$&2.56&ACS&f814w&$19.0$&$ 1.03$&$0.58$&$-126.6$&Arc&DClust63&0.35 P\tabularnewline
~~HAH $J143236.3+344030.4$&$106$&$218.15140$&$ 34.67510$&1.63&ACS&f814w&$20.1$&$ 0.77$&$0.73$&$  30.6$&Arc&DClust64&0.57 P\tabularnewline
~~HAH $J145832.7-002349.9$&$107$&$224.63646$&$ -0.39719$&0.71&UVIS&f475x&$23.1$&$ 0.40$&$1.00$&$ -34.0$&Arc&DClust65&0.6 S\tabularnewline
~~HAH $J151841.7+084549.2$&$108$&$229.67379$&$  8.76369$&2.01&ACS&f814w&$19.7$&$ 0.58$&$0.95$&$ 140.9$&Arc&DClust66&0.49 P\tabularnewline
~~HAH $J161311.2+565004.1$&$109$&$243.29650$&$ 56.83446$&0.95&IR&f140w&$19.8$&$ 0.75$&$0.79$&$ 100.9$&Arc&DClust67&0.4 P\tabularnewline
~~HAH $J161420.9+323306.1$&$110$&$243.58700$&$ 32.55170$&1.17&ACS&f814w&$21.4$&$ 0.41$&$0.46$&$   4.5$&Arc& & \tabularnewline
~~HAH $J162122.2+381006.5$&$111$&$245.34262$&$ 38.16847$&0.91&ACS&f814w&$21.7$&$ 0.50$&$1.00$&$ -70.2$&Arc&DClust68&0.2 S\tabularnewline
~~HAH $J173146.2+225237.6$&$112$&$262.94232$&$ 22.87712$&1.14&ACS&f814w&$18.8$&$ 0.84$&$0.75$&$  49.7$&Arc&DClust69&0.44 P\tabularnewline
~~HAH $J174109.0+690236.1$&$113$&$265.28755$&$ 69.04336$&4.53&ACS&f814w&$21.5$&$ 0.31$&$0.87$&$ -61.1$&Arc& & \tabularnewline
~~HAH $J180652.4+292945.3$&$114$&$271.71867$&$ 29.49593$&0.87&ACS&f606w&$21.4$&$ 0.23$&$0.83$&$   8.8$&Arc&DClust70& \tabularnewline
~~HAH $J191700.9-333207.0$&$115$&$289.25409$&$-33.53530$&1.29&ACS&f814w&$19.1$&$ 1.14$&$0.75$&$ -71.3$&Arc&DClust71& \tabularnewline
~~HAH $J191709.5-333127.0$&$116$&$289.28972$&$-33.52418$&0.67&ACS&f814w&$22.5$&$ 0.17$&$0.56$&$  10.2$&Arc&DClust72& \tabularnewline
~~HAH $J204548.7-243833.2$&$117$&$311.45298$&$-24.64258$&1.52&ACS&f814w&$19.6$&$ 0.77$&$0.94$&$-167.5$&Arc& & \tabularnewline
~~HAH $J224324.6-093617.3$&$118$&$340.85260$&$ -9.60480$&2.06&ACS&f606w&$21.1$&$ 1.22$&$0.37$&$ 120.6$&Arc&DClust73&0.44 P\tabularnewline
~~HAH $J233220.8-535909.8$&$119$&$353.08657$&$-53.98605$&1.26&IR&f140w&$17.2$&$ 0.72$&$0.85$&$  -7.2$&Arc&DClust74&0.4 $\star$\tabularnewline
\tabularnewline*
\hline
{\bfseries C grade Lenses}&&&&&&&&&&&&&\tabularnewline*
\tabularnewline*
~~HAH $J000146.0-544026.2$&$120$&$  0.44161$&$-54.67395$&1.15&ACS&f606w&$20.9$&$ 0.54$&$0.92$&$-118.0$&Arc&DClust75& \tabularnewline
~~HAH $J002357.7-244154.3$&$121$&$  5.99142$&$-24.69858$&0.88&ACS&f606w&$23.2$&$ 0.85$&$0.72$&$-146.0$&Arc&DClust76& \tabularnewline
~~HAH $J002616.0-110548.2$&$122$&$  6.56680$&$-11.09673$&2.9&ACS&f814w&$20.9$&$ 0.90$&$0.67$&$  87.6$&Arc& &0.35 P\tabularnewline
~~HAH $J003501.7+023358.6$&$123$&$  8.75690$&$  2.56628$&2.25&ACS&f814w&$18.9$&$ 0.78$&$0.7$&$-144.6$&Arc&DClust77&0.38 P\tabularnewline
~~HAH $J004802.0+402624.1$&$124$&$ 12.00867$&$ 40.44004$&1.4&ACS&f814w&$20.0$&$ 0.67$&$0.36$&$ 133.2$&Arc& &0.54 P\tabularnewline
~~HAH $J013726.6-082747.9$&$125$&$ 24.36080$&$ -8.46330$&2.5&ACS&f814w&$18.4$&$ 1.60$&$0.75$&$-134.7$&Double&DClust78&0.29 P\tabularnewline
~~HAH $J014236.3+144114.3$&$126$&$ 25.65113$&$ 14.68730$&1.5&ACS&f775w&$19.3$&$ 1.18$&$0.96$&$ 174.8$&Arc& &0.5 S\tabularnewline
~~HAH $J015246.7+360750.5$&$127$&$ 28.19450$&$ 36.13070$&0.85&ACS&f435w&$22.8$&$ 1.23$&$0.75$&$ -89.5$&Ring x 2&DClust79& \tabularnewline
~~HAH $J015317.2-135220.6$&$128$&$ 28.32207$&$-13.87241$&0.65&ACS&f814w&$21.4$&$ 0.40$&$0.69$&$-137.8$&Arc&DClust80& \tabularnewline
~~HAH $J022141.7-032038.9$&$129$&$ 35.42410$&$ -3.34414$&1.25&ACS&f850lp&$20.4$&$ 0.79$&$0.67$&$-101.9$&Arc& &0.46 P\tabularnewline
~~HAH $J022253.9+422616.4$&$130$&$ 35.72461$&$ 42.43789$&2.04&ACS&f814w&$20.6$&$ 0.71$&$0.66$&$  57.5$&Arc& & \tabularnewline
~~HAH $J023726.4-262957.6$&$131$&$ 39.35982$&$-26.49932$&2.76&ACS&f606w&$19.3$&$ 0.70$&$0.76$&$-165.5$&Arc&DClust81& \tabularnewline
~~HAH $J024336.7-483319.4$&$132$&$ 40.90317$&$-48.55540$&1.33&UVIS&f200lp&$22.1$&$ 0.45$&$0.75$&$ -98.4$&Ring&DClust82& \tabularnewline
~~HAH $J030901.7+264522.2$&$133$&$ 47.25701$&$ 26.75616$&1.25&ACS&f814w&$18.3$&$ 2.26$&$0.98$&$-136.6$&Arc&DClust83& \tabularnewline
~~HAH $J040411.1-554433.8$&$134$&$ 61.04657$&$-55.74273$&1.64&ACS&f606w&$17.6$&$ 1.40$&$0.64$&$-158.2$&Arc& & \tabularnewline
~~HAH $J041308.3-655338.7$&$135$&$ 63.28420$&$-65.89428$&0.51&ACS&f606w&$24.7$&$ 0.28$&$0.82$&$-156.5$&Ring& & \tabularnewline
~~HAH $J042905.4-101206.7$&$136$&$ 67.27276$&$-10.20188$&1.84&IR&f140w&$17.8$&$ 0.73$&$0.88$&$-166.1$&Arc&DClust84& \tabularnewline
~~HAH $J043916.2-460115.2$&$137$&$ 69.81779$&$-46.02089$&0.97&ACS&f606w&$21.8$&$ 0.17$&$0.91$&$ 175.9$&Arc&DClust85& \tabularnewline
~~HAH $J044632.8-591132.0$&$138$&$ 71.63668$&$-59.19223$&1.5&ACS&f814w&$19.9$&$ 1.03$&$0.78$&$ 120.5$&Arc& & \tabularnewline
~~HAH $J045159.3+000645.9$&$139$&$ 72.99696$&$  0.11276$&1.86&ACS&f814w&$20.9$&$ 0.01$&$0.39$&$-102.3$&Cross&DClust86& \tabularnewline
~~HAH $J051537.3-432514.2$&$140$&$ 78.90505$&$-43.42074$&2.87&ACS&f814w&$18.1$&$ 1.88$&$0.70$&$-170.3$&Arc&DClust87&0.39 P$\star$\tabularnewline
~~HAH $J051856.3-431439.1$&$141$&$ 79.73450$&$-43.24420$&4.61&ACS&f814w&$20.5$&$ 0.39$&$0.66$&$  72.0$&Arc&DClust88& \tabularnewline
~~HAH $J051856.4-431444.4$&$142$&$ 79.73518$&$-43.24569$&4.61&ACS&f814w&$19.5$&$ 1.90$&$0.57$&$  13.4$&Arc&DClust89& \tabularnewline
~~HAH $J052116.2-510409.5$&$143$&$ 80.31740$&$-51.06930$&0.52&ACS&f606w&$23.4$&$ 0.45$&$0.51$&$ -60.9$&Triple&DClust90& \tabularnewline
~~HAH $J052214.2-362408.4$&$144$&$ 80.55929$&$-36.40232$&1.09&ACS&f814w&$21.0$&$ 0.28$&$0.87$&$  23.5$&Arc&DClust91& \tabularnewline
~~HAH $J065327.1-574306.1$&$145$&$103.36301$&$-57.71836$&1.21&UVIS&f200lp&$20.3$&$ 0.89$&$0.41$&$  53.2$&Ring&DClust92& \tabularnewline
~~HAH $J071718.8+374842.3$&$146$&$109.32870$&$ 37.81175$&1.12&ACS&f814w&$20.9$&$ 0.02$&$0.47$&$ 113.9$&Arc&DClust93& \tabularnewline
~~HAH $J071731.9+374449.9$&$147$&$109.38300$&$ 37.74720$&1.04&ACS&f814w&$19.3$&$ 1.47$&$0.42$&$ 153.5$&Arc&DClust94& \tabularnewline
~~HAH $J074704.6+693824.0$&$148$&$116.76942$&$ 69.64002$&1.43&ACS&f606w&$20.5$&$ 0.98$&$0.87$&$  37.8$&Arc&DClust95& \tabularnewline
~~HAH $J080911.4+064341.9$&$149$&$122.29730$&$  6.72820$&2.41&ACS&f606w&$18.2$&$ 1.84$&$0.68$&$ 174.8$&Arc&DClust96&0.36 P\tabularnewline
~~HAH $J083637.2+005332.6$&$150$&$129.15500$&$  0.89240$&3.5&ACS&f850lp&$17.7$&$ 1.53$&$0.53$&$ 128.2$&Arc& &0.22 P\tabularnewline
~~HAH $J084349.0+411635.4$&$151$&$130.95433$&$ 41.27651$&0.68&ACS&f814w&$21.2$&$ 0.50$&$0.90$&$-110.8$&Ring&DClust97&0.44 P\tabularnewline
~~HAH $J084959.8+360340.6$&$152$&$132.49950$&$ 36.06130$&0.82&ACS&f814w&$21.7$&$ 0.85$&$0.65$&$ -16.1$&Arc&DClust98&0.32 P\tabularnewline
~~HAH $J091023.7+021049.2$&$153$&$137.59885$&$  2.18035$&0.9&UVIS&f606w&$22.1$&$ 0.82$&$0.92$&$ -42.7$&Ring&DClust99&0.69 P\tabularnewline
~~HAH $J091044.5+424945.8$&$154$&$137.68550$&$ 42.82940$&1.78&ACS&f814w&$20.2$&$ 0.34$&$0.69$&$  15.2$&Arc&DClust100&0.62 P\tabularnewline
~~HAH $J092141.2-062241.6$&$155$&$140.42165$&$ -6.37810$&0.79&ACS&f606w&$22.7$&$ 0.49$&$0.85$&$ -84.0$&Arc&DClust101& \tabularnewline
~~HAH $J094058.7+074541.1$&$156$&$145.24482$&$  7.76142$&0.87&ACS&f814w&$20.9$&$ 0.96$&$0.50$&$ -40.5$&Arc&DClust102&0.49 P\tabularnewline
~~HAH $J095557.2-095159.0$&$157$&$148.98830$&$ -9.86640$&1.19&ACS&f606w&$21.0$&$ 0.31$&$0.81$&$  67.5$&Arc& & \tabularnewline
~~HAH $J100050.2+013251.5$&$158$&$150.20939$&$  1.54765$&0.9&IR&f140w&$19.2$&$ 0.85$&$0.75$&$ 152.7$&Arc&DClust103&0.69 P\tabularnewline
~~HAH $J100142.8+015448.1$&$159$&$150.42840$&$  1.91330$&1.48&ACS&f814w&$18.6$&$ 0.51$&$0.92$&$ -65.9$&Arc&DClust104&0.32 P\tabularnewline
~~HAH $J100222.3+023220.0$&$160$&$150.59303$&$  2.53890$&1.1&ACS&f814w&$21.5$&$ 1.32$&$0.98$&$-164.3$&Arc& &1.23 P$\star$\tabularnewline
~~HAH $J103943.7-001516.3$&$161$&$159.93217$&$ -0.25453$&1.2&ACS&f814w&$20.7$&$ 1.26$&$0.91$&$ -30.4$&Arc& & \tabularnewline
~~HAH $J111554.8+013015.5$&$162$&$168.97830$&$  1.50430$&1.09&ACS&f606w&$20.6$&$ 0.56$&$0.68$&$-151.8$&Arc&DClust105&0.39 P\tabularnewline
~~HAH $J112001.3-120351.1$&$163$&$170.00540$&$-12.06420$&1.02&ACS&f606w&$22.5$&$ 0.29$&$0.73$&$  84.0$&Arc& &0.51 P\tabularnewline
~~HAH $J113305.2+500840.9$&$164$&$173.27207$&$ 50.14471$&1.81&ACS&f814w&$20.2$&$ 0.37$&$0.42$&$ -56.3$&Double&DClust106&0.4 P\tabularnewline
~~HAH $J114923.6+222926.7$&$165$&$177.34814$&$ 22.49076$&2.61&ACS&f814w&$16.1$&$ 4.58$&$0.75$&$  50.6$&Arc& &0.21 S\tabularnewline
~~HAH $J115054.9-280548.7$&$166$&$177.72889$&$-28.09686$&0.95&ACS&f814w&$20.1$&$ 0.64$&$0.39$&$ 111.6$&Arc&DClust107& \tabularnewline
~~HAH $J115855.4-185941.2$&$167$&$179.73093$&$-18.99480$&1.55&ACS&f606w&$19.3$&$ 1.22$&$0.29$&$  50.2$&Arc& & \tabularnewline
~~HAH $J122027.5+752124.6$&$168$&$185.11459$&$ 75.35685$&1.82&ACS&f850lp&$19.1$&$ 0.91$&$0.85$&$  77.7$&Arc& & \tabularnewline
~~HAH $J122809.5+221743.3$&$169$&$187.03964$&$ 22.29538$&1.59&ACS&f814w&$19.5$&$ 0.83$&$0.77$&$ 158.4$&Arc& &0.37 S\tabularnewline
~~HAH $J123227.1-125100.0$&$170$&$188.11304$&$-12.85002$&1.13&ACS&f814w&$20.2$&$ 0.45$&$0.93$&$ -22.7$&Arc& & \tabularnewline
~~HAH $J123904.4+121231.4$&$171$&$189.76824$&$ 12.20872$&1.73&ACS&f814w&$21.4$&$ 0.39$&$0.81$&$ 118.2$&Double& &0.48 P\tabularnewline
~~HAH $J124355.9+320341.0$&$172$&$190.98280$&$ 32.06140$&1.9&ACS&f814w&$20.5$&$ 0.37$&$0.80$&$ 162.6$&Arc& &0.73 P\tabularnewline
~~HAH $J125801.0-431814.1$&$173$&$194.50418$&$-43.30393$&0.85&ACS&f606w&$20.4$&$ 0.50$&$0.57$&$ -56.1$&Arc& & \tabularnewline
~~HAH $J130154.4+275347.0$&$174$&$195.47657$&$ 27.89638$&1.24&ACS&f475w&$22.8$&$ 0.64$&$0.73$&$  43.8$&Arc&DClust108&0.57 S\tabularnewline
~~HAH $J130236.3+141055.0$&$175$&$195.65149$&$ 14.18195$&1.02&UVIS&f350lp&$21.9$&$ 0.20$&$0.75$&$  76.2$&Arc&DClust109& \tabularnewline
~~HAH $J133422.4+503133.0$&$176$&$203.59323$&$ 50.52586$&1.76&ACS&f775w&$20.1$&$ 0.34$&$0.69$&$   9.2$&Arc&DClust110&0.63 S\tabularnewline
~~HAH $J134711.8-114157.8$&$177$&$206.79914$&$-11.69944$&2.38&ACS&f850lp&$18.5$&$ 0.92$&$0.37$&$ -53.0$&Arc&DClust111& \tabularnewline
~~HAH $J135343.4+050952.1$&$178$&$208.43113$&$  5.16448$&1.7&ACS&f606w&$20.7$&$ 0.53$&$0.80$&$ 159.6$&Arc& &0.4 P\tabularnewline
~~HAH $J141809.5+524058.8$&$179$&$214.53970$&$ 52.68300$&1.89&ACS&f606w&$18.9$&$ 0.99$&$0.76$&$ -61.9$&Arc& &0.23 P\tabularnewline
~~HAH $J142018.4+530157.3$&$180$&$215.07651$&$ 53.03259$&2.07&ACS&f606w&$20.0$&$ 0.88$&$0.77$&$ 130.3$&Arc& &0.37 P\tabularnewline
~~HAH $J142952.3+554752.8$&$181$&$217.46790$&$ 55.79800$&0.93&UVIS&f600lp&$20.9$&$ 0.38$&$0.67$&$ 119.7$&Arc&DClust112&0.6 P\tabularnewline
~~HAH $J143808.8+341940.3$&$182$&$219.53684$&$ 34.32787$&3.22&ACS&f814w&$19.9$&$ 0.43$&$0.72$&$  75.7$&Arc&DClust113&0.54 S\tabularnewline
~~HAH $J144727.3+083001.9$&$183$&$221.86388$&$  8.50054$&0.68&ACS&f606w&$21.7$&$ 0.29$&$0.48$&$  35.3$&Arc&DClust114&0.51 P\tabularnewline
~~HAH $J151832.6-813031.8$&$184$&$229.63600$&$-81.50883$&0.75&ACS&f814w&$20.8$&$ 0.60$&$0.37$&$ -59.2$&Arc&DClust115& \tabularnewline
~~HAH $J160613.2-000039.5$&$185$&$241.55523$&$ -0.01097$&0.91&ACS&f606w&$22.8$&$ 0.51$&$0.87$&$  96.3$&Ring& &0.28 P\tabularnewline
~~HAH $J161543.8-060959.7$&$186$&$243.93268$&$ -6.16660$&3.09&ACS&f814w&$17.5$&$ 1.57$&$0.92$&$  34.9$&Arc&DClust116&0.2 $\star$\tabularnewline
~~HAH $J162308.1+072141.5$&$187$&$245.78380$&$  7.36153$&1.74&ACS&f850lp&$19.5$&$ 0.43$&$0.82$&$ -55.5$&Arc& &0.42 P\tabularnewline
~~HAH $J171721.7+593126.0$&$188$&$259.34068$&$ 59.52392$&1.67&ACS&f814w&$19.1$&$ 0.89$&$0.95$&$-150.2$&Arc& &0.38 P\tabularnewline
~~HAH $J173142.8+225101.0$&$189$&$262.92838$&$ 22.85027$&1.4&ACS&f814w&$20.3$&$ 0.49$&$0.48$&$  75.0$&Arc&DClust117&0.36 P\tabularnewline
~~HAH $J173647.0+461847.1$&$190$&$264.19645$&$ 46.31250$&2.31&IR&f110w&$16.2$&$ 1.58$&$0.91$&$  29.1$&Arc&DClust118& \tabularnewline
~~HAH $J175151.7+443859.0$&$191$&$267.96554$&$ 44.64972$&2.9&ACS&f814w&$20.4$&$ 0.47$&$0.94$&$  46.9$&Arc&DClust119&0.38 P\tabularnewline
~~HAH $J221450.1-140049.7$&$192$&$333.70891$&$-14.01380$&2.08&ACS&f555w&$18.9$&$ 1.29$&$0.76$&$ -35.5$&Arc&DClust120& \tabularnewline
~~HAH $J224154.3+173336.5$&$193$&$340.47639$&$ 17.56015$&0.81&ACS&f814w&$19.2$&$ 0.54$&$0.31$&$  87.0$&Arc&DClust121&0.31 P\tabularnewline
~~HAH $J224901.9-443103.9$&$194$&$342.25819$&$-44.51777$&1.58&ACS&f814w&$21.9$&$ 0.96$&$0.14$&$ -80.8$&Arc&DClust122& \tabularnewline
~~HAH $J230104.7+300726.4$&$195$&$345.26983$&$ 30.12401$&1.8&ACS&f850lp&$17.6$&$ 1.91$&$0.81$&$  26.2$&Arc&DClust123& \tabularnewline
~~HAH $J233813.3+270418.0$&$196$&$354.55585$&$ 27.07173$&0.58&UVIS&f814w&$22.8$&$ 1.83$&$0.18$&$   7.2$&Double& & \tabularnewline
~~HAH $J234232.8-541329.7$&$197$&$355.63659$&$-54.22491$&1.04&ACS&f606w&$22.2$&$ 0.26$&$0.79$&$-130.6$&Arc&DClust124& \tabularnewline
~~HAH $J235747.7+414740.8$&$198$&$359.44843$&$ 41.79490$&4.5&ACS&f814w&$18.3$&$ 2.64$&$0.63$&$  39.9$&Arc& & \tabularnewline
\end{longtable}
{\centering \footnotesize {\bf Notes.} This table summarises a list of characteristics related to each newly discovered lens. The lens objects are named using the standard IAU name, where the acronym HAH stands for Hubble Asteroid Hunter. They are grouped by grades (A, B, C) which were assigned by the authors of this paper as a qualitative assessment of the candidates. RA and Dec: right ascension and declination given in degrees of celestial coordinates J2000.0. $r_{\rm arc}$: the arc radius (in arcsec) corresponding to the angular separation between the lens and the surrounding images. Instru. and filter: the HST instrument (ACS/WFC, WFC3/UVIS or WFC3/IR) and the filter with which the HST image was taken. $Mag$, $r_{\rm e}$, $q$, and $PA$: the apparent magnitude, the effective radius (in arcsec), the axis ratio and the position angle of the foreground lens galaxy (measured from North through East, same orientation as in Figures~\ref{fig:grade_A}-\ref{fig:grade_C}) ; these parameters were retrieved with {\tt Galfit}. class: the morphological class of the source images (Arc, Quad, Ring, Double). clust.: a flag indicating the presence of a foreground or background nearby cluster, details are available in table~\ref{tab:clusters}.  $z$: the photometric (P) or spectroscopic (S) redshift, retrieved from SDSS (no mark) or NED (marked $\star$). } 
}
\clearpage


\setlongtables{\scriptsize
\setlength\tabcolsep{3.6pt}
\onecolumn
\setlength\LTcapwidth{\textwidth}

\begin{longtable}{lrrrlllrrrrllll}
\caption{Rediscovered lenses identified in this study.}
\tabularnewline
\hline
\multicolumn{1}{l}{IAU Name}&\multicolumn{1}{c}{ID}&\multicolumn{1}{c}{RA}&\multicolumn{1}{c}{Dec}&\multicolumn{1}{c}{$r_{\rm arc}\,(\arcsec)$}&\multicolumn{1}{c}{instru.}&\multicolumn{1}{c}{filter}&\multicolumn{1}{c}{$mag$}&\multicolumn{1}{c}{$r_{\rm e}\,(\arcsec)$}&\multicolumn{1}{c}{$q$}&\multicolumn{1}{c}{PA}&\multicolumn{1}{c}{class}&\multicolumn{1}{c}{clust.}&\multicolumn{1}{c}{$z$}&\multicolumn{1}{c}{reference}\tabularnewline
\hline
\endfirsthead
\caption[]{\em (continued.)} 
\tabularnewline*
\hline
\multicolumn{1}{l}{Name}&\multicolumn{1}{c}{ID}&\multicolumn{1}{c}{RA}&\multicolumn{1}{c}{Dec}&\multicolumn{1}{c}{$r_{\rm arc} (\arcsec)$}&\multicolumn{1}{c}{instru.}&\multicolumn{1}{c}{filter}&\multicolumn{1}{c}{Mag}&\multicolumn{1}{c}{$r_{\rm e}(\arcsec)$}&\multicolumn{1}{c}{$q$}&\multicolumn{1}{c}{PA}&\multicolumn{1}{c}{class}&\multicolumn{1}{c}{clust.}&\multicolumn{1}{c}{z}&\multicolumn{1}{c}{reference}\tabularnewline
\hline
\tabularnewline*
\endhead
\hline
\endfoot
\label{tab:rediscoveries}
{\bfseries A grade Lenses}&&&&&&&&&&&&&&\tabularnewline*
\tabularnewline*
~~HAH $J001426.3-302255.3$&$199$&$  3.60962$&$-30.38202$&1&ACS&f814w&$18.9$&$0.55$&$0.72$&$-137.7$&Arc&RClust125&0.24 $\star$& \citet{Pawase2014} \tabularnewline*
~~HAH $J015242.4-135618.2$&$200$&$ 28.17683$&$-13.93838$&0.97&ACS&f775w&$21.4$&$0.30$&$0.92$&$ -96.3$&Arc&RClust126& & \citet{Grillo2008} \tabularnewline*
~~HAH $J032940.8-021318.5$&$201$&$ 52.42010$&$ -2.22180$&0.82&ACS&f435w&$22.2$&$0.67$&$0.94$&$-148.2$&Arc&RClust127& & \citet{Desprez2018} \tabularnewline*
~~HAH $J033238.2-275653.3$&$202$&$ 53.15930$&$-27.94810$&1.67&ACS&f606w&$21.0$&$1.08$&$0.53$&$-124.5$&Arc& & & \citet{More2011} \tabularnewline*
~~HAH $J033304.4-065144.3$&$203$&$ 53.26850$&$ -6.86230$&0.83&IR&f160w&$19.7$&$0.69$&$0.75$&$ 104.2$&Arc&RClust128&0.57 P& \citet{Sharon2020} \tabularnewline*
~~HAH $J041604.1-240522.2$&$204$&$ 64.01720$&$-24.08950$&1.2&ACS&f814w&$20.0$&$0.57$&$0.32$&$  28.6$&Arc&RClust129&0.4 P$\star$& \citet{Diego2015} \tabularnewline*
~~HAH $J045400.6-030833.8$&$205$&$ 73.50260$&$ -3.14280$&1.88&ACS&f814w&$17.0$&$3.45$&$0.69$&$  -8.6$&Arc&RClust130& & \citet{Schirmer2010} \tabularnewline*
~~HAH $J045413.1+025733.8$&$206$&$ 73.55475$&$  2.95939$&2.6&ACS&f814w&$16.5$&$2.55$&$0.96$&$  -3.0$&Ring&RClust131& & \citet{Clowe2012} \tabularnewline*
~~HAH $J074450.9+392736.0$&$207$&$116.21211$&$ 39.45986$&unfit&ACS&f606w&$22.5$&$1.43$&$0.39$&$  21.7$&Arc&RClust132&0.62 P$\star$& \citet{Horesh2010} \tabularnewline*
~~HAH $J091617.5-002405.9$&$208$&139.07306&-0.40164&0.74&ACS&f814w&$ $&$ $&$ $&$ $&Arc&RClust133& & \citet{Clowe2012} \tabularnewline*
~~HAH $J095921.8+020638.5$&$209$&$149.84068$&$  2.11064$&0.71&ACS&f814w&$20.5$&$0.34$&$0.78$&$  49.4$&Quad& &0.58 P& \citet{Faure2008} \tabularnewline*
~~HAH $J095930.9+023427.8$&$210$&$149.87890$&$  2.57440$&0.92&ACS&f814w&$21.9$&$0.38$&$0.61$&$ -69.8$&Arc& &0.89 $\star$& \citet{Lagattuta2010} \tabularnewline*
~~HAH $J095956.0+021901.8$&$211$&$149.98330$&$  2.31719$&2.36&ACS&f814w&$22.9$&$4.65$&$0.21$&$  10.7$&Arc& &0.93 $\star$& \citet{More2012} \tabularnewline*
~~ $ $&$ $&$ $&$ $&2.36&ACS&f814w&$21.3$&$0.74$&$0.88$&$ -64.5$&Arc& &0.93 $\star$&\tabularnewline*
~~HAH $J100012.6+022015.8$&$212$&$150.05258$&$  2.33772$&0.73&ACS&f814w&$19.3$&$0.49$&$0.81$&$   5.9$&Arc&RClust134&0.37 P& \citet{Cabanac2007} \tabularnewline*
~~HAH $J100018.5+023845.6$&$213$&$150.07700$&$  2.64600$&1.51&ACS&f814w&$23.6$&$0.12$&$0.71$&$  10.8$&Arc&RClust135&0.42 P$\star$& \citet{Cabanac2007} \tabularnewline*
~~HAH $J100108.4+024029.9$&$214$&$150.28495$&$  2.67496$&1.41&ACS&f814w&$19.3$&$0.34$&$0.86$&$   1.8$&Arc&&0.32 P$\star$& \citet{Pourrahmani2018} \tabularnewline*
~~HAH $J100133.7+022135.3$&$215$&$150.39058$&$  2.35982$&2.54&ACS&f814w&$20.3$&$0.54$&$0.44$&$ -29.4$&Arc&RClust136&0.36 P& \citet{More2012} \tabularnewline*
~~HAH $J100211.2+021139.5$&$216$&$150.54675$&$  2.19429$&3.45&ACS&f814w&$20.5$&$1.77$&$0.46$&$  13.7$&Arc& &0.7 P& \citet{Faure2008} \tabularnewline*
~~HAH $J114936.7+222612.0$&$217$&$177.40280$&$ 22.43667$&0.859&ACS&f814w&$19.8$&$0.48$&$0.80$&$  60.8$&Cross& &0.5 P& \citet{Desprez2018} \tabularnewline*
~~HAH $J133235.0+503237.3$&$218$&$203.14564$&$ 50.54369$&4.2&ACS&f606w&$18.9$&$1.66$&$0.72$&$-156.6$&Arc& &0.28 S& \citet{Ragozzine2012} \tabularnewline*
~~HAH $J133241.0+503346.3$&$219$&$203.17064$&$ 50.56286$&2.52&ACS&f606w&$18.2$&$1.89$&$0.90$&$ 149.3$&Ring&RClust137&0.28 S& \citet{Ragozzine2012} \tabularnewline*
~~HAH $J135409.1+771557.2$&$220$&$208.53795$&$ 77.26590$&1.79&ACS&f814w&$21.0$&$0.60$&$0.75$&$-179.5$&Arc&RClust138& & \citet{Repp2018} \tabularnewline*
~~HAH $J135409.2+771549.9$&$221$&$208.53816$&$ 77.26385$&2.26&ACS&f814w&$20.5$&$0.91$&$0.65$&$ -90.1$&Arc&RClust139& & \citet{Repp2018} \tabularnewline*
~~HAH $J140237.0+542716.3$&$222$&$210.65450$&$ 54.45455$&1.8&ACS&f814w&$18.1$&$1.14$&$0.74$&$-114.1$&Arc& & & \citet{Pawase2014}\tabularnewline*
~~HAH $J141735.7+522646.3$&$223$&$214.39891$&$ 52.44621$&1.6&ACS&f814w&$20.2$&$0.98$&$0.62$&$-142.5$&Cross& &0.59 P& \citet{Ratnatunga1995} \tabularnewline*
~~HAH $J141820.8+523611.1$&$224$&$214.58664$&$ 52.60309$&0.81&ACS&f814w&$21.2$&$0.62$&$0.76$&$ -91.0$&Arc& &0.58 S& \citet{Moustakas2007} \tabularnewline*
~~HAH $J143703.3+350153.3$&$225$&$219.26352$&$ 35.03145$&0.65&ACS&f814w&$22.5$&$0.36$&$0.76$&$ -78.8$&Ring& &0.56 P& \citet{Pawase2014} \tabularnewline*
~~HAH $J145250.0+580135.3$&$226$&$223.20849$&$ 58.02649$&3.22&ACS&f814w&$18.3$&$1.17$&$0.67$&$  42.3$&Arc&RClust140&0.33 S& \citet{Repp2018} \tabularnewline*
~~HAH $J152745.2+065220.8$&$227$&$231.93830$&$  6.87247$&1.78&UVIS&f606w&$22.0$&$0.30$&$0.33$&$  39.7$&Arc&RClust141&0.32 P& \citet{Koester2010} \tabularnewline*
~~HAH $J213512.7-010143.9$&$228$&$323.80296$&$ -1.02886$&0.98&ACS&f814w&$20.7$&$0.64$&$0.65$&$  98.8$&Arc&RClust142&0.3 P& \citet{Smail2007} \tabularnewline*
~~HAH $J221142.0-035052.4$&$229$&$332.92486$&$ -3.84790$&1.15&ACS&f814w&$21.7$&$0.19$&$0.77$&$ -67.4$&Cross&RClust143&0.42 P& \citet{Bettoni2019} \tabularnewline*
~~HAH $J224851.7-443105.1$&$230$&$342.21568$&$-44.51858$&3.17&ACS&f814w&$20.9$&$0.11$&$0.82$&$  33.3$&Arc&RClust144&0.33 $\star$& \citet{Guzzo2009} \tabularnewline*
~~ $ $&$ $&$ $&$ $&3.17&ACS&f814w&$17.7$&$2.40$&$0.82$&$-103.8$&Arc& &0.33 $\star$&\tabularnewline*
~~ $ $&$ $&$ $&$ $&3.17&ACS&f814w&$19.0$&$3.39$&$0.86$&$ -10.8$&Arc& &0.33 $\star$&\tabularnewline*
~~ $ $&$ $&$ $&$ $&3.17&ACS&f814w&$20.1$&$0.27$&$0.69$&$ -32.1$&Arc& &0.33 $\star$&\tabularnewline*
~~HAH $J235130.6-261500.7$&$231$&$357.87744$&$-26.25008$&3.32&ACS&f814w&$16.3$&$4.47$&$0.74$&$-124.2$&Arc& &0.27 $\star$& \citet{Pawase2014} \tabularnewline
\tabularnewline*
\hline
{\bfseries B grade Lenses}&&&&&&&&&&&&&&\tabularnewline*
\tabularnewline*
~~HAH $J001423.0-302109.7$&$232$&$  3.59578$&$-30.35270$&1.7&ACS&f814w&$19.8$&$0.83$&$0.85$&$  22.5$&Arc&RClust145&0.25 $\star$& \citet{Pawase2014} \tabularnewline*
~~HAH $J033153.2-274619.3$&$233$&$ 52.97181$&$-27.77194$&1.05&ACS&f840w&$20.7$&$0.66$&$0.76$&$ -62.5$&Arc& &0.73 $\star$& \citet{More2011} \tabularnewline*
~~HAH $J095939.1+023044.1$&$234$&$149.91311$&$  2.51224$&1.7&ACS&f814w&$20.0$&$0.76$&$0.69$&$  62.6$&Arc&RClust146&0.72 P& \citet{Faure2008} \tabularnewline*
~~HAH $J100013.9+022249.8$&$235$&$150.05800$&$  2.38050$&1.65&ACS&f606w&$19.3$&$1.90$&$0.79$&$-110.8$&Arc&RClust147&0.35 S& \citet{Faure2008} \tabularnewline*
~~HAH $J100047.7+015023.3$&$236$&$150.19854$&$  1.83981$&1.71&ACS&f814w&$20.6$&$0.75$&$0.72$&$   2.2$&Arc&RClust148&0.89 $\star$& \citet{Faure2008} \tabularnewline*
~~HAH $J100055.7+013821.1$&$237$&$150.23214$&$  1.63919$&1.06&ACS&f814w&$20.9$&$0.35$&$0.95$&$  92.4$&Arc&RClust149&0.55 P& \citet{Faure2008} \tabularnewline*
~~HAH $J100056.8+021225.8$&$238$&$150.23661$&$  2.20719$&1.67&ACS&f814w&$18.8$&$1.01$&$0.66$&$  64.1$&Arc& &0.36 S& \citet{Faure2008} \tabularnewline*
~~HAH $J103751.4-124326.3$&$239$&$159.46400$&$-12.72398$&1.12&ACS&f814w&$19.5$&$1.39$&$0.70$&$ -49.2$&Arc& &0.58 $\star$& \citet{Pawase2014} \tabularnewline*
~~ $ $&$ $&$ $&$ $&1.12&ACS&f814w&$19.8$&$0.66$&$0.86$&$ -53.7$&Arc& &0.58 $\star$&\tabularnewline*
~~HAH $J121228.4-121723.1$&$240$&$183.11829$&$-12.29002$&2.84&ACS&f814w&$21.9$&$0.18$&$0.79$&$-105.5$&Arc& && \citet{Ebeling2019} \tabularnewline*
~~ $ $&$ $&$ $&$ $&2.84&ACS&f814w&$20.1$&$1.03$&$0.84$&$-175.4$&Arc& & &\tabularnewline*
~~HAH $J123729.9+621300.7$&$241$&$189.37456$&$ 62.21685$&2.12&ACS&f814w&$19.9$&$0.81$&$0.91$&$  48.9$&Arc& &0.49 P& \citet{Fassnacht2004} \tabularnewline*
~~HAH $J130042.7+280523.6$&$242$&$195.17804$&$ 28.08989$&1.08&ACS&f814w&$18.2$&$1.70$&$0.61$&$-118.2$&Arc&RClust150&0.37 S& \citet{Pawase2014} \tabularnewline*
~~HAH $J131829.4-010421.6$&$243$&$199.62246$&$ -1.07267$&0.67&ACS&f814w&$19.9$&$0.67$&$0.84$&$ -85.0$&Arc& &0.66 S& \citet{Brownstein2012} \tabularnewline*
~~HAH $J141833.1+524352.5$&$244$&$214.63790$&$ 52.73126$&0.69&ACS&f814w&$20.5$&$1.32$&$0.68$&$  97.2$&Arc& &0.47 P& \citet{Marshall2009} \tabularnewline*
~~HAH $J214014.8-233904.3$&$245$&$325.06171$&$-23.65125$&2.58&ACS&f606w&$20.0$&$0.58$&$0.94$&$ 133.2$&Arc& & & \citet{Fort1992} \tabularnewline*
~~HAH $J230824.7-021213.3$&$246$&$347.10300$&$ -2.20370$&3.24&ACS&f606w&$18.2$&$2.76$&$0.53$&$ -88.3$&Arc&RClust151&0.27 P& \citet{Jacobs2019} \tabularnewline
\tabularnewline*
\hline
{\bfseries C grade Lenses}&&&&&&&&&&&&&&\tabularnewline*
\tabularnewline*
~~HAH $J013152.3-133658.3$&$247$&$ 22.96793$&$-13.61621$&1.9&ACS&f475w&$19.0$&$1.58$&$0.86$&$-170.7$&Arc&RClust152& & \citet{Zitrin2015} \tabularnewline*
~~HAH $J071815.0+374200.3$&$248$&$109.56253$&$ 37.70009$&1.91&ACS&f814w&$20.6$&$0.44$&$0.96$&$  61.4$&Arc&RClust153&0.34 P& \citet{Ebeling2014} \tabularnewline*
~~HAH $J095810.1+022942.0$&$249$&$149.54211$&$  2.49500$&2.21&ACS&f814w&$18.9$&$1.05$&$0.75$&$ -57.8$&Arc& &0.4 P& \citet{Jackson2008} \tabularnewline*
~~HAH $J114934.3+222440.5$&$250$&$177.39310$&$ 22.41127$&0.61&ACS&f606w&$23.1$&$0.67$&$0.93$&$   7.0$&Arc&RClust154&0.47 P& \citet{Morishita2017} \tabularnewline*
~~HAH $J173217.5+193250.3$&$251$&$263.07333$&$ 19.54731$&1.2&ACS&f814w&$19.8$&$0.61$&$0.93$&$  44.6$&Arc&RClust155& & \citet{Ebeling2018} \tabularnewline*
~~HAH $J203146.3-403705.4$&$252$&$307.94293$&$-40.61818$&1.02&ACS&f814w&$22.4$&$0.30$&$0.76$&$   0.2$&Arc&RClust156& & \citet{Richard2015} \tabularnewline
\tabularnewline*
\hline
\end{longtable}
     {\centering \footnotesize {\bf Notes.} This table summarises a list of characteristics related to each rediscovered lens. The lens objects are named using the standard IAU name, where the acronym HAH stands for Hubble Asteroid Hunter. The lenses are grouped by grades (A,B,C) which were assigned by the authors of this paper as a qualitative assessment of the candidates. RA and Dec: right ascension and declination given in degrees of celestial coordinates J2000.0. $r_{\rm arc}$: the arc radius (in arcsec) corresponding to the angular separation between the lens and the surrounding images. Instru. and filter: the HST instrument (ACS/WFC, WFC3/UVIS or WFC3/IR) and the filter with which the HST image was taken. $Mag$, $r_{\rm e}$, $q$, and $PA$: the apparent magnitude, the effective radius (in arcsec), the axis ratio and the position angle of the foreground lens galaxy (measured from North through East, same orientation as in Figures~\ref{fig:grade_A}-\ref{fig:grade_C}); these parameters were retrieved with {\tt Galfit}. class: the morphological class of the source images (Arc, Quad, Ring, Double). clust.: a flag indicating the presence of a foreground or background nearby cluster, details are available in table~\ref{tab:clusters}. $z$: the photometric (P) or spectroscopic (S) redshift, retrieved from SDSS (no mark) or NED (marked $\star$). The reference columns shows the paper where the the strong lensing system is mentioned, as listed by the ESASky portal. Because of the complexity of lens ID~207, we did not manage to measure its arc radius. Finally, We did not manage to obtain a good model for the lens ID~208 and, therefore, left the \texttt{GALFIT} parameters blank. }
}
\clearpage

\setlength\tabcolsep{3.2pt}
\setlongtables{\scriptsize
\onecolumn
\setlength\LTcapwidth{\textwidth}
\begin{longtable}{llclcrrclr}
\caption{Candidate cluster flags.}
\tabularnewline
\hline
\multicolumn{1}{l}{\bfseries Flag}&\multicolumn{1}{c}{\bfseries }&\multicolumn{1}{c}{\bfseries }&\multicolumn{1}{c}{\bfseries HST Observation Target}&\multicolumn{1}{c}{\bfseries }&\multicolumn{2}{c}{\bfseries DESI Catalog}&\multicolumn{1}{c}{\bfseries }&\multicolumn{2}{c}{\bfseries redMaPPer SDSS Catalog}\tabularnewline*
\cline{4-4} \cline{6-7} \cline{9-10}
\multicolumn{1}{l}{}&\multicolumn{1}{c}{IAU name}&\multicolumn{1}{c}{}&\multicolumn{1}{c}{Targeted Cluster ID}&\multicolumn{1}{c}{}&\multicolumn{1}{c}{Candidate Cluster ID}&\multicolumn{1}{c}{Distance (\arcsec)}&\multicolumn{1}{c}{}&\multicolumn{1}{c}{Candidate Cluster ID}&\multicolumn{1}{c}{Distance (\arcsec)}\tabularnewline
\hline
\endfirsthead
\caption[]{\em continued.} 
\tabularnewline*
\hline
\multicolumn{1}{l}{\bfseries Flag}&\multicolumn{1}{c}{\bfseries }&\multicolumn{1}{c}{\bfseries }&\multicolumn{1}{c}{\bfseries HST Observation Target}&\multicolumn{1}{c}{\bfseries }&\multicolumn{2}{c}{\bfseries DESI Catalog}&\multicolumn{1}{c}{\bfseries }&\multicolumn{2}{c}{\bfseries redMaPPer SDSS Catalog}\tabularnewline*
\cline{4-4} \cline{6-7} \cline{9-10}
\multicolumn{1}{l}{}&\multicolumn{1}{c}{IAU name}&\multicolumn{1}{c}{}&\multicolumn{1}{c}{Targeted Cluster ID}&\multicolumn{1}{c}{}&\multicolumn{1}{c}{Candidate Cluster ID}&\multicolumn{1}{c}{Distance (\arcsec)}&\multicolumn{1}{c}{}&\multicolumn{1}{c}{Candidate Cluster ID}&\multicolumn{1}{c}{Distance (\arcsec)}\tabularnewline
\hline
\tabularnewline*
\endhead
\tabularnewline*
\hline
\endfoot
\label{tab:clusters}
{\bfseries Discovery group}&&&&&&&&&\tabularnewline*
\tabularnewline
~~DClust1&HAH $ J002348.6-244149.6$&&&&$3469200139$&$ 84$&&& \tabularnewline*
~~DClust2&HAH $ J002753.2-753730.0$&&SMACS $J0028.2-7537$&&$ $&$ $&&&\tabularnewline*
~~DClust3&HAH $ J004924.3-205729.8$&&ESO $540-31$&&$3316000106$&$ 93$&&&\tabularnewline*
~~DClust4&HAH $ J015949.3-084906.0$&&MACS $J0159.8-0849$&&$2830900043$&$ 64$&&RM $J015949.3-084958.9$&$ 53$ \tabularnewline*
~~DClust5&HAH $ J024329.8-593102.7$&&SPT-CL $J0243-5930$&&$ $&$ $&&&\tabularnewline*
~~DClust6&HAH $ J024533.6-530203.5$&&ACO S$295$&&$4427400112$&$ 36$&&&\tabularnewline*
~~DClust7&HAH $ J025241.6-150025.5$&&&&$3087900009$&$ 30$&&&$ $\tabularnewline*
~~DClust8&HAH $ J025659.9-163059.5$&&&&$3164700123$&$ 32$&&&$ $\tabularnewline*
~~DClust9&HAH $ J034445.2-642133.8$&&SPT-CL $J0345-6419$&&$ $&$ $&&&$ $\tabularnewline*
~~DClust10&HAH $ J054707.0-390516.3$&&MACS $J0547.0-3904$&&$3986900096$&$ 67$&&&$ $\tabularnewline*
~~DClust11&HAH $ J061345.7-562750.4$&&SPT-CL $J0613-5627$&&$4505800076$&$  0$&&&$ $\tabularnewline*
~~DClust12&HAH $ J072221.9+072326.7$&&PSZ2 $G209.79+10.23$&&$ $&$ $&&&$ $\tabularnewline*
~~DClust13&HAH $ J083420.3+452506.9$&&&&$ 692800146$&$ 70$&&RM $J083416.3+452420.0$&$ 63$\tabularnewline*
~~DClust14&HAH $ J102914.8+262207.0$&&&&$1355200029$&$ 92$&&&$ $\tabularnewline*
~~DClust15&HAH $ J105722.7+580046.5$&&&&$ 369000033$&$ 92$&&&$ $\tabularnewline*
~~DClust16&HAH $ J111337.4+221249.2$&&A$1300$&&$ $&$ $&&&$ $\tabularnewline*
~~DClust17&HAH $ J114938.4+222107.7$&&&&$1510200105$&$161$&&RM $J114935.7+222354.6$&$172$\tabularnewline*
~~DClust18&HAH $ J131953.6+700420.8$&&MACS $J1319.9+7003$&&$ 157400151$&$ 75$&&&$ $\tabularnewline*
~~DClust19&HAH $ J133525.9+405952.4$&& ACO $1763$&&$ 846500034$&$ 14$&&RM $J133520.1+410004.1$&$ 67$\tabularnewline*
~~DClust20&HAH $ J141139.2+565200.9$&&&&$ 406600086$&$ 80$&&&$ $\tabularnewline*
~~DClust21&HAH $ J214817.6-484350.5$&&SPT-CL $J2148-4843$&&$ $&$ $&&&$ $\tabularnewline*
~~DClust22&HAH $ J234106.5-000007.5$&&&&$2470000044$&$141$&&&$ $\tabularnewline*
~~DClust23&HAH $ J001419.9-302600.7$&&Abell $2744$&&$ $&$ $&&&$ $\tabularnewline*
~~DClust24&HAH $ J002706.3+170313.5$&&&&$1728400102$&$ 48$&&&$ $\tabularnewline*
~~DClust25&HAH $ J003350.8-074959.9$&&MACS $J0033.8-0751$&&$2778200022$&$ 75$&&RM $J003353.1-075210.4$&$135$\tabularnewline*
~~DClust26&HAH $ J004426.2-403941.4$&&SPT-CL $J0044-4037$&&$4058400124$&$140$&&&\tabularnewline*
~~DClust27&HAH $ J005141.6+272001.5$&&MACS $J0051.6+2720$&&$1319300120$&$ 49$&&RM $J005138.6+271959.9$&$ 40$\tabularnewline*
~~DClust28&HAH $ J011138.3-454015.1$&&&&$4208300009$&$ 70$&&&\tabularnewline*
~~DClust29&HAH $ J013723.7-082852.2$&&WHL $J013719.8-082841$&&$2804900125$&$ 78$&&RM $J013725.0-082722.7$&$ 92$\tabularnewline*
~~DClust30&HAH $ J023951.1-013205.9$&&&&$2498800056$&$177$&&RM $J023952.7-013418.9$&$135$\tabularnewline*
~~DClust31&HAH $ J024535.7-530125.5$&&ACO S$295$&&$4427400112$&$ 36$&&&$ $\tabularnewline*
~~DClust32&HAH $ J025735.5-220928.5$&&&&$3369500071$&$ 55$&&&$ $\tabularnewline*
~~DClust33&HAH $ J025740.9-221027.3$&&&&$3369500071$&$ 55$&&&$ $\tabularnewline*
~~DClust34&HAH $ J032030.9+003242.8$&&&&$2422800030$&$ 88$&&&$ $\tabularnewline*
~~DClust35&HAH $ J033530.7-540734.7$&&&&$4447600038$&$ 41$&&&$ $\tabularnewline*
~~DClust36&HAH $ J033645.9-535536.7$&&&&$4428100085$&$ 63$&&&$ $\tabularnewline*
~~DClust37&HAH $ J035851.7-295418.7$&&MACS $J0358.8-2955$&&$3677800030$&$ 79$&&&$ $\tabularnewline*
~~DClust38&HAH $ J041631.5-240512.5$&&MACS $J0416.1-2403$&&$ $&$ $&&&$ $\tabularnewline*
~~DClust39&HAH $ J044208.3-281503.7$&&RCS $J044207-2815.0$&&$3601800082$&$150$&&&$ $\tabularnewline*
~~DClust40&HAH $ J045430.7-030133.2$&&&&$2578000137$&$128$&&&$ $\tabularnewline*
~~DClust41&HAH $ J045441.7-030319.9$&&&&$2578000137$&$142$&&&$ $\tabularnewline*
~~DClust42&HAH $ J051850.3-431442.0$&&&&$4114300083$&$ 60$&&&$ $\tabularnewline*
~~DClust43&HAH $ J055326.6-334254.0$&&MACS $J0553.4-3342$&&$3807900148$&$ 98$&&&$ $\tabularnewline*
~~DClust44&HAH $ J072218.2+072314.8$&&PSZ2 $G209.79+10.23$&&$ $&$ $&&&$ $\tabularnewline*
~~DClust45&HAH $ J072333.8-732550.9$&&SMACS $J0723.3-7327$&&$ $&$ $&&&$ $\tabularnewline*
~~DClust46&HAH $ J084712.7+344910.9$&&&&$1046200011$&$ 65$&&&$ $\tabularnewline*
~~DClust47&HAH $ J094949.9+170614.4$&&RXC $J0949.8+1707$&&$1738400144$&$ 51$&&RM $J094951.8+170710.6$&$ 62$\tabularnewline*
~~DClust48&HAH $ J095803.9+653409.1$&&&&$ 216600094$&$ 17$&&&$ $\tabularnewline*
~~DClust49&HAH $ J100030.1+023716.7$&&&&$2353000132$&$ 98$&&&$ $\tabularnewline*
~~DClust50&HAH $ J100124.2+555407.2$&&&&$ 421600140$&$ 67$&&&$ $\tabularnewline*
~~DClust51&HAH $ J104132.0+000125.0$&&&&$2456200026$&$114$&&&$ $\tabularnewline*
~~DClust52&HAH $ J105726.9+575944.0$&&&&$ 369000033$&$ 92$&&&$ $\tabularnewline*
~~DClust53&HAH $ J111507.3+531955.8$&&MACS $J1115.2+5320$&&$ 499300020$&$ 79$&&RM $J111514.8+531954.6$&$ 67$\tabularnewline*
~~DClust54&HAH $ J111822.1+033802.1$&&&&$ $&$ $&&RM $J111814.9+033931.6$&$141$\tabularnewline*
~~DClust55&HAH $ J122944.7+112238.3$&&&&$1971700080$&$ 99$&&&$ $\tabularnewline*
~~DClust56&HAH $ J123618.8+260905.0$&&&&$1383000036$&$ 68$&&&$ $\tabularnewline*
~~DClust57&HAH $ J125907.8-014439.0$&&Abell $1650-9-13-0$&&$ $&$ $&&&$ $\tabularnewline*
~~DClust58&HAH $ J130152.7+275147.7$&&&&$1306700027$&$151$&&&$ $\tabularnewline*
~~DClust59&HAH $ J132217.5+464630.4$&&&&$ 674100042$&$ 62$&&RM $J132226.8+464630.2$&$ 95$\tabularnewline*
~~DClust60&HAH $ J132331.9+302155.8$&&&&$1204700022$&$ 63$&&RM $J132334.1+302249.2$&$ 60$\tabularnewline*
~~DClust61&HAH $ J133240.6+503315.5$&&Abell $1758N-P2$&&$ 563100075$&$ 67$&&RM $J133238.4+503336.0$&$ 29$\tabularnewline*
~~DClust62&HAH $ J133252.7+503026.9$&&Abell $1758N-P1$&&$ 563100075$&$142$&&&$ $\tabularnewline*
~~DClust63&HAH $ J135953.3+623118.4$&&&&$ 275000026$&$ 34$&&RM $J135950.6+623105.5$&$ 23$\tabularnewline*
~~DClust64&HAH $ J143236.3+344030.4$&&&&$1052300092$&$ 59$&&&$ $\tabularnewline*
~~DClust65&HAH $ J145832.7-002349.9$&&&&$2486400029$&$ 55$&&&$ $\tabularnewline*
~~DClust66&HAH $ J151841.7+084549.2$&&&&$2102700082$&$ 55$&&RM $J151846.8+084550.5$&$ 75$\tabularnewline*
~~DClust67&HAH $ J161311.2+565004.1$&&&&$ 408100043$&$ 44$&&&$ $\tabularnewline*
~~DClust68&HAH $ J162122.2+381006.5$&& MACS $J1621.3+3810$&&$ 926200081$&$ 35$&&RM $J162124.8+381008.9$&$ 30$\tabularnewline*
~~DClust69&HAH $ J173146.2+225237.6$&&&&$1516300115$&$ 87$&&&$ $\tabularnewline*
~~DClust70&HAH $ J180652.4+292945.3$&&$ $&&$1260900074$&$ 29$&&&$ $\tabularnewline*
~~DClust71&HAH $ J191700.9-333207.0$&&PLCK ESZ $G004.5-19.5$ ACS&&$ $&$ $&&&\tabularnewline*
~~DClust72&HAH $ J191709.5-333127.0$&&PLCK ESZ $G004.5-19.5$ IR&&$ $&$ $&&&$ $\tabularnewline*
~~DClust73&HAH $ J224324.6-093617.3$&&MACS $J0717+3745$ ACS&&$2878600004$&$ 43$&&RM $J224319.8-093530.9$&$ 85$\tabularnewline*
~~DClust74&HAH $ J233220.8-535909.8$&&SMACS $J2332.4-5358$&&$4463600107$&$ 84$&&&$ $\tabularnewline*
~~DClust75&HAH $ J000146.0-544026.2$&&SPT-CL $J0001-5440$&&$4444800055$&$ 89$&&&$ $\tabularnewline*
~~DClust76&HAH $ J002357.7-244154.3$&&&&$3469200139$&$ 84$&&&$ $\tabularnewline*
~~DClust77&HAH $ J003501.7+023358.6$&&MACS $J0034.9+0234$&&$2343000008$&$ 64$&&RM $J003457.8+023331.8$&$ 64$\tabularnewline*
~~DClust78&HAH $ J013726.6-082747.9$&&WHL $J013719.8-082841$&&$2804900125$&$ 78$&&RM $J013725.0-082722.7$&$ 35$\tabularnewline*
~~DClust79&HAH $ J015246.7+360750.5$&& NGC $708$&&$ $&$ $&&&$ $\tabularnewline*
~~DClust80&HAH $ J015317.2-135220.6$&&&&$3035600139$&$ 99$&&&$ $\tabularnewline*
\newpage
~~DClust81&HAH $ J023726.4-262957.6$&&RXC $J0237.4-2630$&&$3548400119$&$ 30$&&&$ $\tabularnewline*
~~DClust82&HAH $ J024336.7-483319.4$&&SPT-CL $J0243-4833$&&$4279300045$&$ 50$&&&$ $\tabularnewline*
~~DClust83&HAH $ J030901.7+264522.2$&&MACS $J0308.9+2645$&&$ $&$ $&&&$ $\tabularnewline*
~~DClust84&HAH $ J042905.4-101206.7$&&&&$2884800097$&$ 71$&&&$ $\tabularnewline*
~~DClust85&HAH $ J043916.2-460115.2$&&SMACS $J0439.2-4600$&&$4211700086$&$ 25$&&&$ $\tabularnewline*
~~DClust86&HAH $ J045159.3+000645.9$&&MACS $J0451.9+0006$&&$2449900135$&$ 66$&&&$ $\tabularnewline*
~~DClust87&HAH $ J051537.3-432514.2$&&&&$4139400014$&$ 36$&&&$ $\tabularnewline*
~~DClust88&HAH $ J051856.3-431439.1$&&&&$4114300083$&$ 60$&&&$ $\tabularnewline*
~~DClust89&HAH $ J051856.4-431444.4$&&&&$4114300083$&$ 60$&&&$ $\tabularnewline*
~~DClust90&HAH $ J052116.2-510409.5$&&SPT-CL $J0521-5104$&&$4368600053$&$ 42$&&&$ $\tabularnewline*
~~DClust91&HAH $ J052214.2-362408.4$&&&&$3909700155$&$ 34$&&&$ $\tabularnewline*
~~DClust92&HAH $ J065327.1-574306.1$&&SPT-CL $J0653-5744$&&$ $&$ $&&&$ $\tabularnewline*
~~DClust93&HAH $ J071718.8+374842.3$&&MACS $J071718+374841$&&$ $&$ $&&&$ $\tabularnewline*
~~DClust94&HAH $ J071731.9+374449.9$&&&&$ 942200072$&$ 56$&&&$ $\tabularnewline*
~~DClust95&HAH $ J074704.6+693824.0$&&&&$ 154800014$&$ 33$&&&$ $\tabularnewline*
~~DClust96&HAH $ J080911.4+064341.9$&&&&$2171900062$&$158$&&&$ $\tabularnewline*
~~DClust97&HAH $ J084349.0+411635.4$&&&&$ 841300099$&$ 63$&&&$ $\tabularnewline*
~~DClust98&HAH $ J084959.8+360340.6$&&&&$ 995000060$&$103$&&RM $J085007.9+360413.7$&$103$\tabularnewline*
~~DClust99&HAH $ J091023.7+021049.2$&&&&$2377800058$&$ 84$&&&$ $\tabularnewline*
~~DClust100&HAH $ J091044.5+424945.8$&&&&$ 790800062$&$ 36$&&&$ $\tabularnewline*
~~DClust101&HAH $ J092141.2-062241.6$&&&&$2736400046$&$119$&&&$ $\tabularnewline*
~~DClust102&HAH $ J094058.7+074541.1$&&MACS $J0940$ ACS&&$2122300012$&$118$&&RM $J094053.7+074425.4$&$107$\tabularnewline*
~~DClust103&HAH $ J100050.2+013251.5$&&&&$2378700125$&$131$&&&$ $\tabularnewline*
~~DClust104&HAH $ J100142.8+015448.1$&&&&$2378700108$&$ 99$&&&$ $\tabularnewline*
~~DClust105&HAH $ J111554.8+013015.5$&&MACS $J1115.8+0129$&&$2380000092$&$ 49$&&&$ $\tabularnewline*
~~DClust106&HAH $ J113305.2+500840.9$&&MACS $J1133.2+5008$&&$ 561400014$&$ 31$&&RM $J113313.2+500840.5$&$ 76$\tabularnewline*
~~DClust107&HAH $ J115054.9-280548.7$&&PLCK ESZ $G287.0+32.9$&&$ $&$ $&&&$ $\tabularnewline*
~~DClust108&HAH $ J130154.4+275347.0$&&&&$1306700027$&$105$&&&$ $\tabularnewline*
~~DClust109&HAH $ J130236.3+141055.0$&&&&$1869900048$&$112$&&&$ $\tabularnewline*
~~DClust110&HAH $ J133422.4+503133.0$&&RX $J1334.5+5030$ POS1&&$ 563100033$&$ 34$&&&$ $\tabularnewline*
~~DClust111&HAH $ J134711.8-114157.8$&&RX $J1347-1145$ ACS&&$ $&$ $&&&$ $\tabularnewline*
~~DClust112&HAH $ J142952.3+554752.8$&&&&$ 425100125$&$106$&&&$ $\tabularnewline*
~~DClust113&HAH $ J143808.8+341940.3$&&&&$1078000014$&$143$&&&$ $\tabularnewline*
~~DClust114&HAH $ J144727.3+083001.9$&&MACS $J1447.4+0827$&&$2102100021$&$ 88$&&RM $J144726.0+082825.1$&$ 99$\tabularnewline*
~~DClust115&HAH $ J151832.6-813031.8$&&PLCK ESZ $G308.3-20.2$&&$ $&$ $&&&$ $\tabularnewline*
~~DClust116&HAH $ J161543.8-060959.7$&&Abell $2163$&&$ $&$ $&&&$ $\tabularnewline*
~~DClust117&HAH $ J173142.8+225101.0$&&MACS $J1731.6+2252$&&$1516300115$&$ 87$&&&$ $\tabularnewline*
~~DClust118&HAH $ J173647.0+461847.1$&&&&$ 678200065$&$  0$&&&$ $\tabularnewline*
~~DClust119&HAH $ J175151.7+443859.0$&&MACS $J1752.0+4440$&&$ 750100068$&$ 78$&&&$ $\tabularnewline*
~~DClust120&HAH $ J221450.1-140049.7$&&MACS $J2214-1359$&&$3057300114$&$117$&&&$ $\tabularnewline*
~~DClust121&HAH $ J224154.3+173336.5$&&MACS $J2241.8+1732$&&$1726600046$&$ 86$&&RM $J224158.1+173303.7$&$ 63$\tabularnewline*
~~DClust122&HAH $ J224901.9-443103.9$&&&&$4206000045$&$167$&&&$ $\tabularnewline*
~~DClust123&HAH $ J230104.7+300726.4$&&&&$1240500156$&$ 73$&&&$ $\tabularnewline*
~~DClust124&HAH $ J234232.8-541329.7$&&SPT-CL $J2342-5411$&&$ $&$ $&&&$ $\tabularnewline*
\hline
{\bfseries Rediscovery group}&&&&&&&&&\tabularnewline*
\tabularnewline
~~RClust125&HAH $ J001426.3-302255.3$&&&&$3673800147$&$109$&&&$ $\tabularnewline*
~~RClust126&HAH $ J015242.4-135618.2$&&&&$3035600098$&$ 64$&&&$ $\tabularnewline*
~~RClust127&HAH $ J032940.8-021318.5$&&&&$2550900117$&$118$&&&$ $\tabularnewline*
~~RClust128&HAH $ J033304.4-065144.3$&&&&$2730200015$&$ 38$&&&$ $\tabularnewline*
~~RClust129&HAH $ J041604.1-240522.2$&&&&$3473300094$&$120$&&&$ $\tabularnewline*
~~RClust130&HAH $ J045400.6-030833.8$&&&&$2578000063$&$118$&&&$ $\tabularnewline*
~~RClust131&HAH $ J045413.1+025733.8$&&Abell $520$ P1&&$ $&$ $&&&$ $\tabularnewline*
~~RClust132&HAH $ J074450.9+392736.0$&&&&$ 891500113$&$ 54$&&&$ $\tabularnewline*
~~RClust133&HAH $ J091617.5-002405.9$&&MACS $J0916.1-0023$&&$2480300042$&$110$&&RM $J091617.6-002405.7$&$  0$\tabularnewline*
~~RClust134&HAH $ J100012.6+022015.8$&&&&$2353000151$&$104$&&&$ $\tabularnewline*
~~RClust135&HAH $ J100018.5+023845.6$&&&&$2353000132$&$ 98$&&&$ $\tabularnewline*
~~RClust136&HAH $ J100133.7+022135.3$&&&&$2353000126$&$140$&&&$ $\tabularnewline*
~~RClust137&HAH $ J133241.0+503346.3$&&&&$ 563100075$&$ 87$&&RM $J133238.4+503336.0$&$ 26$\tabularnewline*
~~RClust138&HAH $ J135409.1+771557.2$&&MACS $J1354.6+7715$&&$ $&$ $&&&$ $\tabularnewline*
~~RClust139&HAH $ J135409.2+771549.9$&&MACS $J1354.6+7715$&&$  58300091$&$ 50$&&&$ $\tabularnewline*
~~RClust140&HAH $ J145250.0+580135.3$&&MACS $J1452.9+5802$&&$ 371800109$&$112$&&&$ $\tabularnewline*
~~RClust141&HAH $ J152745.2+065220.8$&&&&$2179700063$&$ 16$&&RM $J152745.8+065233.6$&$ 16$\tabularnewline*
~~RClust142&HAH $ J213512.7-010143.9$&&&&$2519000027$&$169$&&RM $J213512.2-010255.8$&$ 72$\tabularnewline*
~~RClust143&HAH $ J221142.0-035052.4$&&&&$2622000023$&$115$&&RM $J221145.9-034944.5$&$ 90$\tabularnewline*
~~RClust144&HAH $ J224851.7-443105.1$&&&&$4206000045$&$ 58$&&&$ $\tabularnewline*
~~RClust145&HAH $ J001423.0-302109.7$&&&&$3673800147$&$109$&&&$ $\tabularnewline*
~~RClust146&HAH $ J095939.1+023044.1$&&&&$2353000051$&$ 48$&&&$ $\tabularnewline*
~~RClust147&HAH $ J100013.9+022249.8$&&&&$2353000156$&$124$&&&$ $\tabularnewline*
~~RClust148&HAH $ J100047.7+015023.3$&&&&$2378700136$&$ 94$&&&$ $\tabularnewline*
~~RClust149&HAH $ J100055.7+013821.1$&&&&$2378700129$&$169$&&&$ $\tabularnewline*
~~RClust150&HAH $ J130042.7+280523.6$&&&&$1306700012$&$111$&&&$ $\tabularnewline*
~~RClust151&HAH $ J230824.7-021213.3$&&&&$2571800113$&$ 28$&&RM $J230822.2-021131.8$&$ 56$\tabularnewline*
~~RClust152&HAH $ J013152.3-133658.3$&&&&$3009600047$&$ 58$&&&$ $\tabularnewline*
~~RClust153&HAH $ J071815.0+374200.3$&&&&$ 942200005$&$ 24$&&&$ $\tabularnewline*
~~RClust154&HAH $ J114934.3+222440.5$&&&&$1510200105$&$ 63$&&RM $J114935.7+222354.6$&$ 50$\tabularnewline*
~~RClust155&HAH $ J173217.5+193250.3$&&&&$1644300126$&$114$&&&$ $\tabularnewline*
~~RClust156&HAH $ J203146.3-403705.4$&&&&$4079500042$&$96$&&&
\tabularnewline
\end{longtable}
{\centering {\bf Notes.} Candidate clusters from Tables~\ref{tab:unpublished} and~\ref{tab:rediscoveries}. As the presence of nearby clusters could provoke excess lensing substructures, we report the confirmed and targeted clusters from the HST observations, as well as candidate clusters identified by cross-matching within a 3\arcmin radius with DESI \citep{Zou2021} and redMaPPer on SDSS \citep{Rykoff2014, Rykoff2016} catalogues. }
}

\newpage
\section{\texttt{GALFIT} modelling of light profiles of lenses}

\begin{figure*}[!htb]
    \centering
    \includegraphics[width=0.93\textwidth]{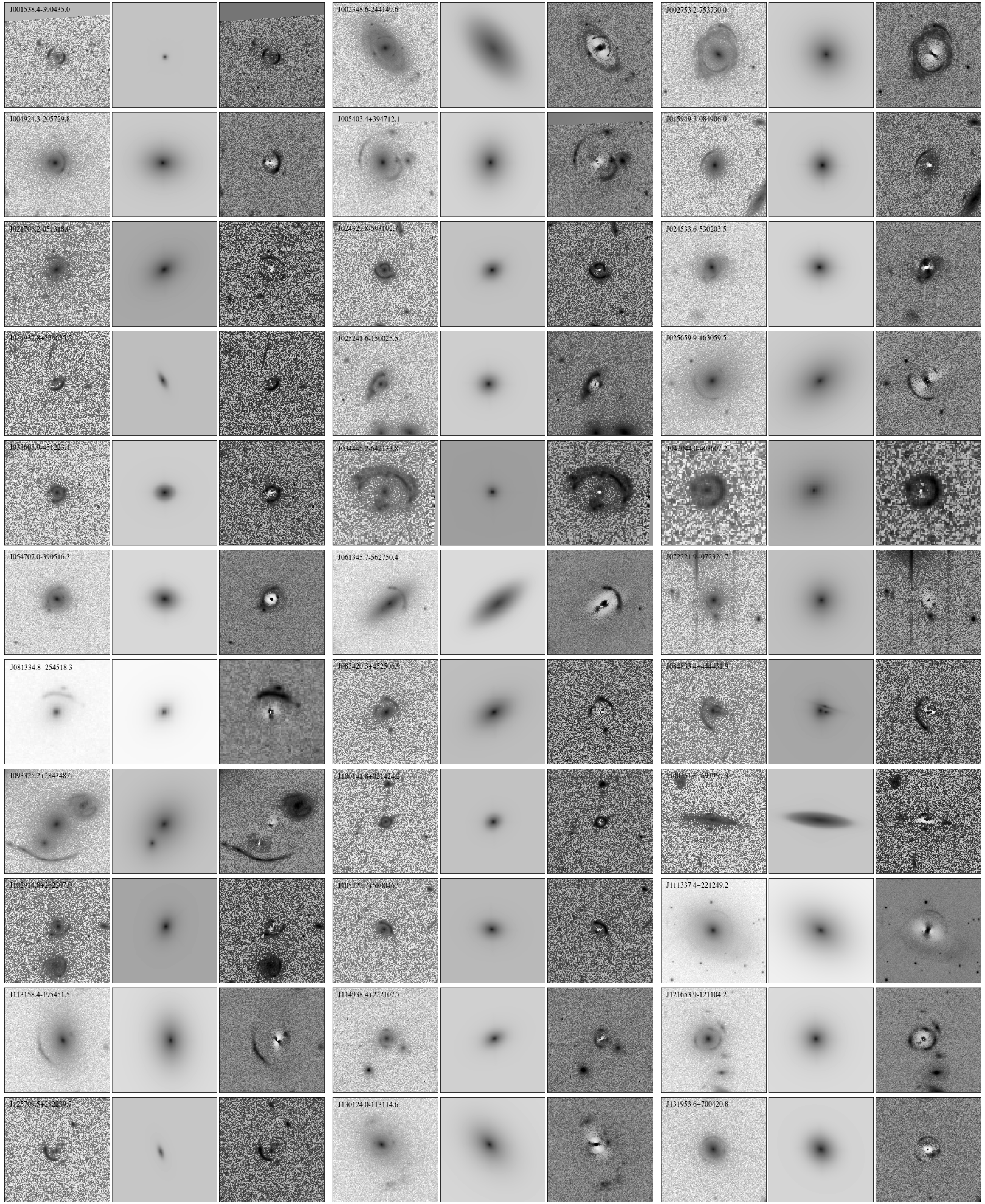}
    \caption{Postage stamps showing the lens galaxy fitting with \texttt{GALFIT} for the newly-discovered HAH grade A candidates. We present three lens candidates per line, where each HAH lens is indicated by the sequence part of the name. Each postage stamp is composed by (1) the original image used for the fit, using the HST instruments and filters indicated in Tables~\ref{tab:unpublished} and~\ref{tab:rediscoveries}, (2) the fitted model using chi-square minimisation in \texttt{GALFIT}, (3) the residuals obtained by subtracting the fitted model from the original image. The arcs and background galaxies were masked using \texttt{SExtractor} during the fitting process, such that only the light of the lens was fitted. The postage stamps show the default orientation of the HST images, which can be different from the one shown in Figure~\ref{fig:grade_A}.}
    \label{fig:galfitB1}
\end{figure*}

\begin{figure*}[!htb]
    \centering
    \includegraphics[width=0.93\linewidth]{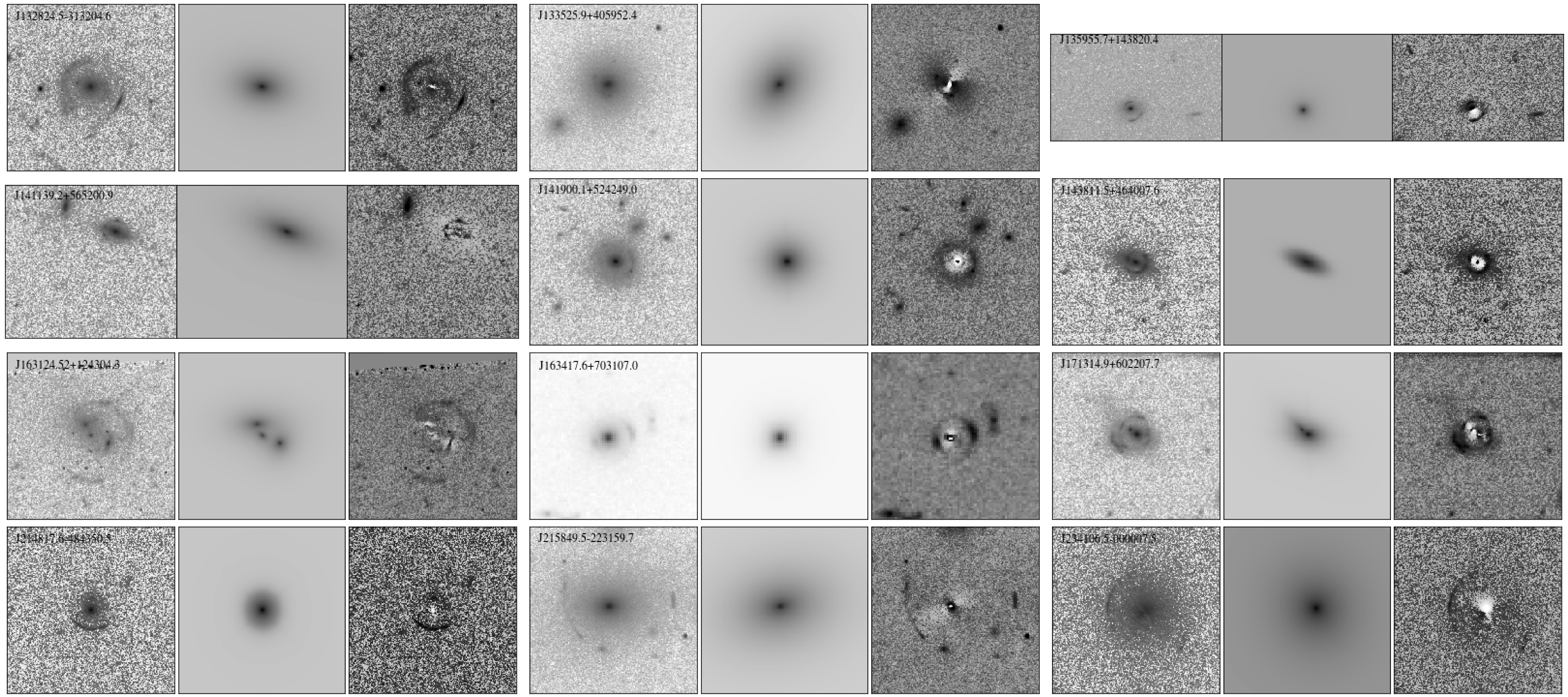}
    \caption{\em continued.} 
    \label{fig:galfitB2}
\end{figure*}

\begin{figure*}[h]
    \centering
    \includegraphics[width=\textwidth]{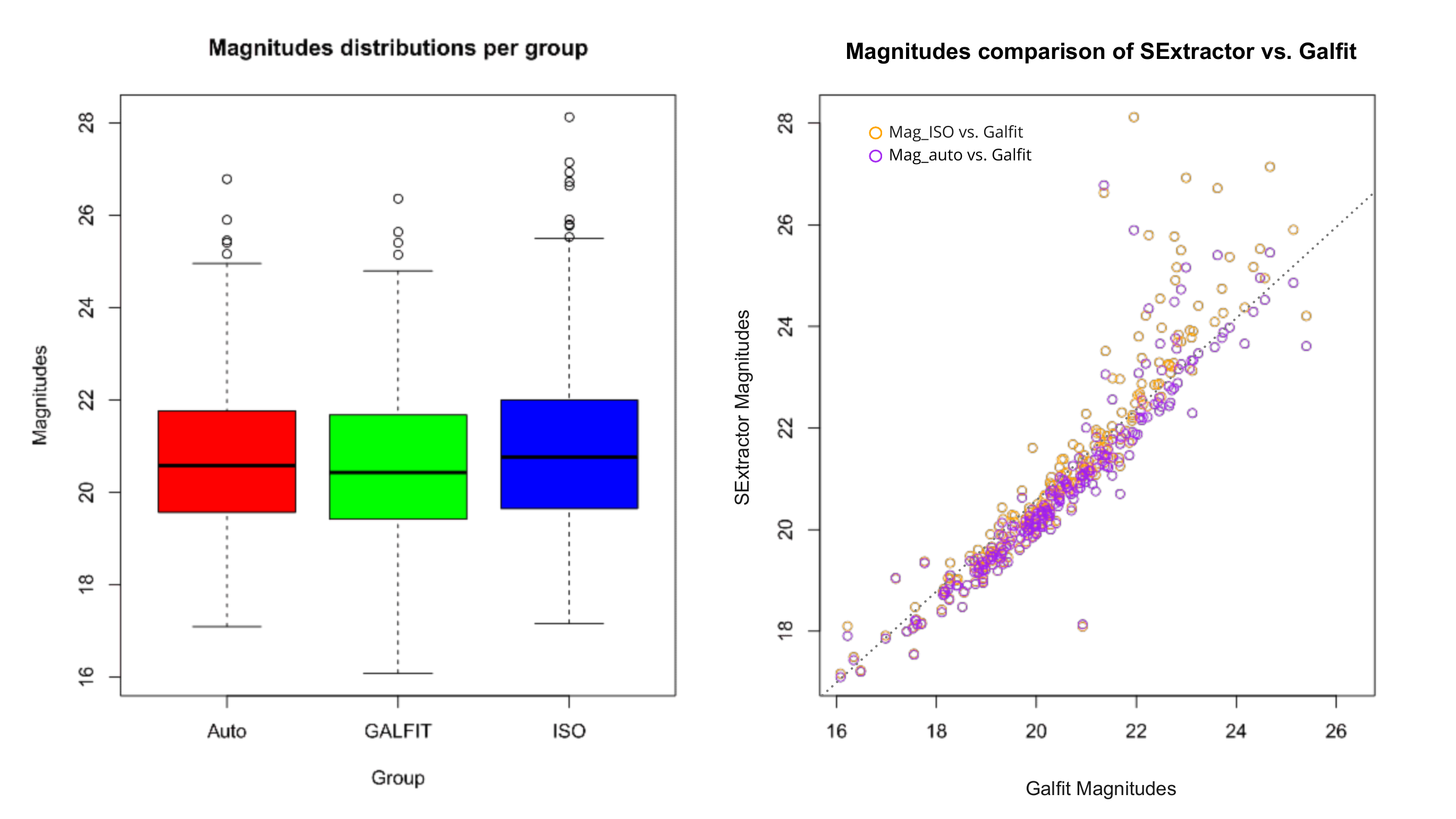}
    \caption{{\it Left:} Boxplots representing the distributions of the magnitudes measured with {\tt Galfit} and the Auto and ISO magnitudes measured with {\tt SExtractor}. For each boxplot, the middle line in the box represent the median of the distribution. The box itself extends from the lower quartile to the upper quartile, covering the interquartile range. The points which fall outside the whiskers situated on each side of the box are considered to be extreme values. {\it Right:} Comparison between the magnitudes measured with {\tt Galfit} (abscissa) against the 'ISO' and 'Auto' magnitudes (ordinates) from {\tt SExtractor}. The grey dotted line is a 45° guide line. These two plots do not show systematic differences in the photometry of the foreground lens galaxies between the different measurements.}
    \label{fig:compare_magnitudes}
\end{figure*}

\clearpage
\section{Additional lens candidates}
\label{recent}

\begin{figure}[!htb]
 \centering
 \includegraphics[width=0.5\columnwidth]{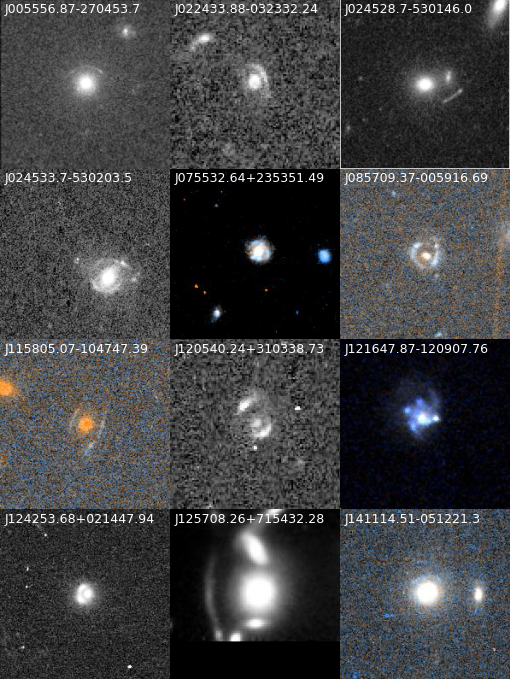}
 \caption{Serendipitous, non-targeted lens candidates identified after the end of the Hubble Asteroid Hunter citizen science project, not included in the main analysis of the paper. The HAH lenses from the mosaics are indicated using only the sequence part of their name.}
 \label{fig:recent}
\end{figure}

After the end of the Hubble Asteroid Hunter citizen science project, we found twelve serendipitous non-targeted lens candidates from recent imaging observations released in the HST archives after June 2020 (Table~\ref{tab:recent} and Fig.~\ref{fig:recent}). These candidates were identified by citizen scientist Claude Cornen, by inspecting the 'Daily Data Reports'\footnote{\url{https://archive.stsci.edu/hst/daily/}}, while discarding calibrations as well as observations of asteroids, comets, stars or clusters. 

\begin{table*}[!htb]
\caption{New strong lens candidates identified after the end of the Hubble Asteroid Hunter citizen science project.}
\small
\begin{center}
\begin{tabular}{llllllll}
\hline
\hline
Candidate (IAU name)            & Obs. id   & Obs. date & Instrument & Apertures & Filters & Proposal id & Grade \\
\hline
\tabularnewline
HAH J005556.9-270453.7   & ie6jy6010 & 2020-06-01  & WFC3       & IR        & F140W            & 15347   & B \\
HAH J022433.88-032332.24 & jeco01010 & 2021-09-15  & ACS       & WFC   & F475W            & 16300   & C \\
HAH J024528.7-530146.0   & iepl29020 & 2022-02-04  & WFC3       & UVIS      & F606W            & 16729   & C \\
HAH J024533.7-530203.5   & iepl29020 & 2022-02-04  & WFC3       & UVIS      & F606W            & 16729   & A \\
HAH J075532.64+235351.49 & jece46020 & 2021-03-12  & ACS        & WFC      & F814W    & 16257   & B \\
HAH J085709.37-005916.69 & jea302010 & 2021-04-07  & ACS        & WFC      & F814W    & 16186  & A  \\
HAH J115805.07-104747.39 & jeh903010 & 2021-04-08  & ACS        & WFC  & F814W    & 16429  & A  \\
HAH J120540.24+310338.7  & jdrz1h010 & 2021-05-26  & ACS        & WFC       & F606W    & 15446  & A \\
HAH J121647.87-120907.76 & jc8h07020 & 2014-06-05  & ACS        & WFC  & F814W    & 13393  & C \\
HAH J124253.68+021447.94 & jece02010 & 2021-04-12  & ACS        & WFC      & F606W    & 16257  & B \\
HAH J125708.26+715432.28 & ieer07010 & 2021-01-02  & WFC3       & IR        & F110W            & 16243  & B \\
HAH J141114.51-051221.3  & jece78020 & 2021-04-18  & ACS        & WFC      & F814W    & 16257  & C \\
\tabularnewline
\hline
\end{tabular}
\end{center}
{\small {\bf Notes.} IAU names, observing programs, HST instruments and filters corresponding to new strong lens candidates identified after the end of the Hubble Asteroid Hunter citizen science project, from data released in the HST archives after June 2020.}
\label{tab:recent}
\end{table*}

\label{lastpage}
\end{document}